# Why we live in hierarchies: a quantitative treatise

by

Anna Zafeiris[1] and Tamás Vicsek[1, 2]

[1]Statistical and Biological Physics Research Group of HAS
[2]Department of Biological Physics, Eötvös University

# Contents

# 1 Introduction

This book is concerned with the various aspects of hierarchical collective behaviour which is manifested by most complex systems in nature. From the many of the possible topics, we plan to present a selection of those that we think are useful from the point of shedding light from very different directions onto our quite general subject. Our intention is to both present the essential contributions by the existing approaches as well as go significantly beyond the results obtained by traditional methods by applying a more *quantitative* approach then the common ones (there are many books on qualitative interpretations). In addition to considering hierarchy in systems made of similar kinds of units, we shall concentrate on problems involving either dominance relations or the process of collective decision-making from various viewpoints.

## 1.1 General considerations

Since hierarchy is abundant in nature and society, but many of its quantitative aspects are still unexplored, the main goal we intend to achieve is the systematic interpretation and documentation of new unifying principles and basic laws describing the most relevant aspects of hierarchy (being perhaps the most widespread organizing principle in the Universe). To do so we shall discuss recent experiments and models that are both simple and realistic enough to reproduce the observations and develop concepts for a better understanding of the complexity of systems consisting of many organisms. We shall cover systems ranging from flocks of birds to groups of people.

The related research goes beyond being interdisciplinary and can be rather described as multidisciplinary, since it involves many kinds of systems (both living and non-living), various techniques and technologies typically used in different branches of science and engineering. The topics we address might look too diverse. However, one can always think of these research directions as facets of a single, to be explored idea.

Although we shall concentrate on hierarchical collective behaviour in general, there will be two aspects of it which will pop up in the majority of cases: collective motion and dynamically changing partially directed networks (and the natural relation of the two). A few of the many possible examples are visualized in Fig. 1. In addition, we give a brief description of the most relevant concepts which hierarchy is related to.

*Organisms versus agents, entities or "particles"*

Throughout of this book we shall consider systems made of many (from a few dozens to several thousands) organisms, i.e., living entities. Of course, hierarchy is present in the non-living world as well; starting from elementary particles through the solar system up to the whole universe, but that is a beautiful and long story which is not the subject of the present work.

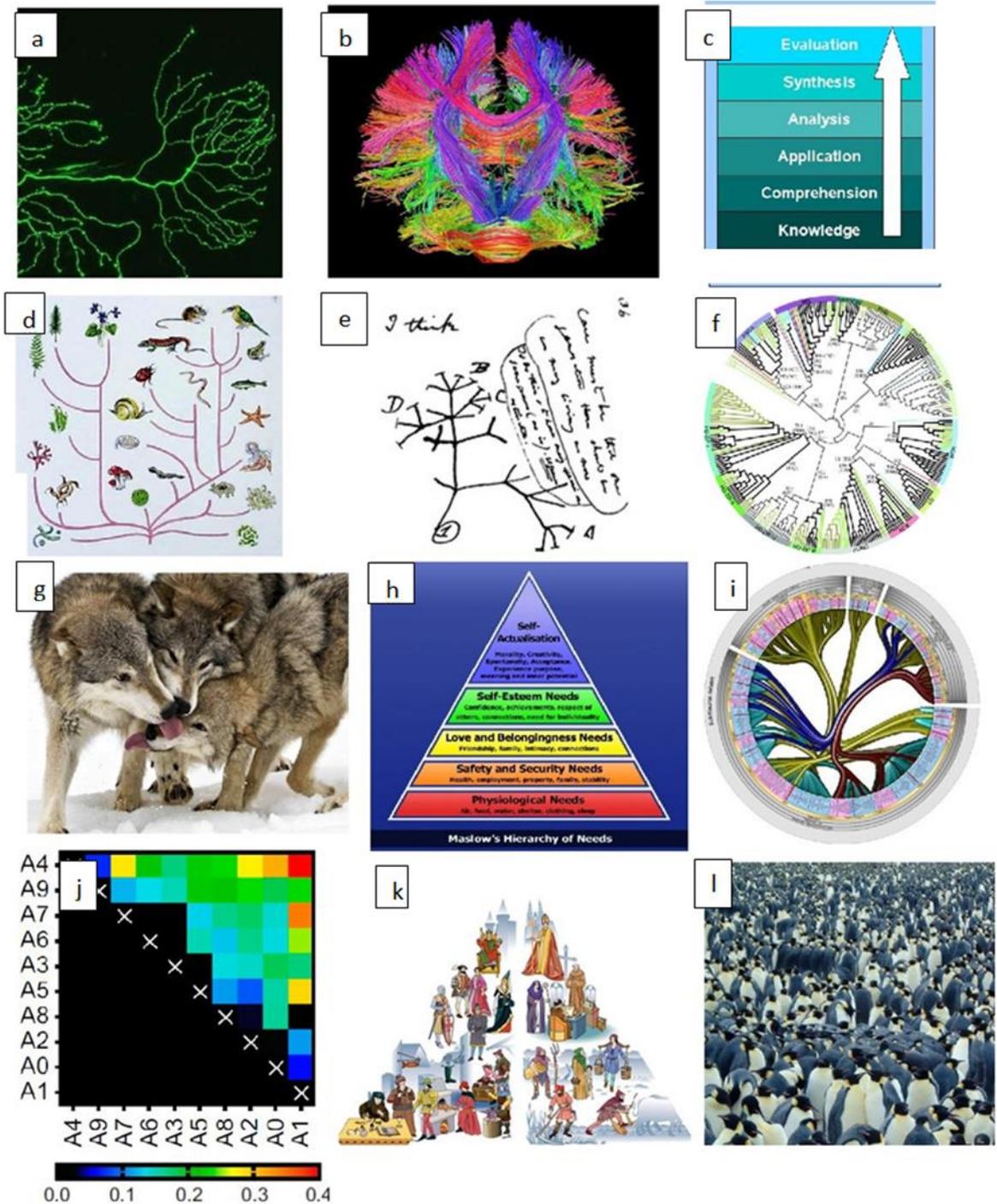

**Fig. 1 a** Axon arborisation (the end part of a major kind of neuronal cells) shows typical hierarchical tree-like structure in space. **b** The wiring of a human brain. Hierarchy is not obvious, but closer inspection and additional MRI images indicate hierarchical functional operation. **c** And this is a possible interpretation of how we think (thoughts being one of the end products of a functioning brain. **d** The visualization (of the by today commonplace) idea of the evolutionary tree. **e** The famous first drawing about the branching of the phylogenetic tree with the "I think" note by Darwin. **f** This complex tree with its hundreds of branches shows the birth of new variants (associated with new plant species) of a single protein! **g** The well-known hierarchy of wolfs indicated by who is licking who (subordinates do this with those above them). The same behaviour can be observed between a dog and her owner. **h** Perhaps the only hierarchy named after a person. This pyramid is called the "Maslov's hierarchy of needs". **i** Visualization of the connections (call relations) between the various parts of a C+ software system (containing many thousands of entities and relations: more closely related parts are colour coded and bundled). **j** The strength of the directional correlations between pairs of pigeons in a flock (individuals being denoted by A0,…,9. The asymmetric structure of the dominant part of the matrix (whole matrix minus its symmetric components) indicated strictly hierarchical leader-follower relations. **k** The picturesque representation of the two pyramids of medieval relations among the member s of a society: left corresponding to social, the right side corresponding to the religious organization. **l** And finally: we show a huge community of relatively simple animals. Where is here the hierarchy? Nowhere, the groups of many thousands of animals (large flocks of birds, schools of fish) typically do not display the signs of hierarchy (and, and, indeed, are assumed not to be hierarchically organized.) (All pictures are freely available from the internet except **j** which is from one of our papers)

Hierarchy in life can be understood in several ways. For example, one may rank a quality as more important than another type of quality. However, in most of the cases hierarchy involves many "units" which are related to each other in relatively simple ways. The stress is on "many" and on "simple". Perhaps the best way to demonstrate this point is to consider a group of people. The interactions (relations) among them can be extremely complicated (just think of two people being in love with each other). Instead of considering such interactions, we assume that two people, let us say, in a large organisation are either working in the same kind of unit or one of them has a job of a leader (of a group, a department, a division, etc.). In this case, we assume that there is a directed link between the two which is pointing from the leader to the regular member of the company. When accounting their relation, this will be the aspect we shall consider and all of the other, extremely complex features of the two persons (they are made of cells, they feel the smell of the other person, etc.) will be neglected.

This is how "particles" can be defined even for a system of people: particles are units whose interactions can be - in the given context (!) - assumed to be very simple.

"Agents" are a bit more complicated than particles. Although their interactions are assumed to be also relatively simple, these units have a "purpose". The purpose is usually also simple and can be interpreted as optimizing/maximizing some sort of advantageous quantity. In its most typical form this quantity is the difference between the "benefit" and the "cost" usually called fitness. Fitness can be defined for a whole group of agents as well.

To summarize the above: *hierarchy is typically defined for systems of agents and can be advantageous to a varying degree*. One of the main messages of our text is that the *main reason for the hierarchical structure of the relations among organism is that such a structure is more advantageous* than a fully regular or a random or any other arrangement.

### Collective behaviour

Collective behaviour applies to a great variety of phenomena in nature, which makes it an extremely useful notion in many contexts. Examples include collectively migrating bacteria, insects or birds; or phenomena where groups of organisms or non-living objects synchronize their signals — think of fireflies flashing in unison or people clapping in phase during rhythmic applause. The main features of collective behaviour are that an individual unit's action is dominated by the influence of its neighbours — the unit behaves differently from the way it would behave on its own. On one hand such systems show interesting ordering phenomena as the units simultaneously change their behaviour to a common pattern (Camazine 2003, Sumpter 2010) and on the other hand can form structures that are capable of exhibiting much more complex functions than a single unit (consider, e.g., a single neuron versus a complete brain).

The world is made of many highly interconnected parts over many scales, whose interactions result in a complex behaviour needing separate interpretation for each level. This realization forces us to appreciate that new features emerge as one goes from one scale to another, so it follows that the science of complexity and the closely related hierarchy is about - following a classification based on major analogies - is expected to reveal the principles governing the ways by which these new properties appear.

Over the past decades, one of the major successes of statistical physics has been the explanation of how certain patterns can arise through the interaction of a large number of similar units. Interestingly, the units themselves can be very complex entities, too, and their internal structure has little influence on the patterns they produce. It is much more the way they interact that determines the large-scale behaviour of the system. It has been found that not only interacting spins or atoms, but also assemblies of molecules or granular particles, and even large groups of complex biological structures (bacteria, ants, birds, etc.) can be examined by statistical physics models (Vicsek 2001). It has been demonstrated that the collective behaviour of units has a number of features typical for many different systems. From the point of statistical physics these could be considered as "universality classes" or major types of behavioural patterns.

It is, however, very important to note that in the above context the hierarchical nature of interactions has been largely neglected, especially for the directed (or asymmetric) case (except a few network theory papers). Our basic assumption is that by observing and *quantitatively* interpreting the patterns of behaviour in hierarchically organized systems is likely to lead to a unified picture of hierarchical collective behaviour, and, in an ideal case, to the discovery of a number of basic relations or "laws" describing them.

### Collective motion

The actions of moving individual organisms add together creating patterns of motion, so complex that they seem to have been choreographed from "above". Flocks and schools have a distinctive style of behaviour - with fluidity and a seeming intelligence that far transcends the abilities of their members. Vast congregations of birds, for example, are capable of turning sharply and suddenly en masse, always avoiding collisions within the flock. It has turned out over the two decades that computer models and sophisticated techniques to collect data about a large number of animals have been very useful for establishing a significantly better understanding of such systems than before (Vicsek and Zafeiris 2012).

### Networks

When "generating" life as we perceive it today, nature "made use of" the existence of the above mentioned hierarchical levels by spontaneously separating them as molecules, macromolecules, cells, organisms, species and societies. The big question is whether there is a unified theory for the ways elements of a system organize themselves to produce such a highly hierarchical structure of behaviour typical for wide classes of systems. Interesting principles have been proposed, including self-organization, simultaneous existence of many degrees of freedom, self-adaptation, rugged energy/fitness landscapes and scaling, etc. Physicists are learning how to build relatively simple models producing complicated behaviour. At the same time researchers working on inherently very complex systems (biologists or economists, say) are uncovering the ways how their infinitely complicated subjects can be interpreted in terms of interacting, well-defined (i.e., simpler) units (such as proteins) with the interactions corresponding to links (which can be directed and weighted) and the units to nodes (having attributes) in a complex network (Albert and Barabási 2002, Newman 2010, Barabási 2016).

Most of the networks in life and technology are dynamically changing and are highly structured. For example, a dynamically changing network can be associated with a flock of collectively moving organisms or robots interacting as a function of their positions.

## *1.2 Motivation*

It is widely accepted that we do not understand deeply enough the reasons behind the abundance of multi-level hierarches. However, there must be an advantage of such an organization, because of the permanent evolution of the corresponding systems preferring more efficient variants. But where is this advantage? Better adaptability? A more efficient, robust or stable structure? A faster spreading of relevant information? Or, perhaps, better controllability (think of, e.g., an army)? On a more abstract level: What are the conditions for a hierarchical organization to emerge? Are there any general (valid for many systems) necessary and/or sufficient condition for this emergence?

These are challenging questions and if we can answer them it could bring us to designing and producing much more efficient devices or perhaps, more importantly, creating much better functioning industrial, educational or many more kinds of organizations.

Motivated by the above reasons, in this book will be centered around topics and answers related to questions like:

*What is our subject?*

We shall consider primarily systems (structures, processes, phenomena) that are common in the living world. The related, practical questions are: what are the conditions under which hierarchy emerges? What kinds of mathematical tools are appropriate for describing the various aspects of hierarchy?

*Why do we study?*

We use a quantitative approach to interpreting realistic situations in life because most of the presently available experimental and theoretical treatments of hierarchical organization are predominantly qualitative so a need arises in presenting results involving numerics. On the other hand, the interest in the topic seems to be increasing quickly. Understanding leadership and further aspects of hierarchy are expected to be very useful from the point of optimizing economy-related structures. On a less applied level, getting a deeper insight into the collective behaviour of groups has also been attracting growing interest.

*How do we study?*

As mentioned above (and explored here in a bit more detail) there can be several methods to treat the various quantitative aspects of hierarchy. First, it is possible – but far from being trivial – to design experiments for studying how a hierarchical set of leader-follower relationship emerges from an originally disordered set of living entities. Second, one can design models and study them either analytically or using computer simulations. The two major quantitative approaches have been: game theory and agent-based modelling. In this book we treat the second alternative, since the game theoretical works we know of allow a less straightforward comparison with actual, real life observations and experiments. A rare but important exception is the very recent book by Boix (2015) delivering an impressive mixture of calculations, facts and ideas to treat large scale (political) hierarchy. Our work, concerned with hierarchies on a smaller scale of groups or collectives can be looked at as complementing the book of Boix.

## 1.3 Hierarchical structures in space and in networks

There exist a few fields in sciences which are closely related to the general notion of hierarchy, but fall beyond the scope of our work (they represent the self-similar aspect of hierarchy). This is mainly so because these areas represent a research field of their own. In addition, in most of the present book we consider hierarchy as a set of related entities, such that the relation between two connected entities is directed (one is, in ways later to be specified, plays a role being superior/leading/embedding etc. considering the other entity). Thus, here we only briefly touch upon the topic of spatially hierarchical objects (called fractals) and undirected (symmetric relations) but still hierarchical networks (called scale free). For further details about such self-similar aspects of hierarchy we suggest that the readers use as a source the following books (Falconer 2003, Feder 1988, Vicsek 1992 - about fractals, and Barábasi 2009, Newman 2010, Newman et al. 2006, Dorogovtsev and Mendez 2003, Pastor-Satorras and Vespignani 2007) – about networks and scale free networks).

*Fractals* are objects for which the topological dimension (the number of independent directions one can move into from a given point of the fractal) is smaller than the dimension of the Euclidean space they can be embedded into. They also possess a self-similar geometry which means that a small part of a fractal has the same statistical features than the whole. Here by the expression "same statistical features" we typically understand that the density correlations are the same. This is equivalent to saying that scaling up (blowing up) a small part of a fractal results in a structure which is statistically identical to the full fractal itself. This is a non-trivial feature and involves the fact that the dimension of the fractals is a non-integer number as opposed to regular objects having dimensions 1, 2 or 3.

Interestingly enough, a large variety of living systems involve fractal geometry in one way or another. As one proceeds from simpler to more complex manifestations of life, it is possible to encounter fractal bacteria colonies (Matsuyama and Matsushita 1993), ant trails (Jun et al. 2003) or the network of blood vessels in higher order organisms described by - among other important features - by the so called allometric scaling laws in biology in general (West et al. 1997) and, in particular, in mammalian metabolism (see, e.g., White and Seymour 2005). Perhaps on the largest scale built by organisms are the cities we live in display fractal-like features as well (Batty and Longley 1994).

The so-called *scale-free networks* can also be considered as manifestations of a self-similar structure. Such a structure is not realized in space but shows up in the specific way the entities of a system are connected to each other. Using the language of network theory, the degree of a node (entity) is the number of edges (connections) this node has leading to its neighbours in the network. The degrees may follow all sorts of distributions, but if this distribution is a power law then the degree distribution is invariant under scaling: a smaller part of the network will possess the same power law distribution as the whole network.

The possible examples for systems which can be characterised in terms of scale-free networks are numerous. Most of these are not assumed to exist in real space. Going from smaller to larger scale, examples include networks corresponding to the interactions among proteins in a cell, then, with a large jump, many human made systems (internet, web pages, airlines, etc.) or the various networks of social interactions (friendships, collaborations, industrial relations, etc.).

There are, however some spatial structures that can be best interpreted in terms of hierarchical networks. Louf et al. (2013) introduced a generic model for the growth of a spatial network based on a general concept of cost-benefit analysis. Their model leads to a wide variety of hierarchical spatial structures (trees) minimizing a conditions-dependent fitness function. The work by Daqing et al. (2011) connects the fractal and the network aspects of a structures by calculating the dimensions of spatially embedded networks.

## *Reference list*

Albert R, Barabási A-L (2002) Statistical mechanics of complex networks. Rev. Mod. Phys. 74(1):47

Barabási A-L (2009) Scale-Free Networks: A Decade and Beyond. Science 325:412-413

Barabási A-L (2016) Network Science. Cambridge Univ Press, Cambridge

Batty M, Longley P (1994) Fractal Cities: A Geometry of Form and Function. Academic Press, Cambridge

Boix C (2015) Political Order and Inequality: Their Foundations and their Consequences for Human Welfare (Cambridge Studies in Comparative Politics). Cambridge Univ Press, New York

Camazine S, Deneubourg J-L, Franks N et al (2001) Self-Organization in Biological Systems. Princeton Univ Press, Princeton

Daqing L, Kosmidis K, Bunde A et al (2011) Dimension of spatially embedded networks. Nat Phys 7:481-484

Dorogovtsev SN, Mendes JFF (2003) Evolution of Networks: From Biological Nets to the Internet and WWW (Physics). Oxford Univ. Press, New York

Falconer K (2003) Fractal Geometry: Mathematical Foundations and Applications. Wiley, Chichester

Feder J (1988) Fractals (Physics of Solids and Liquids). Springer, New York

Jun J, Pepper JW, Savage VM et al (2003) Allometric scaling of ant foraging trail networks. Evol Ecol Res 5:297-303

Louf R, Jensen P, Barthelemy M (2013) Emergence of hierarchy in cost-driven growth of spatial networks. PNAS 110(22):8824-8829

Matsuyama T, Matsushita M (1993) Fractal morphogenesis by a bacterial cell population. Crit Rev Microbiol 19(2):117-35

Newman MEJ (2010) Networks: An Introduction. Oxford Univ Press, Oxford

Newman M, Barabási A-L, Watts DJ (2006) The Structure and Dynamics of Networks: (Princeton Studies in Complexity). Princeton Univ. Press, New Jersey

Pastor-Satorras R, Vespignani A (2007) Evolution and structure of the Internet: A statistical physics approach. Cambridge Univ Press, Cambridge

Sumpter DJ (2010) Collective Animal Behavior. Princeton Univ Press, Princeton

Vicsek T (2001) A question of scale. Nature 411(6836):421-421

Vicsek T (1992) Fractal Growth Phenomena, 2nd edn. World Scientific, Singapore

Vicsek T, Zafeiris A (2012) Collective motion. Phys Rep 517(3-4):71-140

West GB, Brown JH, Enquist BJ (1997) A General Model for the Origin of Allometric Scaling Laws in Biology. Science 276(5309):122-126. doi: 10.1126/science.276.5309.122

White CR, Seymour RS. (2005) Allometric scaling of mammalian metabolism. J Exp Biol 208: 1611-9

# 2 Definitions and Basic Concepts

As we indicated in the introduction, the notion of hierarchy applies to a great variety of topics and contexts, let it be the social structure of animal groups, human virtues, psychological needs or the structure of a computer program. Accordingly, it does not have a compact, precise, widely accepted definition that would be applicable for all cases. Available definitions usually by-pass the problem of clarification by using synonymous words – which are, unfortunately, similarly unclear. For example, according to the Cambridge dictionary, hierarchy is "a system in which people or things are arranged according to their importance." Here "importance" is the keyword, but importance is highly subjective: something that is important in a given context might not be important at all from another point of view. Here we also find that hierarchy corresponds to "the people in the upper levels of an organization who control it". So we learn that it is about *control*, but according to this definition, hierarchy is restricted to people in an organization – which is a very narrow interpretation. Checking a very popular cite, Wikipedia, we find that "A hierarchy (from the Greek hierarchia, "rule of a high priest", from hierarches, "leader of sacred rites") is an arrangement of items (objects, names, values, categories, etc.) in which the items are represented as being "above," "below," or "at the same level as" one another". However, this interpretation does not inform us about the basic aspects of the arrangement, which represent, on the other hand, the heart of the problem.

As we shall see, it turns out from more strict investigations that usually we talk about hierarchy if *entities of a system can be classified into levels in a way that elements of a higher level **determine or constrain the behaviour and/or characteristics** of entities in a lower level*. That is, in the heart of hierarchy we find control of behaviour.

**Definition**: A system is hierarchical if it has elements (or subsystems) that are in dominant-subordinate relation with each other. *A unit is **dominant** over another unit to the extent of its ability to influence behaviour of the other. In this relation, the latter unit is called **subordinate***.

**Fig. 2.1** An example for flow hierarchy. The feeding-queuing hierarchical structure of a pigeon flock. Each square represents an individual. The edges point from the higher ranked bird towards the subordinate one with edge widths corresponding to the ability to influence the behaviour of the lower ranked individual. For the sake of better visibility, higher ranked notes are depicted higher on the picture. Reproduced from Nagy et al. (2013).

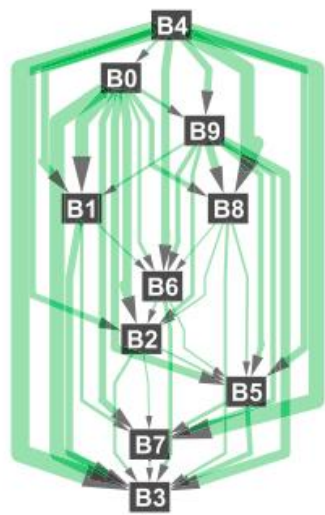

A typical hierarchical structure can be seen in Fig. 2.1 depicting the ranks within a pigeon flock. The inner structure of the group has been established by observing and measuring the feeding-queuing behaviour of its members (Nagy et al. 2013)

Note that this definition does not tell how hierarchical the *system* is. Instead, it expresses whether its elements (or subsystems) are in hierarchical relation or not (manifesting itself in a dominant-subordinate relation). Furthermore, it tells the origin (reason) and extent of the dominant-subordinate relation. Consider for example the Rock–paper–scissors game. According to the rules,

- The rock blunts the scissors (and hence "disarms" it, beats it)
- The scissors cut the paper, and
- The paper covers the stone.

Figure 2.2 shows how the elements overpower each other. Based on the above definition, the hierarchical (dominant-subordinate) relation among the units is clear, but the hierarchical nature of the whole system is not: is this network hierarchical at all?

**Fig. 2.2** The graph representation of the rock-paper-scissors game. The dominant-subordinate relationship among the elements is clear, but the hierarchical nature of the entire system is not.

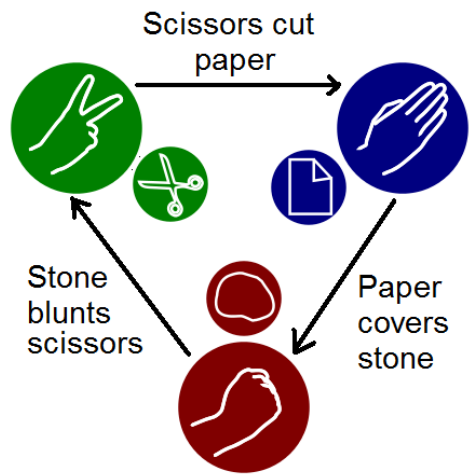

In other words, from a graph-theoretical point of view, the above definition gives a lead regarding the *arrows* (where they should be and what is their deeper meaning) but it does not tell us how hierarchical the entire system is. At this point, we choose to keep it this way, mainly because the extent of hierarchy within a system has subjective aspects: for some, the rock-paper-scissors game is "fully" hierarchical, since its elements are clearly in hierarchical relation. For others it is not, because no source (leader) can be determined.

Many approaches have been proposed to measure the hierarchy of a network, but none of them is "universal", or accepted by everyone for all cases. Sect. 2.1.2 "Measuring the level of hierarchy", gives an overview of these measures and algorithms.

A few comments related to the definition:

- During different group activities the influence of the members might vary. In other words, *hierarchy is context/task sensitive*, even in the same group! For example, as we shall see it in Sect. 3.1.3, "Leadership versus dominance", the members of the same pigeon flock arrange themselves into different hierarchies according to the actual activity: when they feed, the ranks are entirely different from the ones that can be observed during flight. This phenomenon is even more expressed in human groups.

- Hierarchy might vary over time. As certain characteristics of the group members change over time (for example the physical strength of the individuals in a pack of wolves) so do their ranks.

- This definition implies that the units *behave* somehow, or have some observable characteristics. In other words, entities without observable behaviour or characteristics cannot form a hierarchical structure.

- The influence can be either forced by the higher ranked individual (e.g., when a higher ranked pigeon does not let a lower ranked one near to the food source), or it can be voluntary (for example leader-follower relationships during flight).

- A higher ranked unit, by influencing the behaviour of other units more extensively, has a larger effect on the collective (emergent) group behaviour as well.

Hierarchical systems can by classified into the following subtypes:

1. **Order** hierarchy is basically an ordered set, in which a value is assigned to each element characterizing one of its arbitrarily chosen features. This assigned value defines the rank of the entity within the hierarchy. An example for this can be the ranking of artists, e.g. painters or sculptors, based on the average price of their artworks. In this example the "set" is composed by the artists, and the feature is the average price of their artwork. Another example can be a hierarchy of firms, ordered by, say, the number of employees. In this type, the network behind the system is neglected or it does not exist. More formally, this type of hierarchy is "equivalent to an ordering induced by the values of a variable defined on some set of elements" (Lane 2006).

2. **Nested** (or **embedded, containment, inclusive**) hierarchy is a structure in which entities are embedded into each other. Higher level entities consist of and contain lower level entities, or, as Wimberley (2009) has formulated it, "larger and more complex systems consist of and are dependent upon simpler systems and essential system-component entities". (According to some categorizations, a *nested* hierarchy can contain only one entity at each lower level, a bit like in case of the Russian Matryoshka dolls, while a *generalized* nested hierarchy allows multiple objects.) Uncovering nested hierarchy structure within a system is closely related to community detection in graphs. Containment hierarchy has two sub types:
   - A *subsumptive* **containment** hierarchy (a.k.a. *taxonomic hierarchy*) is a structure in which items are classified from specific to general. For example domestic cats, lions, tigers and cheetahs (gepards) belong to the family of cats called "*Felidae*", dogs, foxes and wolfs belong to the family of carnivorans a.k.a. "*Canidae*",

*Canidae* and *Felidae* both belong to the order of *Carnivora*, etc (See Fig. 2.3 **a**). Entities are containers, containing other containers.

Mathematically this arrangement can be formulated as:

Foxes ⊂ Canidae ⊂ Carnivora (and Carnivora ⊂ Mammals ⊂ Animals, to go further on). Each entity in a lower level "is an" entity of a higher level: a fox "is a" Canidae, a Canidae "is a" Carnivora, a fox "is a" mammal, etc. It is assumed that entities on a lower level are proper (or strict) subsets of the entities on a higher level.

- *Compositional* **containment** hierarchy (a.k.a. *level hierarchy*) describes how a system is composed of subsystems, which are also composed of subsystems, etc. The "hierarchy of life" is the best example for this structure, describing how organisms are composed of organ systems, which are composed of organs, which are composed of tissues, which are composed of cells, etc., see Fig. 2.3 **b**. Two important features often (but not always) appear in this type of hierarchy: firstly, there is a "scalar quality", meaning that entities on higher levels are often bigger than entities on lower levels (a cell is bigger than a molecule). Secondly, emergent properties – properties that are not present on lower levels, but due to interactions among the units, appear on higher levels – also often accompany this structure. For example consciousness appears on the level of the brain (which is an organ), but it originates from the interactions of the neuron cells. Emergent properties are of prime importance, since they are a fundamental characteristic of "complex systems".

3. **Flow** (or **control**) hierarchy: "intuitively" it is an acyclic, directed graph. The nodes are layered into levels in a way that nodes on higher levels influence nodes on lower levels, and the influence is represented by edges. Layers refer to power, that is, an entity on a higher level gives orders or passes on information to entities on lower levels. This is where the name is coming from: such a structure represents the flow of orders, or, equivalently, how entities control other entities. Armies, churches, schools, political parties and institutions are typically organized in this way. Downwards orders flow on the edges, upwards pointing edges correspond to requests or sending information. Technological systems are also often organized in this way. In this case a central unit controls devices which control lower level devices, etc. At the bottom-most level sensors do not control anything directly, but they send information upwards, which are used to refine the decision making process done by devices on higher levels. (See Fig. 2.1)

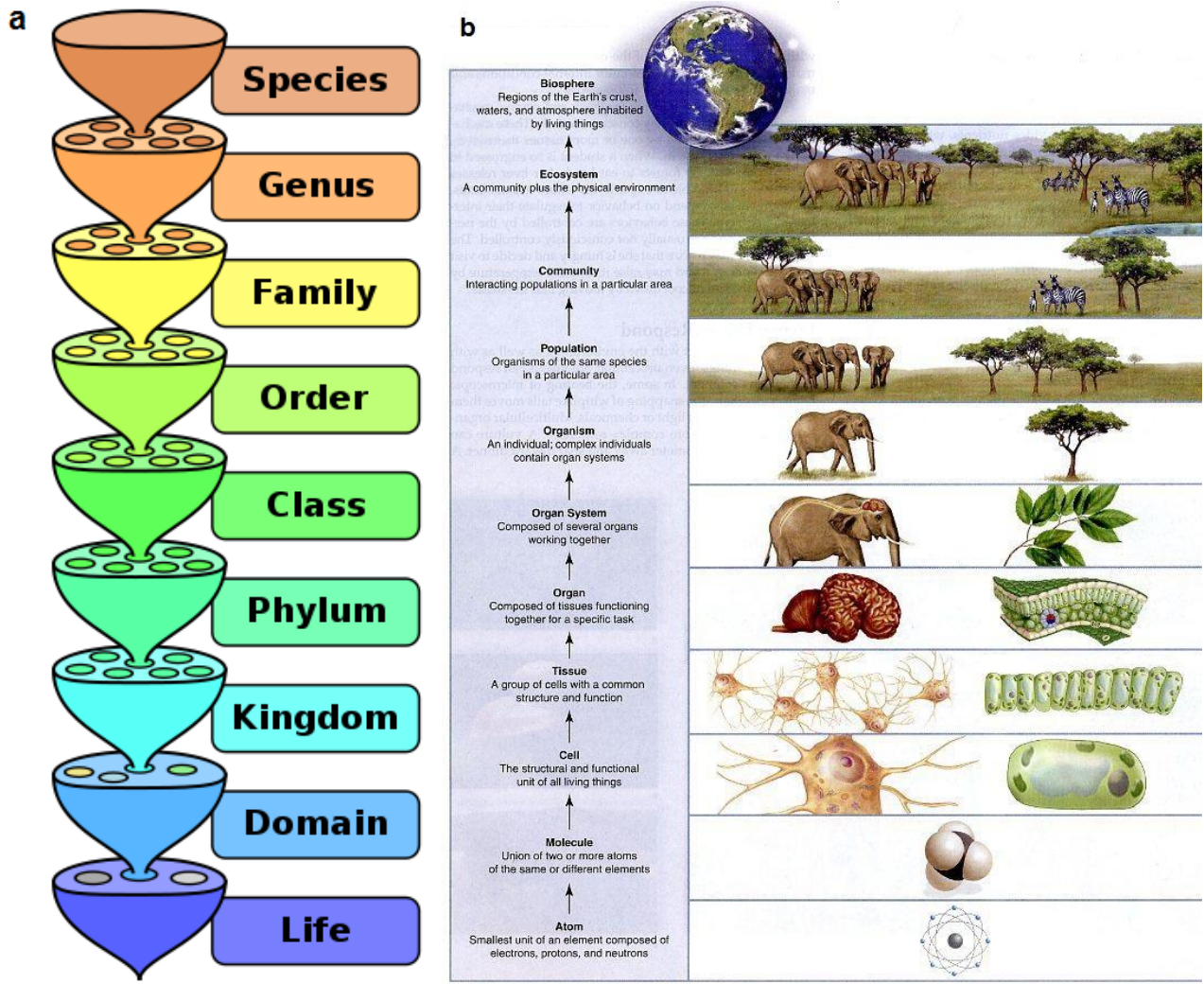

**Fig. 2.3** The two types of containment hierarchies: "taxonomic" and "compositional". **a** In a *taxonomic* (or *subsumptive*) containment hierarchy entities are containers, containing other containers. **b** A *compositional* containment (or *level*) hierarchy describes how a system is composed of subsystems, which are also composed of subsystems, etc. The best known example for this type of hierarchy is the "hierarchy of life". **b** is Reproduced from Mader (2010).

Importantly, these hierarchy types are not independent of each other. On the one hand, many systems can be described by more than one type. For example, members of an army form control hierarchy in a way that people having higher rank give orders to lower-rank soldiers, but, at the same time, the very same army forms a compositional containment hierarchy as well. This is so since an army is composed of various divisions (infantry divisions, motorized divisions, airborne divisions, etc.) which are also composed of smaller contingents, all the way down to the soldiers, who are the "units" in this structure.

On the other hand, both order and nested hierarchies can be converted to flow hierarchy. In an order hierarchy, a directed edge can be assigned to each pair of adjacent members in the hierarchy and this produces a chain of directed edges. In a nested hierarchy, a virtual node is assigned to every sub-graph, and if a sub-graph contains another, then the two corresponding virtual nodes are connected with a directed link, which produces a flow hierarchy on the network of virtual nodes.

Thus, flow hierarchy is the most important variant and we shall mainly concentrate on its manifestations.

## *2.1 Describing hierarchical structures*

In this chapter we shall briefly summarize the basic concepts related to *graphs*, the mathematical object most often used in relation to hierarchy. It is important to highlight that graphs and networks are only the *models* of the real-life systems, not the systems themselves. It is a mathematical representation of the system under investigation, used because they, using graph theoretical methods and algorithms described in subsequent chapters, can reveal many important characteristics. An important further comment is that – as it is done in the literature – we shall use the term graphs for abstract mathematical constructions, while the term networks will be associated with the underlying interactions within a real-life structure. Readers familiar with graphs may skip this chapter.

### 2.1.1 Graphs and networks

As mentioned above, the most commonly used mathematical tool for describing hierarchical systems are graphs. Primarily, but not exclusively, they are connected to systems embodying flow (or control) hierarchy. Such systems and their graph representations go so much hand in hand, that when trying to assign a "hierarchy value" to a system (describing "how hierarchical" the given structure is), usually it is the hierarchy level of the *graph* (representing the system) that is measured.

The concept behind this representation is rather straightforward: the entities of the systems are the nodes of the graph, and if a pair of entities is in a subordinate-dominance relation, then there is a directed edge between them.

In the followings, we give a short overview of the basic graph theoretical concepts.

- A *graph* is a mathematical tool which is appropriate to handle a set of *objects* with *connections* among them. The objects are represented by *nodes* and the connections between them by *edges*. Formally, $G = \{V, E\}$ with a function $f : E \rightarrow V \times V$. The elements of $V$ are the nodes (or *vertices*, or *points*), and the elements of $E$ are the *edges* of the graph. The nodes are usually denoted by small Latin letters (e.g. *i, j, k*) or by Arabic numbers (1, 2, ..., *N*). Formally, *f* sends edges to pairs of vertices (which are the "endpoints" of the edge), but in practice we usually forget about the function *f* and simply think of $E$ (the set of edges) as a subset of $V \times V$. Accordingly, edges are usually given by the starting and nodes, such as $e = (i, j)$, for any $e \in E$. The word *network* is often used as synonym for graph in the case it stands for actually observed data.

- A graph can be either *directed* or *undirected*. In case of a directed graph (or *digraph*) the relation has a special direction as well. For example, in case of a hierarchy network, the direction can show which element dominates which other. In contrast, in an undirected graph the connections do not have special directions, like in the network representing the flight connections among cities. Informally speaking, in case of an undirected graph the edges are just "lines", and in case of digraphs, they are "arrows".

- A simple *loop* is an edge that connects a node to itself. (An edge whose starting and endpoint is the same vertex.)

- A *path* in a graph is a sequence of connected vertices. (Most definitions specify that the nodes within a path have to be distinct from each other.)

- A *cycle* is a closed path, that is, a path whose beginning and endpoint is the same vertex. Many times cycles are also referred to as loops.

- A *tree* is a graph in which there are no loops, cycles or multiple edges. In other words, it is a graph in which any two nodes are connected by exactly one path. There are two special kinds of vertices: (i) the root node, which does not have parents, and the leaves (or end-nodes), which do not have children. Accordingly, in a tree, nodes can be layered into levels.

- A *cluster* (a.k.a. *module*, *community* or *cohesive group*) is a part of the graph in which the units are more densely connected to each other than to the rest of the graph. We will use this elastic description, since the concept does not have a well-defined, widely accepted definition. Importantly, in real-life networks, the presence of such modules is a signature of the hierarchical nature of the structure (see, e.g., Vicsek 2002, Ravasz et al. 2002, Palla et al. 2005).

- A *directed* community is simply a community in a directed graph. Here the nodes can be related to each other based on the number of their incoming and outgoing links connecting them to other nodes within the same module. A node having more outgoing edges towards other members of the module is more like a "source"-node, whereas a node with mostly incoming links from these members is more like a "drain". (Palla et al. 2007)

- Vertices can be characterised by the number of links they have, reflecting how "strongly" they are connected to other nodes. Accordingly, the *degree* of a node in an undirected graph is simply the number of its edges. In a directed graph vertices can be characterised by their *in-degree* and *out-degree* values: the *in-degree* value refers to the number of links pointing *towards* the given node, whereas the *out-degree* value refers the number of links going outwards from the vertex.

## 2.1.2 Measuring the level of hierarchy

In this section we shall focus on measures for *flow* hierarchies. More precisely, we consider measures *for graphs representing flow hierarchy*. We have two main reasons to do so: (i) observations and experiments, as well as results of computer simulations are likely to return flow hierarchy, (ii) all other hierarchy types can be transformed into flow hierarchy in a rather straight forward way. For example, considering a containment hierarchy, its clusters can be identified with the nodes of a graph in which the directed edges will indicate the containment relation. That is, in the graph there will be an edge pointing from node *A* to node *B*, if cluster *B* fully contains cluster *A* in the original structure (Nepusz 2013).

Most of the proposed measures take values on the [0, 1] interval, returning nearly 0 for a completely hierarchy-less structure, like a full graph or a circle, and returning a value close to 1 for "completely hierarchical" structures, like a directed tree. Values for transient structures are up to "intuitions", and intuitions differ from person to person. This is one of the main reasons why there is no "most appropriate" measure serving all needs. The measures reviewed in the present book have values on the [0, 1] interval, with higher values representing higher degree of hierarchy.

This section of the book is relatively extensive for two reasons: (i) it is about an obviously central quantitative characteristic of a hierarchical structure, (ii) in spite of its essential importance there is no unique definition of the level of hierarchy of a system.

This latter situation is analogous to that of the definition of a community in a network. The notion itself is so complex that, depending on the aspect that we are interested in, a suitable definition should be chosen. For example, a community (cluster) in a network can be defined as a sub-network of nodes that have relatively more connections among them than with the other nodes. However, we can require this "relatively more" in various ways. Directed, weighted and connections specified according to further criteria make the problem of defining clusters in a network an open problem even more .

To introduce this aspect of the problem of finding the best measure of hierarchy, the reader is asked to consider the following question: please decide which structure is more hierarchical. A set of nodes arranged into layers connected by directed edges all directing from an upper to a lower layer or a "star" consisting of a central node from which a number of directed edges lead to the other nodes of the network? To us, the right answer is: it depends on the context, on the function, etc. Next we account for a number of relevant possible angles from which such a question can be approached.

*Global Reaching Centrality*

The central idea of this approach is to give a rank to each node by measuring its "impact" on other nodes. Impact is defined by the ratio of vertices that can be reached from the focal node *i*. *Local reaching centrality,* $C_R(i)$ defines exactly this quantity: in a directed, un-weighted graph, $C_R(i)$ is the maximum number of vertices that can be reached from node *i*, divided by $N - 1$. Then, the level of hierarchy is inferred from the distribution of the local reaching centralities: the more heterogeneous the distribution is, the more hierarchical the corresponding graph/network is. In order to demonstrate this statement (namely, that the distributions of the local reaching centralities reveal the hierarchical nature of a network), three different graph types are compared in Fig. 2.4: an Erdős-Rényi (random) graph (which is not hierarchical), a tree (which is highly hierarchical), and a scale free graph (which is "moderately" hierarchical). The most homogeneous $C_R(i)$ distribution belongs to the Erdős-Rényi (ER) graph: the $C_R(i)$ values are either 0 or close to 1, marked by the two narrow spikes at these values with a solid black line. In contrast, we find all kinds of $C_R(i)$ values in a tree, as it is indicated by the red line in Fig. 2.4 (note the log-log scale).

This distribution follows a power law that is distorted due to the random branching numbers. The blue dashed line belongs to the "moderately hierarchical" scale free graph, marking a "moderately heterogeneous" distribution.

These curves represent distributions, while for a measure we expect a number. The definition proposed by Mones et al. (2012) grasps the heterogeneity of the $C_R(i)$ distribution as follows: Let $C_R^{max}$ denote the highest local reaching centrality in a graph $G = (V,E)$. Then, the Global Reaching Centrality, $GRC$, is defined as:

$$GRC = \frac{\sum_{i \in V}[C_R^{max} - C_R(i)]}{N-1} \tag{2.1}$$

where $V$ is the set of nodes, and $N$ is the number of nodes in $G$. The $GRC$ values for our three example graphs (Tree, Scale-free and Erdős-Rényi), are the following:

Tree: $0.997 \pm 0.001$, which is the highest.
Scale-free: $0.127 \pm 0.008$, that is, SF networks are slightly hierarchical,
Erdős-Rényi: $0.058 \pm 0.005$, that is, these are not hierarchical at all.

These values, the means and variances, are calculated for an ensemble of 1000 graphs, and they demonstrate that the returned values are close to our "intuitions". Eq. (2.1) applies to directed, un-weighted graphs. Its generalized version is suitable for analysing weighted and/or undirected graphs by an appropriate modified definition of the local reaching centrality (Mones et al. 2012).

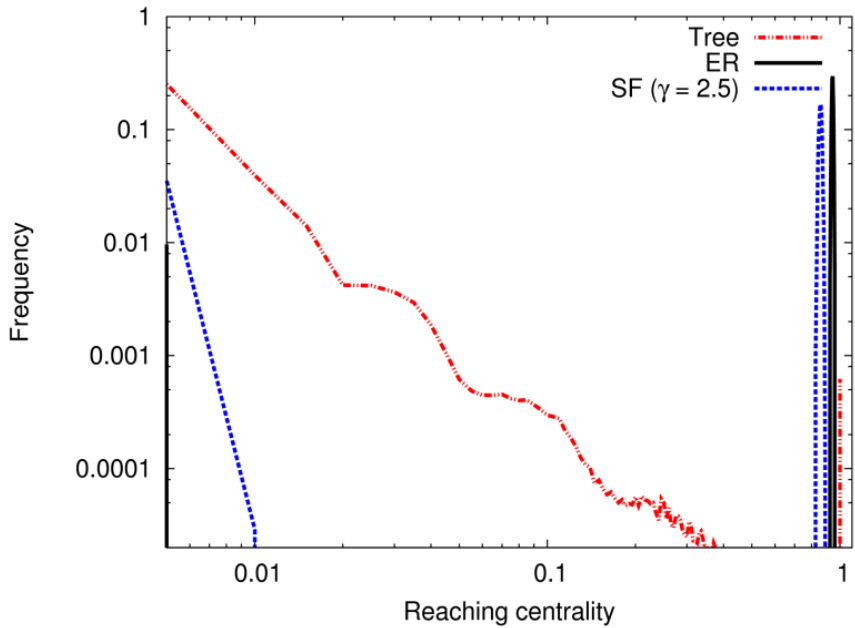

**Fig. 2.4** Distributions of the local reaching centralities for three kinds of directed networks: Tree, Erdős-Rényi (ER) and scale-free (SF). All the curves are averages of 1000 graphs with $N$=2000, of the appropriate graph type. Reproduced from Mones et al. (2012).

### *Random Walk Measure*

The main motivation of this approach is the claim that – in contrast to the assumptions behind most of the proposed methods – it is not correct to treat all directed acyclic graphs as already maximally hierarchical, independently of their inner structure. This observation is based on the common intuition that a hierarchical structure often corresponds to a multi-level pyramid in which the levels become more and more wide as one descends from the higher levels towards the lower ones.

The measure proposed by Czégel and Palla (2015) is based on properties of random walks within the graph, and, in accordance to the above mentioned claim, directed trees corresponding to multi-level pyramidal structures obtain higher hierarchy values than directed stars or chains.

Intuitively, the method is based on the assumption that there is information flow coming from the high-ranking nodes towards to ones at the bottom, similarly as in the case of an army or company, where the leaders send instructions downwards the links. In order to track the sources of the instructions/information, etc., random walkers are dropped onto the nodes who then move *backwards* on the links. Once a steady state is reached, the density of such random walkers (the number of them visiting a given node) can be interpreted as being proportional to the rank of this node: high random walker density indicates that the vertex is a source of information, low density indicates the vertex is more likely to be just a "receiver" of orders – that is, low in rank. The hierarchical nature of the network is then estimated based on the distribution of these random walker densities: if the distribution is homogeneous, the source of information/order cannot be pinpointed, thus, the network is not hierarchical. In contrast, inhomogeneous distribution indicates clear information sources: the network is hierarchical. This homogeneity/inhomogeneity is measured with a value called $H$, with higher values reflecting more hierarchical structures (bigger inhomogeneity), and lower values less hierarchical networks.

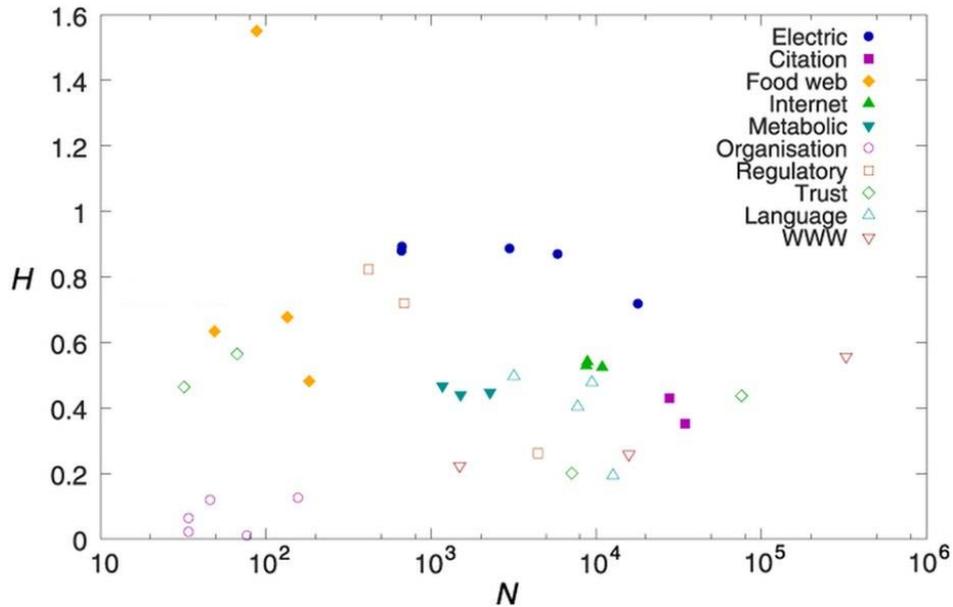

**Fig. 2.5** Hierarchy scores as a function of the network size. The different symbols correspond to different networks. The $x$ axis marks the size of the network ($N$, number of nodes) on a logarithmic scale, whereas the $y$ coordinate shows the hierarchy value ($H$) of the graph. Reproduced from Czégel and Palla (2015)

The largest $H$ values belong to regulatory networks, electric circuits and food webs, whereas the lowest ones belong to the informal networks of acquaintances in different organizations (Fig. 2.5). Moderately hierarchical are the Internet, various citation-, metabolic- language and trust networks, which results are in good accordance to our intuitive expectations.

An even clearer picture regarding the hierarchical nature of a network can be obtained by "normalizing" the hierarchy measure $H$ against the hierarchy measure of the same network, but under the assumption of random connections. This is the "$z$-score", defined as:

$$z = \frac{H - \langle H \rangle}{\sigma(H)} \qquad (2.2)$$

where $H$ is the hierarchy score, $\langle H \rangle$ is the expected $H$ value of the randomized graph, and $\sigma(H)$ is the standard deviation of $H$ in the randomized ensemble.

*An overview of further useful measures*

In the rest of the section we shall give an overview of some further measures, focusing on the main ideas behind them. Here our aim is not to give detailed description of the techniques but rather to flip through the type of concepts that have been proposed so far regarding the problem of measuring the hierarchy level of a graph.

### A measure for undirected networks

The measure proposed by Trusina et al. (2004) quantifies the flow hierarchy of undirected networks. It is based on the assumption that every vertex already has a rank associated with it by denoting its place in the global hierarchy. This estimate for the rank can be the degree of the node (originally proposed by the authors) but can be other conceivable measures as well, such as betweenness centrality or eigenvector centrality. With these assumptions, the hierarchy measure is given by the fraction of directed shortest paths going strictly upwards in the hierarchy.

More precisely, this method assumes that the shortest paths in the network consist of a part going upward the hierarchy (towards more important nodes), followed by a part going downward the hierarchy (towards less important nodes). Either part may be empty of course, but one should not turn back upwards after the downward part again. Paths of this type are said to be hierarchical, and the measure simply calculates the fraction of vertex pairs that are connected by a hierarchical shortest path.

### Determining the levels of organizations

One of the first methods was proposed by Krackhardt (1994), whose main motivation was to measure the levels of hierarchy of organizations. He defined four measures that can be used together as an estimate to the extent of flow hierarchy in networks. These measures are:

- _Hierarchy_: The fraction of unordered vertex pairs $(i, j)$ such that vertex $i$ is reachable from vertex $j$ but vertex $j$ is _not_ reachable from vertex $i$, or vice versa. It works on directed graphs only.
- _Connectedness_: The fraction of unordered vertex pairs $(i, j)$ such that vertex $j$ is reachable from vertex $i$ via a directed path _or_ vertex $i$ is reachable from vertex $j$.
- _Efficiency_: One minus the proportion of possible "extra" edges that are not needed to maintain connectedness of the components. It is assumed that each component should be an out-tree (as an archetype of perfect hierarchy) and thus a component of size $N$ must have at most $N$-1 links; any more than that is a violation of efficiency. This measure obviously penalizes cases when there are two separate paths leading upwards the hierarchy from a node $A$ to its superior $B$; one of the paths is not required to maintain connectedness, hence the structure is inefficient.
- _LUBness_: For each unordered pair of vertices $(i, j)$, the lowest upper bound (LUB) is a vertex $k$ such that both $i$ and $j$ are reachable from $k$. LUBness is the fraction of pairs having a LUB. This definition can be explained by Krackhardt's assumption of an out-tree being the perfect hierarchy one can achieve.

Each of these metrics may take values from zero to one, and each metric measures some kind of a "deviation" from the perfect hierarchy Krackhardt assumed, i.e., a directed out-tree. (It also applies for in-trees if we reverse the edge directions in the definition of LUBness).

However, these measures (with the exception of efficiency) can be calculated only for directed networks.

*Concept for containment hierarchies*

Unlike the measures presented so far, the concept of Ravasz and Barabási (2003) addresses the notion of containment hierarchies. They observed that log $k$ and log $C$ are correlated in many real-world networks (where $k$ is the vertex degree and $C$ is the local clustering coefficient).

They argue that this is due to a containment hierarchy in the network (although they have not used the word "containment"). In order to determinate this, they proposed a simple recursive generation process that creates graphs with a power-law degree distribution, a linear dependence between log $k$ and log $C$ and multiple levels of hierarchies contained within each other. The bottom line of their argument is that hierarchy in undirected networks can be quantified by looking at the log $k$ vs. log $C$ plot and fitting a straight line to the data; the larger the slope of the line is, the more hierarchical the network is.

*Layout-motivated measure*

Carmel et al. (2002) proposed a layout-based metric for measuring the amount of hierarchy in a directed graph. They have conceived a layout algorithm that places the nodes of the graph in 2D space such that a set of constrains related to the target level differences are taken into account as much as possible. More formally, this means the following. For each *i-j* edge, we assign a measure that describes the desired difference between the $y$ coordinates of vertex $i$ and vertex $j$. The graph is then laid out using their algorithm, and the difference between the maximal (*maxY*) and minimal $y$ coordinates (*minY*) is compared to the diameter of the graph. A strictly hierarchical graph with no cycles can be laid out in a way that the distance between levels is 1, thus the difference between *maxY* and *minY* is equal to the diameter, while a cycle (i.e. a perfectly un-hierarchical graph) would be laid out with equal $y$ coordinates, yielding a hierarchy measure of zero.

The disadvantages of this method are twofold:

- In the general case, it is not possible to assign desired target level differences to the edges. We could simply say that the desired difference is 1 for all the edges, but this would work only if none of the edges span more than one layer. Edges skipping layers but otherwise pointing to the right direction would skew the layout and decrease the hierarchy measure
- This measure is not applicable to undirected graphs.

.

*Measures for structures "from down to top"*

Next in contrast to the way we assumed above, we shall consider the edges of directed networks to be oriented *upwards* (i.e. from lower to higher levels), like on a *who-reports-to-whom* organization diagram. We do so in order to follow the terminology of the related literature. It is usually straightforward to apply the definitions to directed networks that use the opposite convention.

Sometimes we will talk about layers or levels (sets of nodes with the same rank). Layers are indexed from 1 upwards, and a lower layer index corresponds to a higher rank.

Some of these measures will work on networks where the ranks of individual nodes are not known in advance; others are defined for a network and a corresponding ranking of nodes, and therefore must be optimized by some optimization procedure when the ranks are unknown.

*Fraction of edges participating in cycles*

Here the main idea is to reveal somehow the possible asymmetry between nodes by assuming some sort of flow on the links, and then check if these flows exhibit any kind of overall directionality or not. One way to do so is to find all of the elementary cycles in the network, count the edges participating in them, and divide this number by the total number of edges. This approach works for undirected and directed graphs as well; in directed graphs, only directed cycles matter. (A cycle is elementary if no vertex appears in it twice).

All the elementary cycles in a directed graph can be found simply using Johnson's algorithm (Johnson 1975), which is $O((N+E)(c+1))$ where $N$ is the number of nodes, $E$ is the number of edges and $c$ is the number of elementary cycles. The case of undirected graphs is a bit more tricky as the union of two elementary cycles with at least one shared edge is also an elementary cycle (after removing the shared edges from the union), thus we can expect a lot more cycles than for directed graphs where this condition does not hold. It is therefore common to search for a *cycle base* instead, i.e., a set of cycles such that every other cycle can be reproduced from selected base cycles by taking their disjoint unions. Since every edge that participates in a cycle must also participate in one of the base cycles, finding a cycle base is enough for our purposes.

Luo and Magee (2011) proposed the opposite of this measure (i.e., the fraction of edges not participating in cycles) as a hierarchy measure for directed networks. A big advantage of this approach is its simplicity.

*Minimum fraction of edges to be removed to make the graph cycle-free*

This approach is slightly different from the one called "fraction of edges participating in cycles". For instance, consider a graph consisting of two interlocking directed links sharing an edge. In this graph, all the edges participate in cycles (hence the previous measure would be 1.0), but removing the shared edge would make the graph entirely cycle-free. We call a set of edges whose removal makes the graph cycle-free *feedback arc set*.

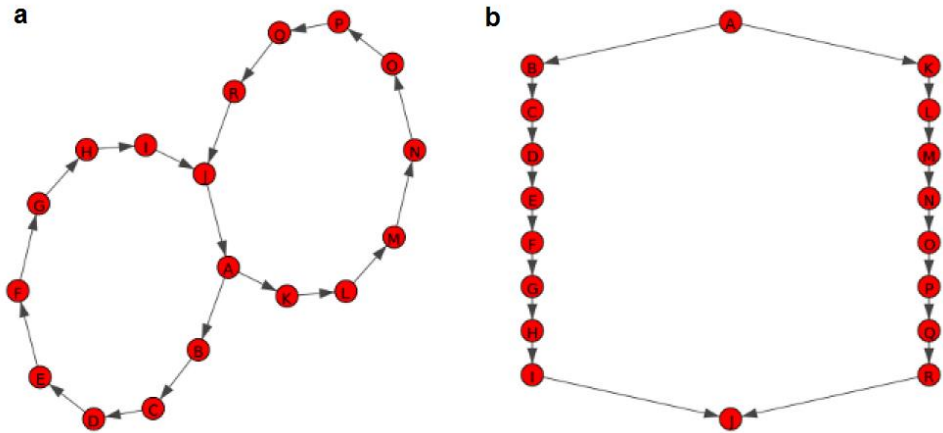

**Fig. 2.6** Illustration of the difference between "the fraction of edges participating in cycles" and the "fraction of edges to be removed to make the graph cycle-free". Subfigure **a** shows a graph where all the edges participate in cycles. However, as it can be seen in **b**, it is enough to remove a single edge (from *J* to *A*) to break both cycles and obtain a perfect hierarchy.

Note that although Fig. 2.6 shows a directed graph, this measure works just as well for undirected graphs– but the number of edges to be removed may be different! For instance, the graph with the two rings on the left of Fig. 2.6 becomes cycle-free by removing one single edge if the edges are directed, but one has to remove *two* edges to make it cycle-free in the undirected case.

This measure is very easy to calculate for connected undirected simple graphs. Since the graph is connected, the minimum number of edges required to connect $N$ vertices is $N$-1. Adding any extra edge on top of these $N$-1 edges necessarily creates a cycle, thus the number of edges one has to remove from an undirected simple connected graph with $N$ vertices and $M$ edges is $M$-$N$+1, and the fraction of such edges is therefore 1-($N$-1)/$M$.

For directed graphs, finding a minimum feedback arc set is an NP-hard problem (Healy and Nikolov 2013), but heuristic procedures exist to find an approximation. One such procedure is the greedy cycle removal algorithm by Eades et al. (1993) Namely:

1. Create an empty "deque" (double-ended queue).
2. If the graph is empty, we are done.
3. While there are sink vertices in the graph, remove them one by one and add them to the *beginning* of the deque.
4. While there are source vertices in the graph, remove them one by one and append them to the deque (add them to the *end* of the deque).
5. If no sinks and sources remain, find a vertex where the difference between the out-degree and the in-degree is as large as possible, remove it from the graph, append it to the deque and return to step 2.

At the end of the algorithm, the deque contains a possible ordering of vertices where ordinary edges point "forward" in the ordering and feedback arcs point "backward". The cardinality of the feedback arc set found by this heuristic is at most $M/2-N/6$ where $M$ is the number of edges and $N$ is the number of vertices.

Another heuristic is as follows. Scan each edge of the graph one by one and maintain two sets, $S$ and $T$. In each step, check whether edge $e$ forms a cycle with the edges already in $S$. If not, add $e$ to $S$, otherwise add $e$ to $T$. In the end, both $S$ and $T$ are acyclic and the smaller of the two sets gives a feedback arc set with at most half of all the edges. More sophisticated approximations are to be found in (Even et al. 1995) and (Saab 2001).

For graphs up to a couple of hundred nodes, one can use the following strategy as well:

1. If the graph is undirected, break it down into components, and calculate the sum of $M$-$N$+1 for each component, where $M$ is the number of edges in the component and $N$ is the number of vertices. This is the total number of edges to be removed to make the graph cycle-free; the fraction follows by a straightforward division.
2. If the graph is directed, break it down into weakly connected components and estimate the number of edges to be removed from each of the components as follows:
- If the component is acyclic (i.e., it has a topological ordering), no edges have to be removed at all.
- If the component has less than 20 edges, use a brute-force search to find the minimum number of edges to be removed to make it cycle-free.
- Otherwise, find a minimum cut of the component, add the edges of the cut to the feedback edge set and proceed recursively with each side of the cut.

*Fraction of hierarchy-violating edges*

A hierarchy-violating edge is one that originates in a higher level and terminates in a lower level, meaning that someone up in the hierarchy "reports to" someone on the lower level. This is a clear violation. Naturally, this measure requires the ranks to be known in advance as it is otherwise impossible to decide which edges violate the hierarchy.

Another, more strict definition of a hierarchy-violating edge is that it is an edge where subtracting the rank of the origin from the rank of the target yields a result that is not zero and not one. This definition penalizes not only the edges that go "the wrong way" in a hierarchy but also the edges that skip levels.

In the absence of ranks, one has to find the ranking that minimizes the fraction of hierarchy-violating edges, which leads to a problem that may be familiar from community detection. A trivial way to minimize the number of hierarchy-violating edges is to use the same rank for every node, assuming that edges between peers (i.e. nodes with the same rank) are allowed. A possible solution is to disallow edges between peers, which effectively reproduces the feedback arc set problem, since a directed graph minus a minimum feedback arc set is a directed acyclic graph which can then be decomposed into layers. Each feedback arc is then a hierarchy-violating edge.

*Average expected downstream path length*

This measure is based on random walks. More precisely, the expected length of a path a random walker is allowed to take on the graph with the following constraints:

1. The walker is only allowed to step downstream in the graph, i.e. towards lower layers. A path that goes downward in a layered hierarchy is called a *downstream path* (-hence the name of the measure).
2. The transition matrix of the random walk is a usual right-stochastic matrix derived from the weighted adjacency matrix of the graph (loop edges are not allowed).
3. The random walk terminates as soon as the walker ends up in a sink node or in a node that has neighbours in higher layers only.

The measure also requires an *a priori* layer assignment, and it is an open problem to find the optimal assignment given the graph only. When the layers are known, the measure can be calculated very easily: one has to proceed recursively from the lowermost layer towards the uppermost layer and make use of the following two equations:

1. If a vertex $v$ is a sink, then the expected length of downstream paths from $v$ is zero.
2. If $v$ is not a sink, the expected length is one more than the expected length of downstream paths from its lower-level neighbours, weighted by the probabilities of reaching those neighbours from $v$ in a single step. Note that only the expected length of downstream paths for vertices in layers lower than $v$ has to be known, therefore, a single sweep from lower layers to the uppermost layer is enough.

To make graphs with different numbers of layers comparable, it is advised to normalize this measure as follows.

Suppose that vertex $v$ is at layer $l(v)$ and there are $k$ layers. The maximal value of the expected downstream path length originating from $v$ (denoted by $h(v)$) is then $k - l(v)$. The normalized variant of the measure takes the average of $h(v)/(k - l(v))$ for all non-sink vertices, assuming that 0/0 is 0.

The above overview of the "further hierarchy measures" was composed using the working paper by T. Nepusz (2013).

## 2.1.3 Classification of hierarchical networks

The methods overviewed in the previous Sect. (2.1.2) assign a value for each graph, reflecting the extent to which the input network is hierarchical. Now we shall reverse the direction, and show an algorithm that creates a graph based on an input parameter $p$ (taking values on the [0, 1] interval) indicating how hierarchical the output graph should be. $p=0$ refers to non-hierarchical and $p$ close to 1 refers to strongly hierarchical structures. The method was proposed by Mones et al. (2012).

The construction of the graph with tunable levels of hierarchy goes as follows (Fig. 2.7 **a**):

- A level-value ($\ell$) is assigned to every node in a directed tree in the following way:
  - The nodes at the "bottom-level" (that is, the leaves) are assigned $\ell=1$.
  - The level-value of the root node is equal to the number of hierarchical levels in the tree. (for example $\ell=5$ in Fig 2.7**a** of the root node)
  - All children of a node with level-value $\ell$ will have $\ell-1$ as level-value.

- Next a given number of random directed edges are added to the tree according to the following rules:
  - 1-*p* proportion of these edges are added completely randomly by choosing their starting point (*A*) and end-node (*B*) with probability $1/N$ (*N* is the number of nodes in the graph). In case there is no directed edge pointing from *A* to *B*, such an edge is added to the graph.
  - Regarding the rest of the edges (accounting for the *p* proportion of the "extra" edges) they are added only if $\ell_A > \ell_B$.

Figure 2.7 **b** depicts the *GRC* values (see Sect. 2.1.2) for hierarchical graphs created with the above algorithm, for *p*=0.0, 0.2, 0.4, 0.6, 0.8 and 1.0.

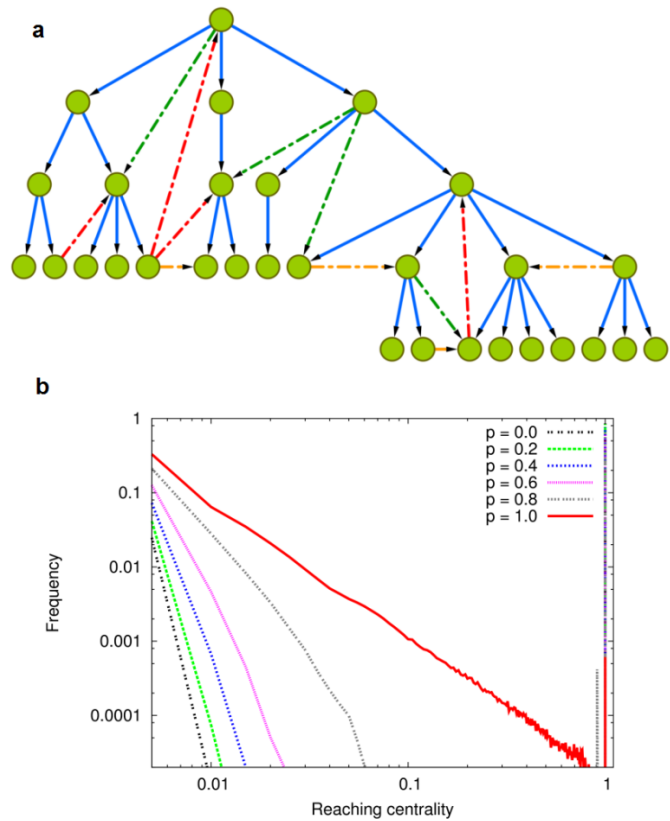

**Fig. 2.7 a** The different types of edges while constructing a hierarchical graph based on an input parameter *p*. Solid blue edges belong to the original tree used as the backbone of the output graph. Edges pointing downwards (green) conserve the hierarchy, horizontal edges (orange) have a slight influence and finally the ones directed upwards (marked with red) make strong change in the structure.
**b** Distribution of the local reaching centrality (see Sect. 2.1.2) values for adjustable hierarchical networks with various *p* values. Each curve is an average of 1000 networks with *N*=2000 nodes for <*k*>=3. Note that from the highly random (*p*=0) to the highly hierarchical (*p*=1) state the distribution changes continuously and monotonously with *p*. Reproduced from Mones et al. (2012).

Similarly to the problem of *measuring* hierarchy, the problem of *classification of hierarchical structures* is not trivial either. Next we shall overview a method proposed by Corominas-Murtra et al. (2013) which is based on three expectations towards hierarchical systems. These are (i) treeness (ii) feed-forwardness, and (iii) orderability. (See later in more details.)

Using the above three features, a 3D morphospace ("phenotype-space") can be defined in which the three axes are the tree quantifiable features. Placing real-life hierarchical and random null-models into such a coordinate system, fundamental characteristics can be revealed. As it turns out, networks do not occupy the entire morphospace, instead they accumulate in four major clusters within the large voids, which most probably results from the constraints under which they evolve.

Let's define the proper position of a network $G(V,E)$ within the morphospace.

First, $G(V,E)=G$ is transformed into its corresponding *node-weighted condensed graph* $G_C(V_C, E_C)=G_C$ which is an acyclic feed-forward structure where the cyclic modules (strongly connected components) of $G$ are replaced by single nodes. Accordingly, in a node-weighted condensed graph $G_C$, each node has a weight $\alpha_i$ indicating the number of nodes it includes from $G$, the original graph. For example, in Fig. 2.8, subfigure **h** depicts the node-weighted condensed graph $G_C$ corresponding to $G$, the one depicted on subfigure **d**. In this, node $S_2$ includes 3 nodes from $G$ and $S_1$ includes 2. (This method, the localization of strongly connected components, is an often used approach to identify subsystems within a graph.)

Then we calculate the three values using both $G$ and $G_C$.

1. "Treeness", $T$, taking values on the [-1, 1] interval, captures how unambiguous the "chain of command" is within $G_C$. In hierarchical networks, like the one in Fig. 2.8 **a** and on its corresponding node-weighted graph depicted on **e**, the chain-of-command is unequivocal, characterized by positive $T$ values. In case the chain of command is ambiguous, the structure is said to be *anti-hierarchical*, marked by negative $T$ values (Fig. 2.8 **b** and **f**). Intuitively, this feature is calculated by comparing the diversity of choices one can make top-down vs. the uncertainty on the way bottom-up, captured by the concepts *forward and backward entropies.*

2. "Feed-forwardness", $F$: Since the paths within cyclic modules (like $S_1$ and $S_2$ in Fig. 2.8 **h**) violate the downstream order within the graph, they are penalized according to their *size* and *position*: larger modules closer to the top of $G$ influence more the overall structure of $G$ than smaller ones close to the bottom. Accordingly, they introduce larger penalty. $F$ is defined on the [0, 1] interval.

3. "Orderabiliy", $O$, is defined as the fraction of nodes that do not belong to any cycle. These nodes are orderable, accordingly, bigger ratio results higher orderability value. $O$ takes values from [0, 1].

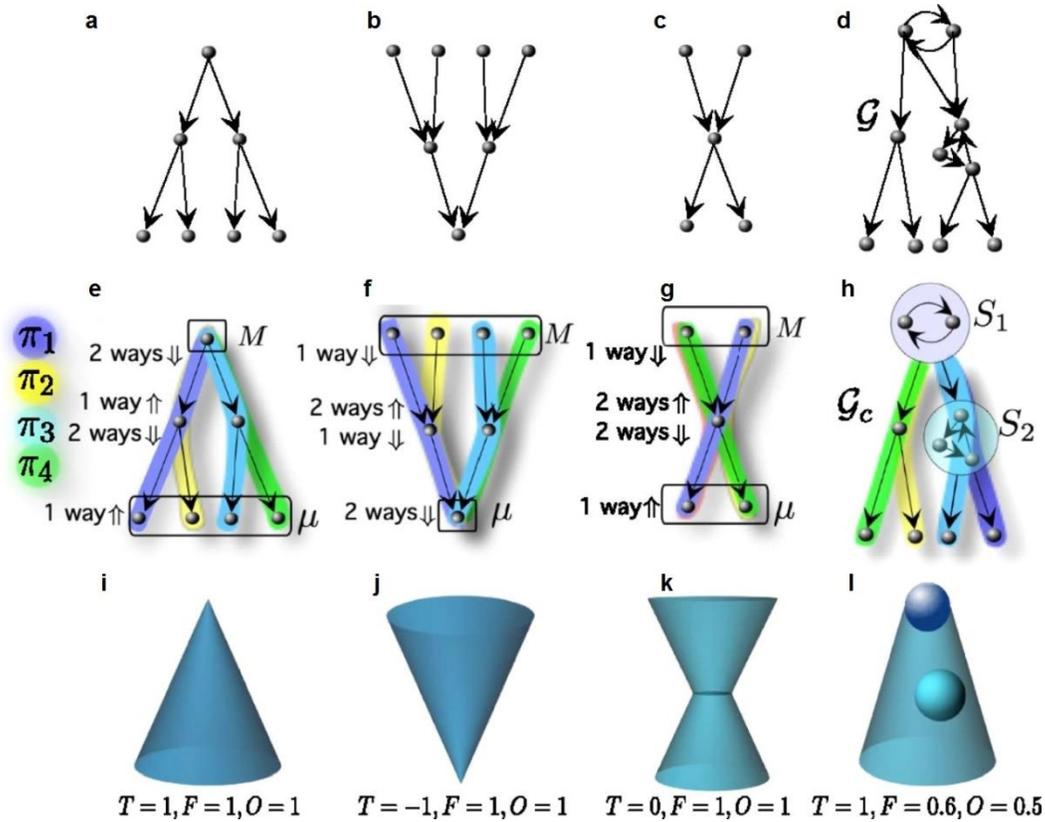

**Fig. 2.8** A Visualization of the tree concepts characterizing hierarchical networks: treeness (*T*), feed-forwardness (*F*) and orderability (*O*). Based on these tree features, a morphospace can be defined in which the similarities and differences (resulting from evolutionary constraints) can be analysed. **a** A perfectly hierarchical graph is tree-like (or pyramidal, *T*=1) with feed-forward edges (*F*=1) and orderable nodes (*O*=1). **b** In anti-hierarchical networks (characterized by negative *T* values and head downwards pyramidal structures) the chain of command is ambiguous. **c** is non-hierarchical (*T*=0) and **d** is a graph with cyclic modules, violating the orderability of the nodes. (**e-h**): the corresponding node-weighted condensed graphs of the networks in the first row, with paths top-down and bottom-up. (**i-l**) : the icon representation of the graphs in the first row, along with their *TFO* values. Reproduced from Corominas-Murtra et al. (2013)

Figure 2.9 depicts the location of random null models (white circles) and 125 real networks within the morphospace. Since random networks are being built without any selection pressure, they are neither hierarchical nor anti-hierarchical, accordingly, they occupy the *T*≈0 segment.

The main observation is that the vast majority of real networks fall into four clusters:

i. Gene regulatory networks (plus a protein kinase NW) occupy the first cluster at the top of the coordinate system (Fig. 2.9), marked as "GRN". These systems are characterized by very high orderabiliy values (*O*) with variable *F* values. The broad range of *F* (feed-forwardness) is caused by various sized modules near to the top of the networks, corresponding to a small fraction of genes, (transcription factors) participating in cycles.

ii. Electronic circuits and software graphs are strictly feed-forward (*F*≈1) with orderable nodes (*O*≈1), biased slightly towards negative *T* values. This cluster (marked as "TECH" in Fig. 2.9) is located on the top right edge of the morphospace.

iii. (ECO) The third cluster is defined by the ecological flow graphs, marked as "ECO" in the Fig. 2.9. Their positions within the morphospace reveal a certain degree of pyramidal structure combined with the important role played by loops. This special, separated position is consistent with the trophic pyramid mingled with recycling.

iv. And finally, the fourth cluster is composed of metabolic, neural, linguistic, and some social networks ("LANG, MET, NEU"), embedded within the cloud of random graphs. These networks display a large central cycle, much larger than their randomized counterparts, which feature is most probably due to the advantage of reusing/recycling molecules.

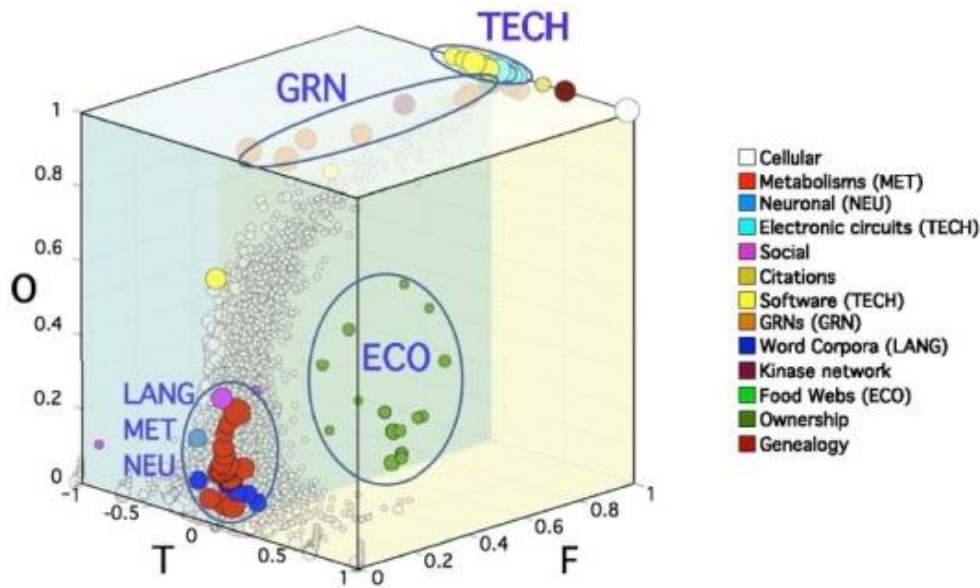

**Fig 2.9** The position of 125 real networks and various random null models within the morphospace defined by the coordinates *T* (treeness), *F* (feed-forwardness) and *O* (Orderabiliy). The random networks are white, while the real networks are colour coded according to their types listed in the key. The size of the circles is proportional to number of nodes the corresponding graph includes. Reproduced from Corominas-Murtra et al. (2013)

Two of these clusters (LANG/MET/NEU and TECH) overlap with random networks with similar connectivity, suggesting that non-adaptive factors shape the topological nature of these graphs. In contrast, the position of the ECO and GRN clusters indicate that the topological features of the ecological and gene networks are the resultant of functional constraints.

## 2.2 Visualization techniques

### 2.2.1 A general overview

The aim of the various visualization techniques is the same: to illustrate the entire network as a single figure in an easily perceptible way, revealing as much information of its hierarchical nature / inner structure as possible. Since (real) hierarchical systems are often complex with many characteristics, the level to which a visualization technique reflects the main features of a network is limited. Different visualization tools highlight different characteristics and different hierarchy types require different visualization tools. There exists no "best method", the appropriate technique depends on the specific characteristics we would like to highlight.

The most simple – and widespread – visualization technique is the *pyramid*, in which each entity is represented by a layer and the higher an entity is in the hierarchy, the higher it is in the diagram. Often (for example in case of social pyramids), but not always (e.g., Maslow's hierarchy), the width of the layer reflects the size of the represented layer. The drawback of this technique is that it can reflect only a linear order (a sequence) of the layers, and – in some cases – their approximate sizes. In other words, this technique reveals only an order hierarchy of the layers, without giving any description about the inner structure of the given system.

In contrast, graphs are applicable to describe not only order hierarchy, but other hierarchy types as well, most importantly flow hierarchy, meanwhile allowing a much more detailed visualization of the inner structure of the system as well. Due to these reasons, visualization of flow hierarchy is the most commonly used technique to represent hierarchical systems.

Because of the lack of loops and cycles, the representation of a "pure" hierarchical system would be a tree. However, in real-life cases, such systems occur only very rarely. Accordingly, trees often correspond to the ideal and/or theoretical case, while graphs that are more complex (have cycles, undirected edges, etc.) are better suitable for representing real-life cases.

This representation is closely connected to the concept of control (or flow) hierarchy, in which the entities (which are represented by nodes in the corresponding graph) are organized into a system of subordinate-superordinate relations, which correspond to the edges of the graph. Accordingly, orders or information flow on the edges (hence the name) from the superior unit(s) towards the inferior element(s), while requests and information flow in the opposite direction. Typical examples are the ranks in armies, various state and church organizations, corporations, etc.

## 2.2.2 Techniques reflecting the overall hierarchy level

Let' have a graph, representing a (real life or artificial) system. The graph can be large, having many communities and sub-communities, therefore difficult to be drawn in a way that is reasonably accessible to overview. However, we would like to know *how hierarchical* the original system is, preferably in a visual form.

The most widely accepted method for visualizing the hierarchical nature of *small networks* is the one proposed by Sugiyama et al. (1981). For such graphs, this technique provides an informative and clear hierarchical layout by layering the vertices into horizontal rows in a way that the edges are directed downwards. This method is often referred to as "*Layered graph drawing*" or "*hierarchical graph drawing*" method.

The main steps are the following (Fig. 2.10):

(i)     *Cycle removal* (a pre-processing step). If the directed input graph is not acyclic, a minimal set of "reversal edges" has to be identified and reversed in order to obtain an acyclic digraph. (Identifying such a minimal edge-set is an NP-complete problem.) (These reversed edges, as well as other changes within the graph will be restored in a later step into their original state.)

(ii)    *Layer assignment*. Partitioning the vertex set of the graph into layers in a way that each edge is directed from a higher level towards a lower one, with the following properties:
    a.   the number of layers is kept small
    b.   as few edges span large number of layers as possible

c. the assignment of nodes into layers is balanced.
(iii) _Insertion of "dummy vertices"_. "Long" edges (edges spanning multiple layers) are chopped up into a series of shorter ones by inserting so called "dummy vertices" into the graph. After this step each edge will connect nodes on adjacent layers.
(iv) _Edge concentration_ (optional step): The aim of this step is to reduce the number of edge crossings and the edge density between adjacent levels. It might reduce the number of dummy vertices as well, but, as important drawbacks, it may increase the number of layers and it also modifies the graph.
(v) _Vertex ordering_ (or "crossing minimization" or "crossing reduction" step). The nodes within the layers are permuted in a way that the numbers of edge-crossings are minimized between the adjacent layers.
(vi) _x-Coordinate assignment:_ The aim of this step is to position the nodes (that is, assigning them an $x$ coordinate) within each layer in a way that the edges become as straight as possible, and the nodes are centred with respect to their neighbours. This positioning should be consistent with the permutation applied in the previous step.
(vii) _Final step:_ Changes that have been introduced to the graph in previous steps are reversed so that the edges return into their original state:
a. edges reversed in the "cycle removal" (first) step are returned into their original direction
b. dummy vertices that have been inserted in step (iii) are removed from the graph and the corresponding "long" edges are drawn back in a way that avoids intersections and crossings. This might be done by drawing the edges as polygonal chains or spline curves.

For a detailed analysis and description of this method see also (Healy and Nikolov 2013). Although this method is very popular for small networks, it has some serious drawbacks as well, which become especially important for large graphs:

• for bigger networks (graphs with more than a few hundred nodes) the generated layout becomes difficult to overview/interpret;
• the steps are NP-complete or NP-hard, which makes the usage of several different heuristics necessary and thus the results become less well-defined.
• independently of the hierarchical nature of the given network, the method provides a hierarchical layout which is often misleading;
• the meaning of the levels is not defined;

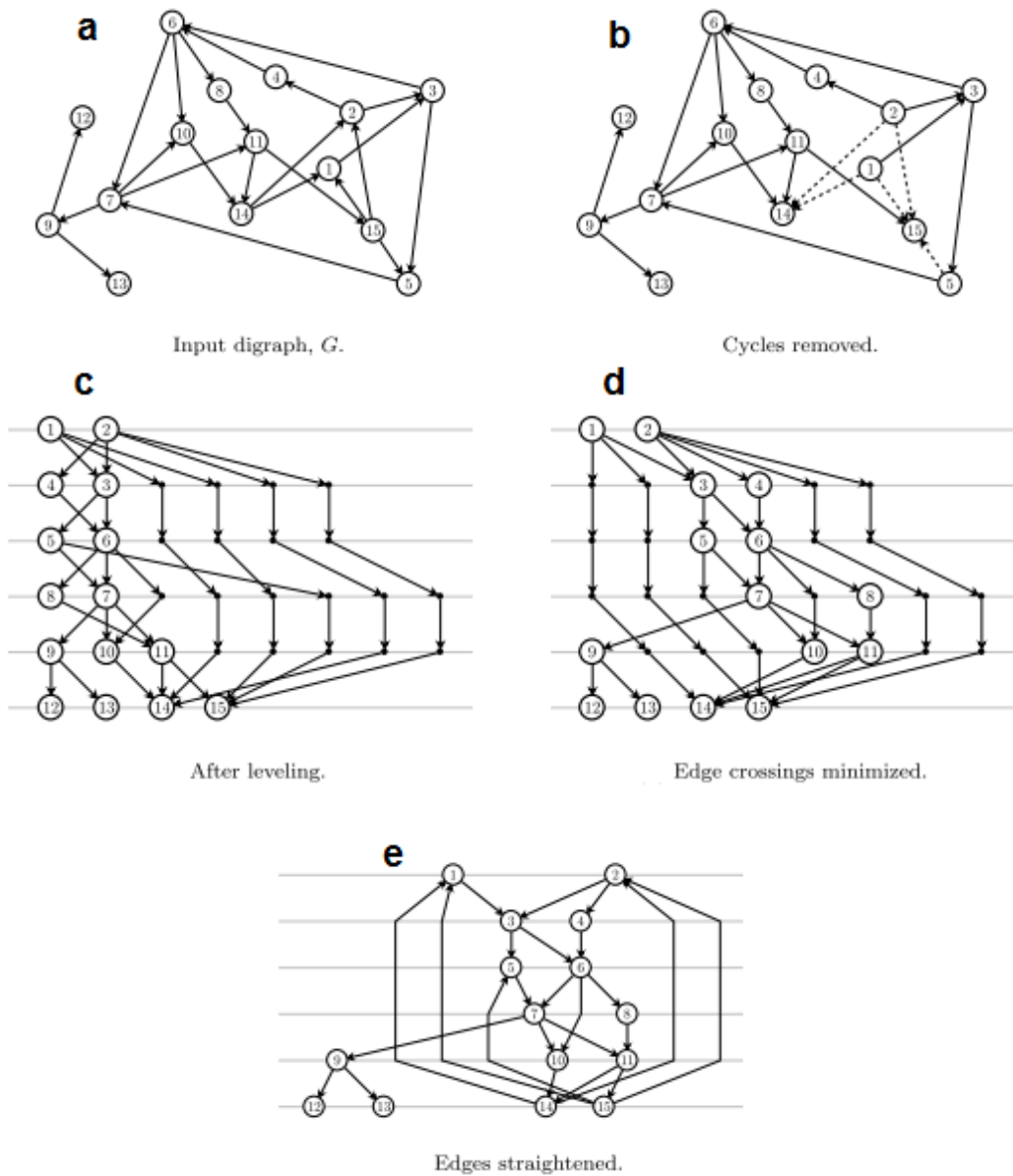

**Fig. 2.10** The main steps of the Sugiyama algorithm. It is hard to see the hierarchical structure of the input graph **a**, whereas it is clearly visible on the output graph **e**. This method is appropriate for relative small graphs (up to a few hundred nodes). Reproduced from Healy and Nikolov (2013)

Next, we discuss a method proposed by Mones et al. (2012) that solves the above problems and is easily applicable even for *complex large networks* (See Fig. 2.11). The algorithm of the proposed method is as follows:

1. Rank the nodes according to their *local reaching centrality* value, $C_R(i)$, where $C_R(i)$ is the ratio of nodes that can be reached from the focal node *i,* reflecting "impact" of *i* on other nodes (see also Sect. 2.1.3). (Importantly, from the viewpoint of the algorithm, instead of $C_R(i)$, other local quantities can be used as well.)

2.  Start to add nodes to the first, bottom-most level of the layout in an increasing order regarding their $C_R(i)$ values, until $\sigma_L < \varepsilon\, \sigma_G$. Here $\sigma_L$ is the standard deviation of the $C_R(i)$ values within the actual level, whereas $\sigma_G$ is that within the entire graph. $\varepsilon$ is an adjustable parameter, defining the "resolution" of the levels.)

3.  Once $\sigma_L \geq \varepsilon\, \sigma_G$, start a new level.

4.  Repeat 2nd and 3rd steps until every node is put in levels. (Step 2 ensures that nodes with similar $C_R(i)$ values will be on the same level.)

5.  In order to get a nice horizontal arrangement, align the centre of mass of each level above one other, that is, to the same vertical line.

6.  The levels are arranged vertically in a way that the distances between adjacent levels are proportional to the logarithm of the differences of the averages inside the certain levels: $(Y_{\ell+1}-Y_\ell) \propto \ln\,[\langle C_R \rangle_{\ell+1} - \langle C_R \rangle_\ell]$. ($Y_\ell$ is the vertical position of level $\ell$ whereas $\langle C_R \rangle_\ell$ is the average of the $C_R(i)$ values within level $\ell$.)
    Next, set the vertical distances of the levels in a way that they become proportional to the differences between their average $x_i$ values. Set the smallest distance to the same value as the horizontal distance between two adjacent nodes. Finally, set the distances to be proportional to the logarithm of the original differences in a way that the height of the graph is kept unchanged.

For large graphs, $\varepsilon$ tunes the vertical extension of the layout

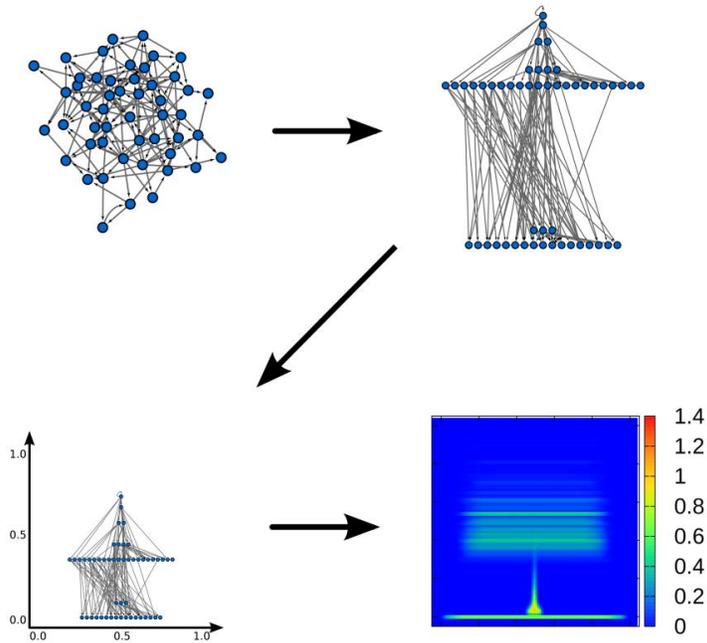

**Fig 2.11** The main steps of the visualization process. Firstly the layout is computed, based on the local reaching centrality, $C_R(i)$, values (top right). Next, the levels are separated with a logarithmic ratio and then each layout is scaled into the unit square (bottom left). Finally, the rescaled layouts are plotted in the unit square with the obtained node-density (bottom right, see also the colour bar as well). In the heat maps, the colour scale shows $\log(\log(\rho(x,y)+1)+1)$, where $\rho(x,y)$ is the average density of the ensemble. Reproduced from Mones et al. (2012).

Figure 2.12 shows the resultant of this method for (**a**) Erdős-Rényi, (**b**) scale-free, and (**c**) directed tree type of graphs.

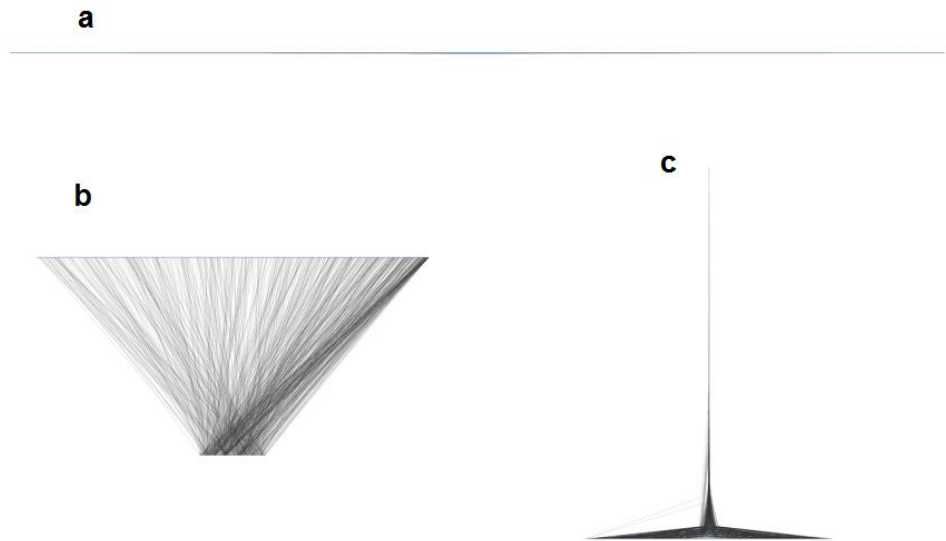

**Fig 2.12** Visualization of the three network types studied in Sect. 2.1.3, based on their local reaching centrality values. **a** An Erdős–Rényi (ER) graph, **b** a scale-free (SF) network, and **c** a directed tree with random branching number between 1 and 5. For all three graphs $N$=1000 with the parameter $\varepsilon$ set to 2/$N$. In case of the ER and the SF graphs $\langle k \rangle$=3. Reproduced from Mones et al. (2012)

## *Reference list*

Carmel L, Harel D, Koren Y (2002) Drawing directed graphs using one-dimensional optimization. In: Graph Drawing 10th International Symposium, Irvine, CA, USA, August 2002. Lecture Notes in Computer Science, vol 2528. Springer, Heidelberg, p 311

Corominas-Murtra B, Goñi J, Solé RV, Rodríguez-Casoa C (2013) On the origins of hierarchy in complex networks. PNAS 110(33):13316-13321. doi: 10.1073/pnas.1300832110

Czégel D, Palla G (2015) Random walk hierarchy measure: What is more hierarchical, a chain, a tree or a star? Sci Rep 5: 17994, doi:10.1038/srep17994

Eades P, Lin X, Smyth WF (1993) A fast and effective heuristic for the feedback arc set problem. Inf.Process.Lett. 47(6):319-323

Even G, Naor J, Schieber B, Sudan M (1995) Approximating minimum feedback sets and multi-cuts in directed graphs. In: Integer Programming and Combinatorial Optimization, Lecture Notes in Computer Science 920:14-28

Healy P, Nikolov NS (2013) Hierarchical drawing algorithms. In: Tamassia R (ed) Handbook of Graph Drawing and Visualization, CRC Press, Boca Raton, USA, p 409-454

Johnson DB (1975) Finding all the elementary circuits of a directed graph. Siam J Comput 4(1):77-84

Lane D (2006) Hierarchy, complexity, society. In Pumain D (ed) Hierarchy in Natural and Social Sciences. Springer, Dordrecht, p 81-119

Luo J, Magee CL (2011) Detecting evolving patterns of self-organizing networks by flow hierarchy measurement. Complexity 16(6):53-61

Krackhardt D (1994) Graph theoretical dimensions of informal organizations. In: Carley K, Prietula M (eds) Computational Organizational Theory. Lawrence Erlbaum Associates, Hillsdale, NJ, p 89-111

Mader S (2010) Biology, 10[th] ed. McGraw-Hill, New York

Mones E, Vicsek L, Vicsek T (2012) Hierarchy measure for complex networks. PlosOne 7:e33799

Nagy M, Vásárhelyi G, Pettit B, Roberts-Mariani I, Vicsek T, Bíró D (2013) Context-dependent hierarchies in pigeons. PNAS 110(32): 13049-13054

Nepusz T (2013) Self-organization of hierarchy in complex networks. Working paper.

Palla G, Derényi I, Farkas I, Vicsek T (2005) Uncovering the overlapping community structure of complex networks in nature and society. Nature 435:814-818

Palla G, Farkas IJ, Pollner P et al (2007) Directed network modules. New J. Phys. 9:186. doi:10.1088/1367-2630/9/6/186

Ravasz E, Barabási A-L (2003) Hierarchical organization in complex networks. Phys Rev E 67:026112

Ravasz E, Somera AL, Mongru DA et al (2002) Hierarchical organization of modularity in metabolic networks. Science 297: 1551–1555.

Saab Y (2001) A fast and effective algorithm for the feedback arc set problem. J. Heuristics 7:235-250

Sugiyama K, Tagawa S, Toda M (1981) Methods for visual understanding of hierarchical system structures. In: IEEE Transactions in Systems, Man and Cybernetics, vol 11. p 109.

Trusina A, Maslov S, Minnhagen P, Sneppen K (2004) Hierarchy measures in complex networks. Phys Rev Lett 92(17):178702

Vicsek T (2002) The bigger picture. Nature 418:131

Wimberley ET (2009) Nested Ecology: The Place of Humans in the Ecological Hierarchy. John Hopkins University Press, Baltimore

# 3 Observations and measurements

## 3.1 Animal groups

### 3.1.1 Dominance

From the viewpoint of an individual, both solitary and social life style has advantages and disadvantages too, mainly defined by the access to resources, such as food or mates. A solitary animal does not have to share anything with others, but such an individual is constantly exposed to a much higher level of danger regarding predators and also faces difficulties in finding mates for reproduction. Living in groups is safer with respect to predator avoidance, ensures the possibility of reproduction and creates an environment in which decision-making is more optimal because of the information transmission among the members. At the same time, it raises competition among the members for the resources and increases the probability of disease and parasite transmissions.

If, on the whole, for a given species the ratio of advantages/disadvantages is higher in case of living in groups than in case of a solitary life style (that is, ensures a higher chance for survival for the individuals), the animals will knit into groups. In such case, effective regulating mechanisms are needed in order to reduce the level of aggression among the members triggered by the competition.

The evolutionary solution for this problem is the emergence of *dominance* hierarchy, a mechanism whose main purpose is to regulate the access to resources. The mechanism is simple: higher ranked individuals have primacy compared to their lower level mates. As one advances in the evolutionary tree, the structure of the dominance hierarchy gets more and more pronounced and complex, accompanied by more and more sophisticated strategies by which individuals try to get higher and higher ranks. When it comes to chimpanzees, *Pan troglodytes*, our closest living relatives, we find that they use such elaborated methods in their everyday fights for positions, which, even just a few decades ago, was believed to be practised only by humans. Such tools, among others include coalition formation, manipulation, and exchange of social favours or adaptation of rational strategies (de Waal 2007).

No surprise that – being embedded into such an evolutionary process – humans are very sensitive for hierarchical positions as well. The unappeasable longing for getting higher and higher in the hierarchy is a basic human characteristic as well (Weisfeld and Beresford 1982). From a physiological point of view, the mechanisms determining the rank of an individual are very similar in primates and humans (Sapolsky 2005) – and in mammals in general: for example the level of *testosterone* in the blood, the principal male sex hormone, is found to be related to the rank: in case of various monkey species, higher testosterone level was measured in higher ranked individuals than in lower level animals (Eibl-Eibesfeldt 1990). Similar correspondence was found in humans as well: in an experiment, the testosterone level of young male tennis players found to be rising in case of victory, but falling in case of defeat, whereas no change was detectable during training. In another experiment, the testosterone level of medical students was measured, before and after exam. For those who passed the exam successfully, the level of the hormone arose, but for those who failed, it fell (Mazur and Lamb 1980). The level of the testosterone hormone and the inclination of behaving dominantly form a positive feedback loop as one intensifies the other.

The other hormone that constantly pops up in relation with dominance rank – primarily in connection with primates – is the g*lucocorticoid* steroid hormone, also known as *stress hormone*. During the last decades many attempts have been made to study their relation, but the picture is still not clear (at least in stratified mammal societies) since various studies report contradictory findings. According to the original view, subordinate individuals must be exposed to a much higher level of stress than their higher-ranking mates, and thus their stress hormone level is higher as well. However, measurements revealed that they are exactly the higher ranking individuals who have higher cortisol level in their blood (Muller and Wrangham 2004). Other studies found support to the original assumption, namely that the glucocorticoid secretion is stronger in lower ranking individuals in general, from which the only exception is the alpha male in the very top of the hierarchy, whose cortisol level is the highest in the whole group (Gesquiere et al. 2011). Furthermore, the correlation between the level of stress hormone and high rank found to be the strongest during periods of social instability, which is no surprise since during transitions in the hierarchical structure it is the highest ranking individuals who are exposed to the highest level of aggression (Sapolsky 1983). The observed differences might be due to the variations of the social organizations in different species and populations (Sapolsky 2005). As such, Creel (2001) identifies the decisive factor determining the relation in the way group members help each other in breeding the offsprings: in those species, among which cooperatively breeding is common, the rank and the hormone level is in direct proportion, while in other species it is in inverse proportion, see Fig.3.1. (Cooperative breeding means that offsprings are taken care of not only by their parents, but also by other group members.)

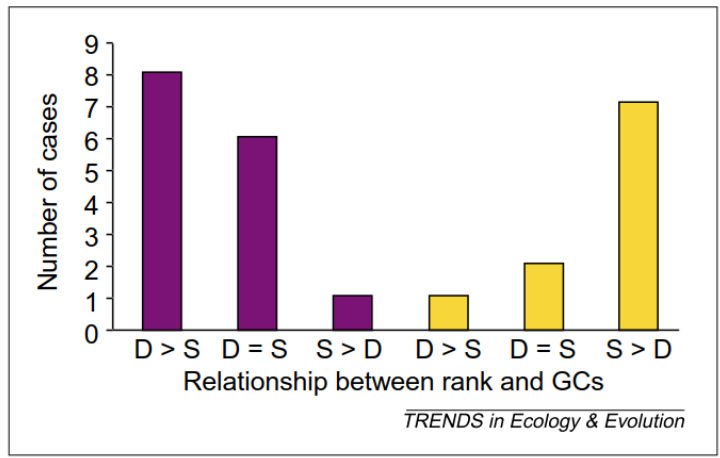

**Fig. 3.1** Relation between the level of stress hormone and rank. Among species with cooperative breeding (purple bars on the left) the level of basal glucocorticoid (GC) is significantly higher in dominant individuals than in subordinates (*D>S*) in contrast with other species (yellow bars on the right), where the relation is vice versa. Reproduced from Creel (2001)

Notably, permanently being exposed to high stress hormone levels has a serious (negative) effect on the individual's health too, especially on the cardiovascular, adrenocortical, reproductive and immune systems (which complaints are often referred to as "stress-related diseases"). For a review on the relation between rank and health in primate societies see (Sapolsky 2005).

In humans, the phenomenon that there is a strong correlation between the *socioeconomic status* and the appearance of various stress-related diseases has been known for a long time and was well documented (Adler et al. 1994). Although the socioeconomic status (among humans) does not cover exactly the concept of rank in the dominance hierarchy, it is still a good approximation to it (Sapolsky 2004), and, more importantly, easier to measure – which explains why human studies use mostly this concept.

These findings refer to the relationship between the physiological state of an individual and its rank in the *dominance* hierarchy. As we shall see in the upcoming chapters, in human societies, besides the dominance hierarchy, another type of hierarchical structure emerges as well, what we shall call "*cultural hierarchy*".

### Measuring dominance

Probably the best known hierarchy type is dominance hierarchy according to which individuals belonging to the same group regulate their access to natural resources such as food, mating partners, nesting locations or a safe lair. While establishing the ranking system, members of the group interact, often aggressively, in the form of repeated fights and threats. The result of each encounter is remembered by both parties and in case of many species by other group members as well. But after the order is set, subordinates will give way to their superiors without further fights or threats. The fundamental advantage of this arrangement is that it minimizes the aggression within the group since individuals do not engage in fights continually, only when creating or altering the dominance structure. In order to maintain such a structure, the individuals have to recognise each other and they also have to remember their mates along with the outcomes of the fights. In other words, they have to be able to create and maintain a mental model of the social structure within their group. Most probably this is the reason why it appears only at certain point of evolution, which is, according to our present knowledge, is at the point when fishes appear (Unfortunately the scope of the present book cannot cover the amazing organizational mechanisms driving the societies of social insects (Hölldobler and Wilson 2008); hence, in the followings we shall focus mainly on the dominance structures determining the social lives of the most various vertebrate species.)

The most simple dominance structure is called *despotism*. In such an order one individual rules over all the others who, on the other hand, have no rank distinctions among each other. For example bumblebees maintain such a structure, as it was recognised and described by Swiss entomologist Pierre Huber in 1802 in a study which is now considered to be the first modern research on the field of dominance hierarchy (Huber 1802). However, real interest started to show only more than a century later when Thorleif Schjelderup-Ebbe described the dominance structure (which he called the "pecking order") of hens (*Gallus domesticus*) in his PhD dissertation of 1922. Later this expression, "pecking order" was extended to the dominance relations of other kinds of birds too, while by now it is often used in a general sense as a synonym for dominance hierarchy.

Since then this topic has yield a lot of attention from biologists, and by now it has a vast literature. However, in the past decades scholars from other fields have also become interested in the *formation* of dominance hierarchies, most prominently economists, computer scientists, theoretical biologists and physicists whose aim has been twofold: (i) to give an account of the *self-organizational process* that is behind the *formation* of dominance hierarchies (in animal and human groups as well), and (ii) to develop techniques allowing the hierarchical relations among group members to be measured.

When trying to come up with such techniques, one immediately runs into the following problem: how should dominance be *quantified*, "measured"? A traditional way to overcome this problem is to sit for a long time in a (preferably hidden) observation point and watch the social life of the observed group, meanwhile making as precise record of their inter-individual interactions as possible. With this approach many fascinating results have been reached, among which the best known ones are probably the ones related to the observations of primate societies, most notably chimpanzee groups (de Waal 2007). Probably the biggest benefit of this approach is that *any* kind of interaction can be recorded (who eats first, who sleeps where, who "beats up" whom, how the conflicts are being solved, etc.). However, this technique also requires an enormous amount of time, special conditions (ensuring *continuous* observation of the group in its *natural environment*, or in an environment in which the human impact is as limited as possible), and the ability to *recognize all* the individuals with high confidence within the observed group. These are requirements that are not easy to fulfil.

Most recent techniques aim to automate somehow the observations: to take video records and analyse the results later with various computer programs (Ballerini et al. 2008, Pérez-Escudero et al. 2014), to put small GPS devices on the individuals and record their motion with these equipment (Ákos et al. 2014, Nagy et al. 2010, Nagy et al. 2013, see Fig. 3.2), or an even more recent technology is to combine the data recorded with various sensors (Gerencsér et al. 2013).

The biggest disadvantage of the video records is that the individuals within the group has to be identified later (at least if the inter-individual interactions are to be analysed). This problem turns out to be extremely difficult, primarily on videos recording animals in their natural environment. For example Ballerini et al. (2008) recorded the free flight of starling flocks (counting 2,600 individuals) and restored the 3D positions of the birds using stereometric and computer vision techniques within the framework of the "Starflag" project, lasting from 2005 to 2007.

A way to overcome the problem of individual recognition on video records is to put some kind of identification marks on the individuals. However, this method can be used only in groups counting top most a few dozens of individuals. For example Nagy et al. (2013) marked pigeons with a colour-bar (a unique combination of three, well-distinguishable colours on the back of each bird) and recorded their activity from above (see Fig. 3.2). In such a way, "only" the colour bar recognition had to be solved in an automated way. (Which also turned out to be a highly non-trivial task, since colours faded on the back of the animals, the birds covered each other from time to time, the efficiency of the recognition depended strongly on the actual lighting conditions, etc. However, even with such difficulties, by using the "colour-bar technique", the problem of individual recognition is still easier to solve than without any crutch.)

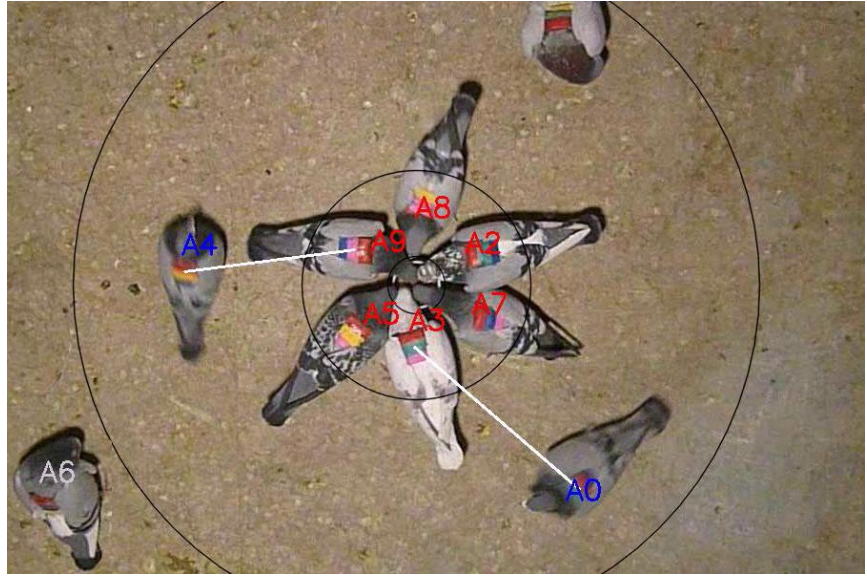

**Fig. 3.2** The "colour bar technique". A snapshot of the processed video sequence. The original video records the feeding-queuing activity of a group of homing pigeons. Each bird is marked with a unique combination of three colours (a "colour bar") serving as an individual code for a computer program designed to identify the individuals automatically, and process their behaviour. Circles divide the different activity regions: birds marked with red colour in the central circle identified as the ones feeding, individuals marked with blue identified as the ones queuing, and pigeons out of the external circle are identified as "not interested". Reproduced from Nagy et al. (2013).

Pérez-Escudero et al. (2014) introduced a method which is based on individual recognition as well, but instead of putting an artificial mark on the individuals, an algorithm (called idTracker) was designed to extract a characteristic fingerprint from each animal which are then used to identify the individuals throughout the video. This technology prevents propagation of errors, and the correct identities can be maintained basically indefinitely. The algorithm has been tested on fish, flies, ants and mice, and was able to distinguish animals even when humans could not.

### 3.1.2 Leadership in motion

As mentioned in Chap. 2, hierarchy is context dependent, that is, the same group often organizes itself into different structures depending on the actual task. In the context of animals *leadership* the hierarchical structure emerges during collective motion. It differs from dominance hierarchy at least in two fundamental ways: (i) followers follow the leader in an unprompted manner, and (ii) in case there is a target, the emerging hierarchy can be related to the way in which the information flows within a group.

Due to the recent technological developments, data on a larger scale and with increasing precision have been gathered within this field of research giving the topic a special importance.

The relation describing "who leads whom" in a group defines a stand-alone, quite well measurable hierarchical structure (assuming that the one who is leading is the "dominant" one and the one being led is the subordinate individual). This is called the *leadership network,* which is *not* directly related to dominance hierarchy (Nagy et al. 2010). Rather, it is probably an interaction among dominance, kinship, the inner state of the individuals (like hunger or fear) and some outer conditions. In this section we shall overview some quantitative results on *leadership*.

From the viewpoint of leadership (and hierarchy, in general) a group whose members are able to identify each other on an individual level differs fundamentally from groups in which the members are basically identical, that is, they cannot recognize each other. In the latter case, the group might reach a collective decision regarding, for example, the direction of motion either with or without a leader. In case when there is no leader, the decision can be made based on some very simple mechanism such as direction alignment or mean value calculation, whereas in case when there *is* a leader, (but still no individual recognition) leadership is still not a well-defined stable structure, but rather a temporal, continually changing network: it is based on temporal differences such as actual level of hunger, fear, spawning inducement, etc. or some pertinent information regarding predator or food location.

The first case (no individual recognition and no leader) can be described accurately with a model proposed by Vicsek et al. (1995). In this approach self-propelled particles move with a fixed velocity on a 2D surface while aligning their direction of motion to that of others being within a given distance.

When leadership emerges from differences in the inner states of the members (and there is still no individual recognition) can be described with a model suggested by Couzin et al. (2005). In this paper they have shown how a few informed individuals can lead the entire group in which the individuals do not know which of them (if any) has information. According to the model, even if the portion of the informed individuals is small, the group as a whole can achieve high accuracy regarding the proper direction towards the food location. In fact, the larger the group is, the smaller the portion of informed individuals is needed.

Entirely different is the case when group members are able to identify each other on an individual level. Most mammals (and some birds, like pigeons) are like this, enabling the emergence of more stable hierarchical leadership structures. According to the biological observations, leadership still depends strongly on the actual inner state of the animals (Fischhoff et al. 2007), but, at the same time, from the point of the motion of the group, dominance hierarchy may play a fundamental role as well.

To study this question, Sárová et al. (2010) GPS devices recorded the motion of 15 cows belonging to the same herd for a period of three weeks. According to their findings, foraging motions and short-distance travels are not lead by a particular individual, but rather they are influenced in a graded manner: the higher position is occupied by an individual in the herd's hierarchy, the bigger influence it exerted on the collective motion. Other observations revealed that Rhesus macaques prefer to join either related or high-ranking individuals, whereas Tonkean macaques exhibited no specific order at departure (Sueur and Petit 2008). In their review paper on this topic Petit and Bon (2010) proposed that the process of collective decision making (regarding the collective motion of a group) can be interpreted as a combination of two kinds of rules: (i) an "individual-based", covering the differences in the inner states of the animals, such as hunger, physiology, energetic state, knowledge, spatial position within the group, position in the group's affiliation network, hierarchical rank, etc. and (ii) "self-organization", referring to the inter-individual interactions among group members.

In order to yield a detailed insight into the leader-follower relationships in a network of a flock of homing pigeons, Nagy et al. (2010) equipped ultra-light GPS devices on members of a flock of 10 birds. In such a way they obtained high precision data of the trajectories which then were analysed using a variety of correlation functions inspired by approaches commonly used in the field of statistical physics. In particular, they analysed the pairwise interactions on the basis of the characteristic delay times between the direction changes ("turns") of the birds. Using this method they have revealed a dynamically changing, but well-defined hierarchy (leadership network) within the flock (Fig. 3.3). According to this study, the average spatial position of a bird within the flock correlates with its hierarchical rank.

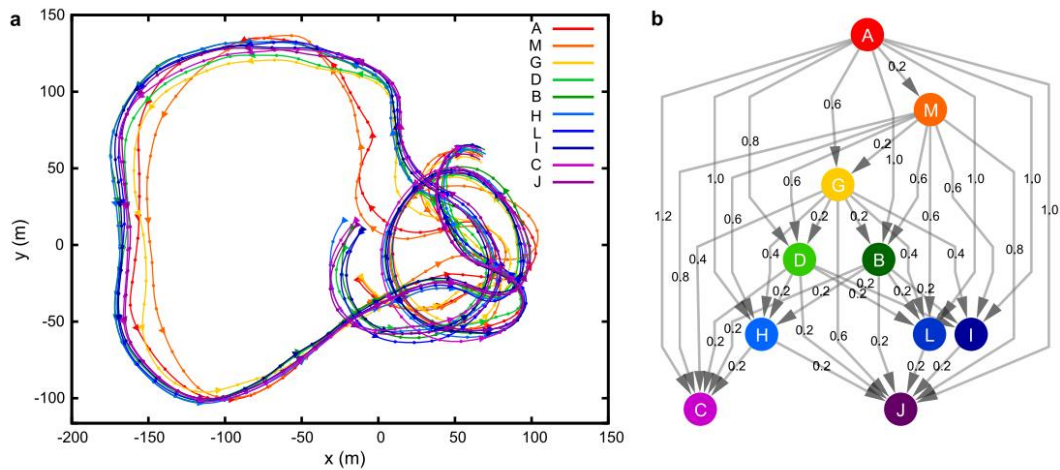

**Fig. 3.3** The analysis of the flight of a flock of homing pigeon, recorded with ultra-light GPS devices. **a** A 2-minute segment from a free flight restored from the GPS data log. Dots and triangles mark every 1s and 5s, respectively. Triangles point in the direction of motion. Different letters and colours refer to different pigeons. **b** The leadership network reflecting a single flock flight shown in **a**. The arrows point from the leader towards the follower. The values on edges mark the time delay (in seconds) in the two birds' motion. Reproduced from (Nagy et al. 2010)

The relationship between the spatial position of an individual within a moving group and its effect on the movement of others is highlighted by Schaerf et al. (2016) as well. By studying moving pairs of fish they found that those being in the front have greater mean changes in their speed and are less likely to move towards their partner than vice versa. Furthermore, the pair moves faster when the front position is occupied by the one who usually leads the pair.

Furthermore, many animal species live in "multilevel modular societies" in which smaller groups of closely related individuals form more coherent communities which are connected in a more loosely connected way creating aggregations on a given level. Among others, primates, elephants, whales and horses live in such "embedded" societies (For the embedded hierarchical structure present in human societies see Sect. 3.2.3). Ozogány and Vicsek (2015) studied the collective motion of a herd of Przewalski horses consisting of many harems (see Fig. 3.4) and found that this group structure has an effect on the collective motion of the herd as well, since the leadership network itself is modular and hierarchical: there is a leadership hierarchy *within* the dense sub-groups (or harems, in case of horses), and there is a hierarchy *among* the harems.

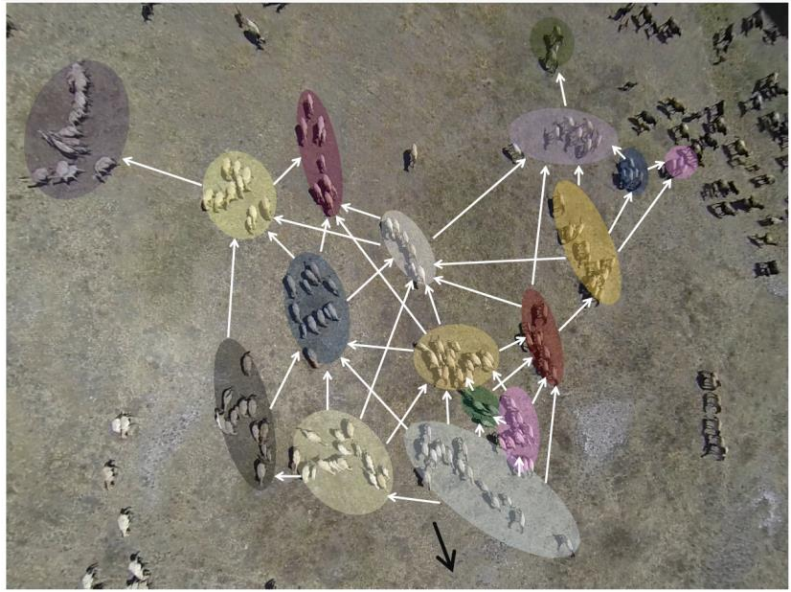

**Fig. 3.4** The motion of a herd of Przewalski horses ($n\approx150$): hierarchy in hierarchy. Different harems, identified by the cohesive motion of its members, are marked by different colours. The white arrows indicate the leadership relations among harems: they point from the leader towards the follower sub-group. Reproduced from Ozogany and Vicsek (2015).

However, these studies do not shed light on the following fundamental question: the so called "individual-based" traits, such as navigational ability or spatial position within the group are a *cause* or a *consequence* of leading? In other words: do they *govern* the self-organizing processes or are they the *consequence* of leadership, arising through another mechanism?

In order to address such question, Pettit et al. (2015) conducted measurements aiming to understand how individual differences structure a flock and affect the information transfer among the birds, and, importantly, *how the leader/follower role affects the learning of navigational skills*. In the experiment, the homing flights of groups of adding up to 40 homing pigeons were tracked using GPS devices with a log rate of 10 Hz. The "level of the leadership" for each bird was defined based on the directional correlation delay time, the method introduced by Nagy et al. (2010).

According to the study, leadership hierarchies can arise from differences in the birds' typical speeds. They also found that leaders learn faster and become better navigators, even if leadership initially did not originate from navigational ability. *In other words, individual differences which originally concerned purely physical abilities (speed) with time turned into individual differences regarding cognitive capacities (navigational skills)*. Figure 3.5 depicts this phenomenon: Speed did not correlate with homing efficiency before the group flight, but faster individuals did tend to gradually become leaders during the collective flight. After the flight these fast pigeons become more efficient as well (their navigational skills improved). This was the first time such a mechanism was reported.

The study also suggests that leadership might be an inevitable consequence of heterogeneous characteristics among individuals within a self-organized group. Furthermore, the role that an individual assumes during collective motion might have a far-reaching effect on the development of its abilities regarding how it learns about the environment and uses social information.

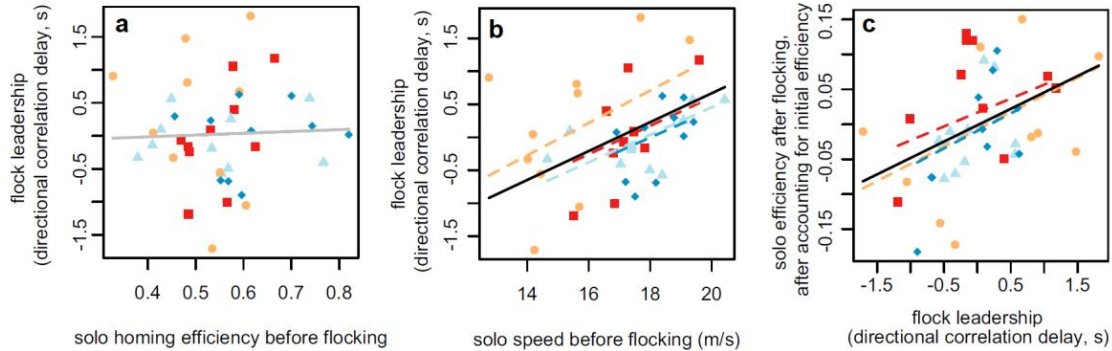

**Fig. 3.5** Solo homing efficiency and speed compared to leadership. Each dot represents the mean value belonging to a bird, with different symbols for the different groups. Fit lines correspond to linear mixed models with group as a random factor. The estimated regression for the fixed effect, if it is significant, is black, or grey if it is not significant, as judged from a likelihood-ratio test against a model without that fixed effect. Dashed coloured lines show the random effects of the groups on slope and intercept. Reproduced from Pettit et al. (2015)

In an ingenious study, Boos et al. (2014) conducted an experiment to test that (i) whether in case of humans, collective motion and coordination can emerge from applying merely simple local rules, as it is believed to be the case in animal groups, (ii) whether an informed minority can lead an uninformed majority to the minority's target, and if so, (iii) how the minority exerts its influence?

In order to hinder all sorts of communications (conscious and unconscious), except for the reading of the movement of others, subjects were playing via their avatars in a multi-client computer game. In order to activate to two basic local rules, "cohesion" and "alignment" without direct instructions, a minority of the players were rewarded higher in case of reaching a pre-defined target, while the players belonging to the uninformed majority were rewarded lower, but equally. ("Cohesion" was one of the local rules expressing that the individuals are attracted towards their neighbours' positions within a local range, and "alignment" was the other, expressing that the individuals align their speed and direction within this range to that of their neighbours.) They found that (i) directed group motion can emerge from simple local rules in case of humans as well, just as in case of animal groups, (ii) within this context, an informed minority is able to lead the group to its target, and (iii) a minority can lead the group effectively if their members are among the first to make a move, with similar initial directions.

### 3.1.3 Leadership versus dominance

The assumption that *dominant* individuals are the ones who at the same time *lead* the group, is very persuasive and intuitive. However, as mentioned in the previous section (Sect. 3.1.2), leadership hierarchy and dominance hierarchy are *not* related to each other in such a straightforward manner. Rather, these hierarchies seem to coexist within the same group without creating any kind of conflict: when it comes to collective travel those will lead the group who have better navigation skills (Nagy et al. 2010) or better information (Couzin et al. 2005), and when it comes to feeding, mating, etc., relations defined by the dominance hierarchy will prevail (Nagy et al. 2013).

The dominance hierarchy and the leadership hierarchy were compared in a more systematic way in a flock of homing pigeons, consisting of ten individuals, by Nagy et al. (2013). The two hierarchies were found to be different from each other (See Fig. 3.6 **b** and **c** and Fig. 3.7). Dominance (pecking order, depicted on subfigure 3.6 **b**) is known to be correlated to aggression and access to food, based on some individual features such as physical strength, in order to strangulate the violence to a low level within the flock. At the same time, the appearance of the stable leadership network during flights is likely to be due to a different set of individual competences.

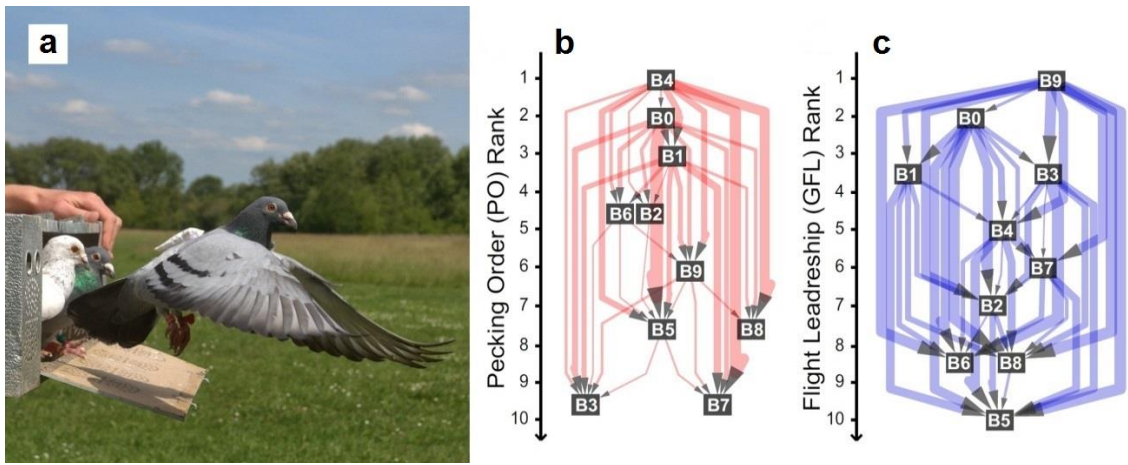

**Fig. 3.6** The comparison of the (**b**) dominance and the (**c**) leadership networks in a flock of homing pigeons. **a** Releasing the pre-trained homing pigeons from their loft. The small white bag on the back of the bird holds the ultra-light, high-precision GPS device (Courtesy of Zs. Ákos). **b** The pecking order (dominance hierarchy) and **c** the leadership network. Directed edges point from the leader towards the follower. The width of the arrow corresponds to the strength of the interaction. Nodes are ordered vertically according to the rank in the hierarchy, with the dominant ones on the top. The two hierarchies are different fundamentally. Reproduced from Nagy et al. (2013).

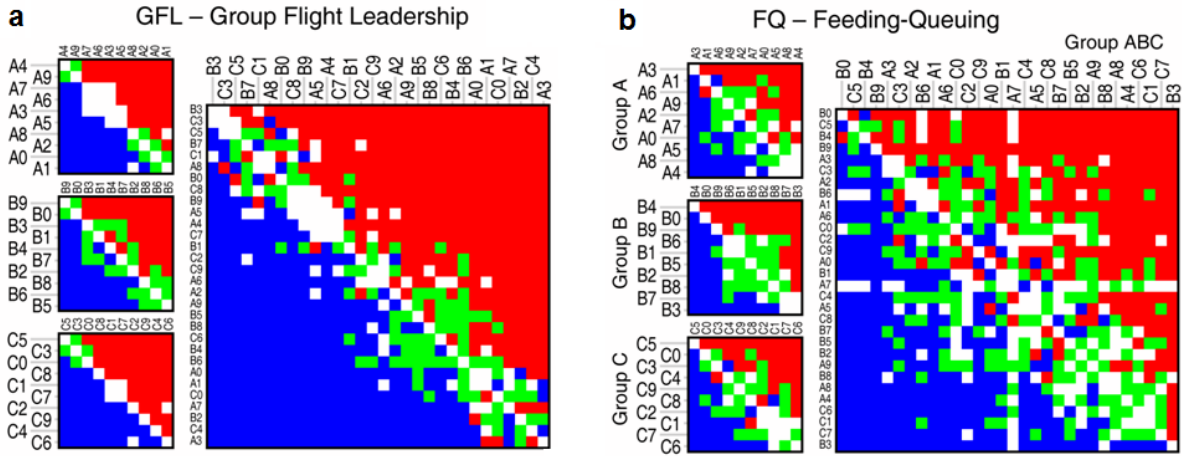

**Fig. 3.7** Adjacency matrices of the mixed graph representation containing both undirected and directed edges for each interaction type (FQ in A, AA in B, PO in C, and GFL in D). The $10 \times 10$ matrices on the left side of each panel show the data for the groups A, B, and C (from Top to Bottom, respectively), and the $30 \times 30$ matrix contains data for the group of 30. Colour indicates the type of the edge: red: directed edge pointing from dominant/leader (in the row) to the subordinate/follower (in the column); blue: directed edge, reverse direction of a red edge; green: undirected edge for mutual interaction; white: no edge. In each matrix the individuals were ordered according to the NormDS scores of that interaction. Reproduced from Nagy et al. (2013).

Using similar techniques, namely the pairwise directional correlation analysis of high-resolution spatio-temporal GPS trajectory data, Ákos et al. (2014) studied the collective motion of six dogs belonging to the same household during more than a dozen of 30 to 40 minutes unleashed walks, accompanied by their owner (see Fig. 3.8). During the walks, dogs adjust their trajectory to that of the owner, but they time to time run away and then turn back to her in a loop. On a shorter time scale, the leader-follower roles in a given pair were changing significantly, whereas on a longer timescale a consistent leadership structure was manifested.

The network constructed from these leader-follower relations is hierarchical, in which the position of a given dog correlates with the rank (dominance), age, trainability, controllability, and aggression. (These values were derived from personality questionnaires.) According to these measures, there are some personality traits which tend to characterize the leader (dominant) dogs: they were found to be more controllable, trainable, aggressive, and also to be the older ones.

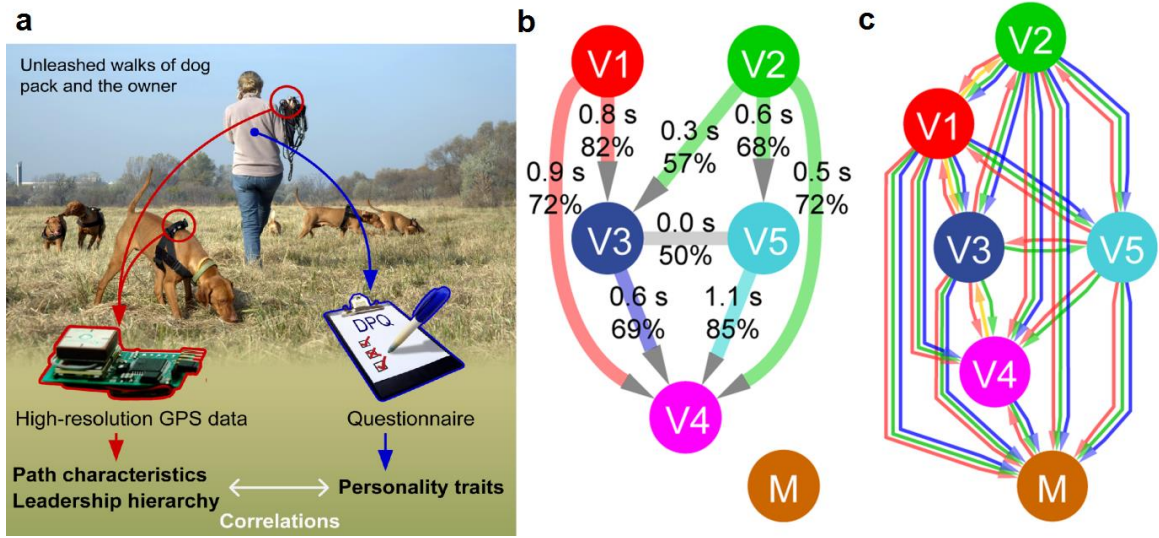

**Fig. 3.8** The collective motion of dogs (belonging to the same household) is influenced by the underlying social structure and by differences in the personalities. **a** The unleashed walks of six dogs belonging to the same household, accompanied by their owner. During these walks high-resolution spatio-temporal GPS trajectory data were collected and then analysed pairwise. **b** The basis of the analysis was the directional delay time, shown on the arrows. The directed links point from the direction of the leader towards the follower. On the lines, the upper values belong to the time delays in seconds whereas the lower values indicate the portion that the leader of that pair was actually leading. All the dogs were 'Vizsla', except for the one marked with "*M*", which was a mixed-breed. This dog did not participate in the "vizsla-network". **c** Dominance network of the dogs, derived from a questionnaire. The arrows point from the direction of the dominant individual towards the subordinate one. The colours represent the context of the dominance: red: barking, orange: licking the mouth, green: eating and blue: fighting. Reproduced f rom Ákos et al. (2014)

### 3.1.4 Collective decision-making

Animals living in groups continually face problems that require collective decision making, such as when and where to rest or to forage, how to defend themselves from predators, how to navigate towards a distant target, etc. The personal notion of the group members depends on many factors, like information, experience or inner state, such as hunger or exhaustion. Many theoretical studies focus on the cost/benefit ratio from the viewpoint of the group members, since if the individuals differ in their preferred outcome (and usually they do defer), some individuals will have to pay higher "consensus cost" than others. ("Consensus cost" is the cost paid by the animal by foregoing its preferred behaviour in order to defer to the common decision (King et al. 2008)).

The first studies addressing the problem of collective decision making mainly focussed on two basic types, both from theoretical and from an experimental point of view. In a *despotic* situation one or a few individuals make the decision, while in an *egalitarian* (or *democratic)* situation the members contribute to the final outcome to about the same degree. In nature, both types have been observed. On one hand, both theoretical and experimental studies show that the egalitarian decision-making process has a smaller average consensus cost than the despotic one (Conradt and Roper 2003), on the other hand, despotic decision-making approach can increase the efficiency of a group (Couzin et al. 2011).

More recently, along with the technical developments applied by the researchers in order to study the collective decision-making techniques, there have been some interesting observations in which egalitarian and despotic methods were alternating according to the circumstances. Using high-precision GPS data on pairs of pigeons, Bíró et al. (2006) studied the behaviour of the birds in case of conflict in the preferred flight direction. If the difference was small (smaller than a certain critical threshold) then the birds averaged their directions (egalitarian decision making), but if the difference rose above the threshold, either one of them became the leader or the pair split (despotic case).

Strandburg-Peshkin et al. (2015) identified similar decision-making methods among wild baboons. They also recorded the movement of the group members with high precision GPS devices over the course of the troop's daily activity. Baboons too, do not follow dominant individuals, rather the majority of the initiators (those starting off in a certain direction). When two groups of initiators (with similar size) heading in different directions, the followings depend on the angle between the motions: in case the angle is small (less than around 90°), the animals compromise (choose a direction in between), but in case the angle is large, they choose one direction over the other, randomly.

Importantly, these animal species live in highly hierarchical social structures, yet – according to the above studies – their collective decisions emerge via shared "democratic" process using simple rules.

In an interesting experiment Couzin et al. (2011) studied the role of uninformed individuals (individuals without any preferences regarding the direction) when two groups of initiators with different preferred directions were influencing the group motion. Their main question was that under what conditions – if any – a strongly opinionated and self-interested minority can exert its influence on the entire group. In the experiments they used a fish species, golden shiner, whose individuals' motion can be predicted with high accuracy by a model using some simple rules: (a) avoidance of collision, (b) attraction and (c) alignment (Couzin et al. 2005). The first rule, avoidance, has the higher priority. In case others are not detected within a certain region, the individual will tend to become attracted towards and aligned with its nearest neighbours within a local interaction range. Using the experimental set-up depicted in Fig. 3.9, the trajectories of the individual fish can be recorded precisely.

Some of the fish were trained to be attracted towards the blue target, some to be attracted towards the yellow target (to which the shiners exhibited a pre-existing bias), and some individuals did not have direction preference at all. Among the trained fish, the *strength* of the preferences was also manipulated. When all the individuals were trained to be attracted towards either the blue or the yellow target, the results were in accordance to the expectations: if the strength of the majority preference was at least as big as the minority preference then the group reached the majority-preferred target with a higher probability. By increasing the strength of the preference in the minority group above the preference-strength of the majority group, the minority was gaining control over the group behaviour.

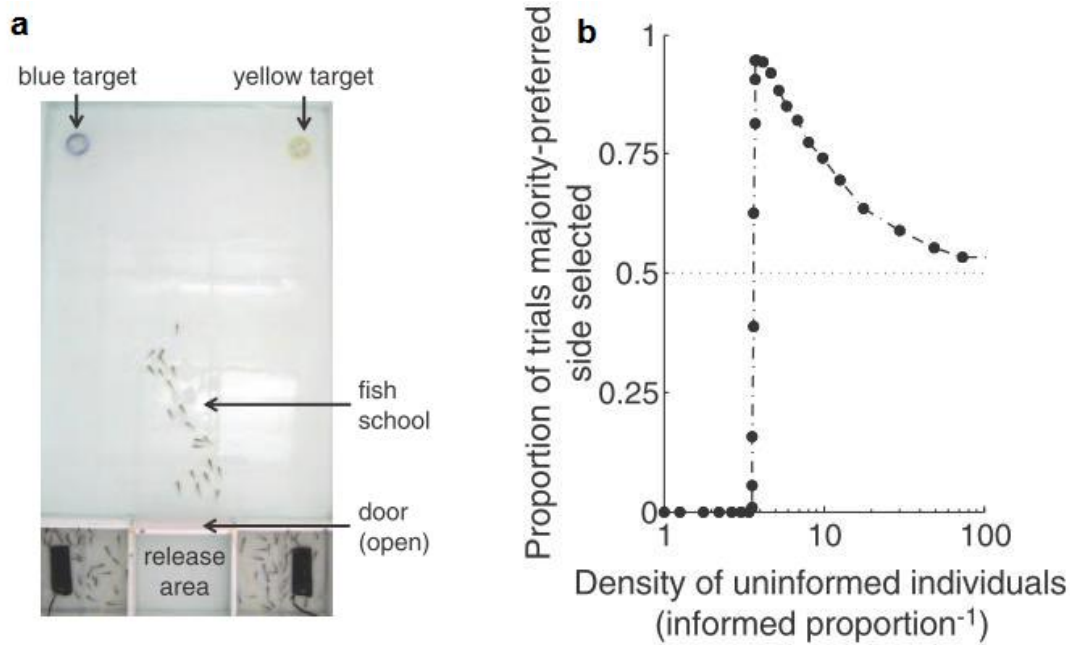

**Fig. 3.9** The experiment conducted by Couzin et al. (2011). **a** An image from an experimental video with 6 and 5 trained, and 10 untrained individuals. **b** The model describing the trajectories of the group members predicts a sharp transition from a minority- to a majority-controlled outcome as the density of uninformed individuals is increased. These simulations are in qualitative agreement with the observed behaviour. Reproduced from Couzin et al. (2011).

The most interesting results were obtained by introducing uninformed individuals into the above setup, namely, when the group was under the influence of a strongly opinionated minority *adding the uninformed individuals to the group returned the control to the numerical majority.* As their number increased, this effect reached a maximum and then slowly began to diminish. This study provided interesting new data for the often observed and widely believed argument that groups with members who are poorly informed or do not have any preferences about the decision to be made are particularly vulnerable to manipulations of determined and self-interested minorities (the latter being on the top of the knowledge-based hierarchy).

Most of the studies analyse the process of collective decision-making from an *informational* point of view, in which individuals make decisions based on their (personal or social) data merely. As Miller et al. (2013) points out, during this process the effort to maintain group cohesion plays a fundamental role as well. Another basic factor in collective decision making – apart from the knowledge, influence and group cohesion – is the structure of the *communication network* defining the way information is spread among the members. A communication network is considered to be effective if it ensures that information can spread among the entire group via a minimum number of connections.

By comparing data gathered from 24 species, Pasquaretta et al. (2014) found correlation between the neocortex ratio and the efficiency of the communication network. Both modularity (showing how strongly the group is clustered) and centralization (the ratio of central individuals) found to be inversely related to the efficiency of the network, from which they concluded that those species which are more "tolerant" have more efficient networks.

As it turns out, leadership, and the way members of various social animal species achieve collective decisions is based on many factors, such as information, actual physiological state, dominance, navigational competence, etc. The resulting leadership network is a complex, multiple level hierarchical structure.

Motivated by the above findings, a computational model was created to identify the *optimal competence*" and *pliancy* distributions within groups facing problems that had to be solved collectively (Zafeiris and Vicsek 2013). In the following part we shall discuss this model along with the results. In our interpretation, "competence" is the ability needed to solve the given task (for example navigational skill in case of collective flight), and "pliancy" expresses the extent to which an individual relies on personal versus social information (learning through the observation of others).

*Order hierarchy for making optimal decisions*

When a group faces a problem, some members have more clues, some others have less idea regarding the solution. Some individuals tend to follow others, while some prefer to rely on their own information. But what are the proper ratios in an optimally performing group, and how these characteristics relate to each other? In the present subsection, optimal performance is associated with finding an accurate solution using the smallest amount of "cost", where the most important cost is competence, since acquiring it requires knowledge, experience, learning, etc. Other aspects can also appear as cost, most importantly the time factor.

In order to address the above question, four estimation problems were considered:

   i. The simplest: an external parameter that can be either -1 or 1 (yes or no, black or white, to follow or not to follow some initiators, etc.). This can be called the "voting model", or "voting GPMM", where GPMM stands for "group performance maximization model",

   ii. A general and abstract, which is at the same time simple and thus widely applicable: Sequence-guessing (SG) GPMM, in which a sequence of values had to be estimated,

   iii. A case study: The direction finding GPMM, where a pre-defined direction had to be found.

   iv. The "Flocking GPMM" in which a group of agents has to reach a pre-defined target by moving on a two dimensional surface. This one is a special case compared to the others because here the communication structure (network) of the agents is not fixed, but changing continuously according to the individuals' actual position.

In order to elicit the possible effects of the communication structure, the first three games (models/GPMMs) are played on different types of fixed communication networks:

   a. Small-world network (SW)
   b. Erdős–Rényi (ER) network
   c. And a real-life social network describing the friendship relations in a school (referred to as "Friendship" or "Frnd" network).

Rather counter-intuitively, it was found that the optimal ability and pliancy distributions were quite independent of the type of the networks used.

The relevance of the fourth ("Flocking") GPMM originates from the fact that among animals, the most often studied collective decision-making scenario is the one during which the group has to navigate collectively towards a(n often pre-defined) target. (See also Sect. 3.1.2)

Accordingly, simulations were designed in which a group had to reach a target in the shortest possible way. In contrast with the other games, the communication network was not fixed, but varied continually as a function of the agents' location: each individual "saw", and thus interacted with others within a given range of interaction, "ROI".

In all cases, the collective decision was achieved via iterative rounds of interactions. In each round, each agent made a guess (decision) in which he/she incorporated his/her own estimate (individual knowledge) with that of his/her neighbours (public information) to a varying extent. We call this disposition to follow others "pliancy", denoted by $\lambda$ in (3.1).

Precisely, the behaviour of agent $i$ is defined by the following equation:

$$Be_i^{(t+1)} = (1-\lambda_i)f(Co_i) + \lambda_i <Be_j^t>_{j\epsilon R}, \tag{3.1}$$

where, $t$ is the time (number of iterations), $Be_i$ is the (observable) behaviour of agent $i$, for example its direction, $Co_i$ is the competence level (between 0 and 1, with higher values denoting better abilities). The estimation of agent $i$ regarding the correct solution is a function of his/her competence: $f(Co_i)$. The (observable) average behaviour of the neighbours of member $i$ at time step $t$ is $<Be_j^t>_{j\epsilon R}$, where $j(\epsilon R)$ denotes the neighbours. $\lambda_i$, the weight parameter, takes on values from the [0, 1] interval.

The optimization was done with genetic algorithm, with the competence and pliancy values evolving. The fitness function depended on the group performance $Pe$ and on the average competition level $<Co>$, according to (3.2)

$$F = Pe - K<Co> \tag{3.2}$$

where $K$ is a parameter reflecting the "cost of learning".

The best group characteristics were then approached by varying the distribution of the competence and pliancy values of the members. The process of problem solving is stopped after some simple criteria are satisfied, e.g., a given number of steps were reached, the guesses converged or they achieved a pre-defined accuracy. The optimal distribution was associated with the average distribution of the pliancy and competence values belonging to the 500 best performing (most optimal, having the largest $F$) groups.

Because of the simplicity of the model, many real-life cases can be mapped on it. The performance of the group, $Pe$, is quantifiable and characterized by a parameter taking values on the [0, 1] interval. Higher $Pe$ values mean better performance. The contribution of member $i$ to the collective problem solving depends on its competence level $Co_i$ which also takes values on the [0, 1] interval. Here too, larger values correspond to better abilities. Some noise, explicitly or implicitly, was also incorporated into the models.

In the present subsection we focus on the first three models in which the interactions were defined according to pre-defined networks.

In the Voting GPMM – having some analogy with the widely used Ising model - , the group had to find the correct answer by choosing from two options (yes/no, -1/1, etc.). This minimal model consists of two steps only:

1. First, all actors make a guess being correct with the probability proportional to their competences.

2. Then the agents count the guesses of their neighbours, based on which the group casts a vote.

The second step is done according to (3.1). Figure 3.10 **c** shows the result for the optimal competence distribution when all $\lambda_i=1$, that is, when the choices of the neighbours determine entirely the vote of an individual. This distribution ensures the highest rate of correct votes, and thus, the highest group performance as well.

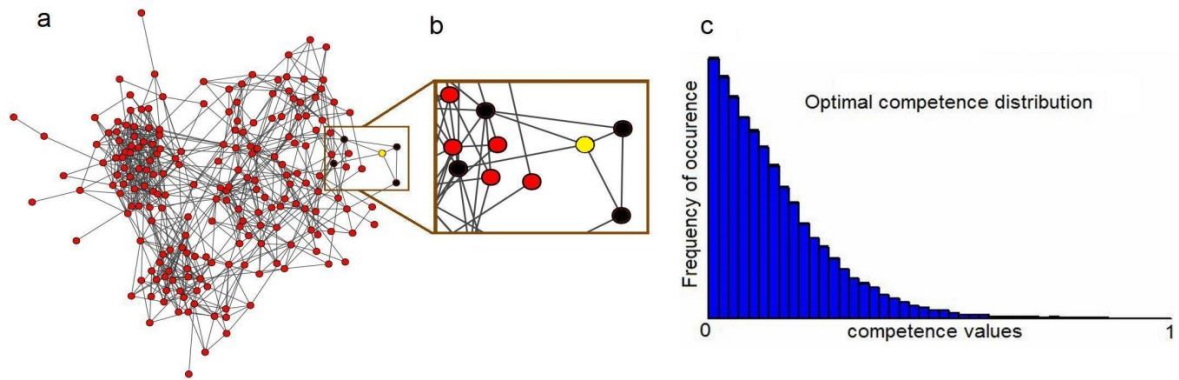

**Fig. 3.10** The Voting GPMM model. **a** "Friendship graph", the pre-defined communication network which was a real-life social network reflecting the amity relations in a high school among 204 students (after Harris et al. 2009). **b** An enlarged part of the network showing the relations of a randomly selected node coloured yellow. **c** The optimal competence distribution: a highly skewed function with a fat tail. Reproduced from Zafeiris and Vicsek (2013)

The optimal competence distributions presented in Fig. 3.10 **c** and 3.11 **d** are typical in the sense that in all cases (that is, in all models with any parameter set and network type) the optimal values form a hierarchically ordered distribution, with progressively fewer members having high competence values than low. The only difference is that in the case of the Direction finding GPMM and Sequence guessing GPMM, the fat tails are structured, having a smooth "hump" on them, while the tail belonging to the Voting GPMM is "smooth" (Fig. 3.10 **c**). Furthermore - somewhat counter-intuitively - the particular structure of the underlying communication network does not have a relevant effect on the optimal distribution. For more details see (Zafeiris and Vicsek 2013).

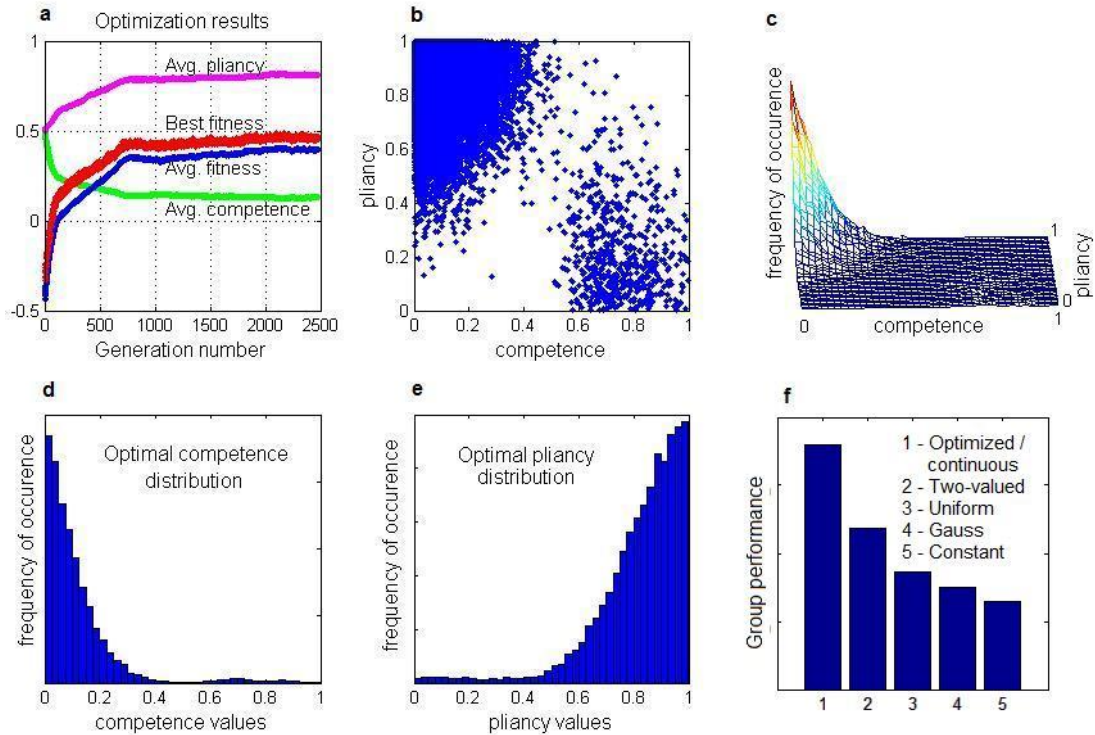

**Fig. 3.11** Results for the Direction finding model on the Friendship graph. $K=2$, the parameter describing the cost of knowledge. **a** The course of the optimization with the relevant values: pliancy, competence and fitness. **b** the competence and pliancy values depicted on a unit square. Each dot represents an individual: **c** the same as **b** but here axis $z$ depicts the density of the points. Considerably more agents have low competence values coupled to high level of pliancy than vice versa. **d** Optimal competence distribution. **e** Optimal pliancy distribution. **f** comparison of the group efficiencies $Pe$ after 20 steps of iteration for the one identified as optimal and a selection of commonly assumed distributions. From left to right: Optimized/continuous, two valued (allowed competence values were 0.1 and 0.9), uniform, Gaussian, and constant. In order to emphasize the effect of the distribution of the competence values, the pliancy values were set to be antagonistic for all five cases according to $\lambda_i=(1-Co_i)+\xi$, where $\xi$ is noise. The average competence level is identical in all cases. Reproduced from Zafeiris and Vicsek (2013).

During the optimization process, the pliancy values ($\lambda$) and the competence values ($Co$) were evolved simultaneously and independently from each other. The first and foremost conspicuous result was that the pliancy values – in analogy with the competence values – form a highly skewed distribution as well. However, in this case agents with high pliancy values made up the majority (Fig. 3.11 **e**).

Figure 3.11 **b** and **c** grants a deep insight to the *relationship* between the competence and pliancy values in an optimized group, and sheds light onto the origin of the "hump" as well. The location of each point in Fig. 3.11 **b** is defined by the corresponding individuals' competences ($x$ axis) and pliancy ($y$ axis) values. Two kinds of agents appear: one kind clusters in the top left corner, corresponding to small competence and high pliancy values (these actors have "sheep mentality", and significantly outnumber the rest of the group), while the others have considerably higher competence values mostly coupled with small pliancy. The hump – observable in most of the competence histograms – is due to the second kind of agents. It can be concluded that the simultaneous choice of both the competence and the pliancy values are essential, and the optimal choice results in a strong improvement of the efficiency of the group.

*Continuous vs. bimodal competence distributions*

We believe that the reason behind the finding that continuous competence distribution eventuates higher group performance than the "bimodal" distribution (in which the competence values can be either 0 or 1) is due to a phenomenon that we call "information spreading or mixing", which can be summarized as follows:

*Multi-level hierarchical interactions make the spreading (mixing) of the information between the individuals much more efficient than in a "two-level" system.*

This conceptual statement is based on the following assumptions: (i) the individuals do not know the competence level of the others, (ii) the pliancy values are approximately proportional to the inverse of the competence values (which is the general assumption in two-level systems), and finally, (iii) not all members interact with all others, but according to an underlying network – which is a natural assumption for groups beyond a certain size. Given these conditions, the two-level competence distribution can often result in temporarily or even permanently segregated groups maintaining different "opinions" or estimates, while the continuous distribution performs better.

The reason behind this possibility of segregation is that uninformed individuals have a strong tendency to follow the others (since they have high pliancy values). Sub-groups of the whole group can thus easily come to a conclusion corresponding to a wrong estimate or solution which they will maintain until a better estimate "diffuses" to their community from other groups having highly competent agents.

These results provide a framework for a wide selection of phenomena including several recent observations, such as: how a few well-informed individual is able to lead a group efficiently (Whallon et al. 2011, Conradt and List 2009, King and Cowlishaw 2009, Reebs 2000), the surprising observation made by company managers that a group of skilled workers do not outperform a group with diverse abilities (Surowiecki 2005, Hamilton et al. 2003) and the results of models optimizing the strategies of individuals performing a specific task as part of a collective. These findings emerge from the interaction dynamics within the collective.

## 3.2 Hierarchy in Humans

### 3.2.1 Our biological and social heritage

In the previous sections of this chapter we presented results regarding dominance hierarchy which seems to be ubiquitous in the animal world. Since humans can be regarded as a particular representative of the animal kingdom, the above mechanisms apply to us as well, although accompanied by some new features. Anatomically modern humans – with whom our bodies are indistinguishable – appeared around 200,000 years ago. About 70,000 years ago an even more important event happened, what anthropologists and historians often call "*Cognitive revolution*" (Harari 2013) – a shift that launched human culture. These changes had fundamental impact on the self-organizational processes of human groups as well, and, accordingly, on the way hierarchies were formed.

The most important elements of this new ability are that (i) humans have become able to create *formal roles* (such as chieftain, king, pharaoh, colonel, etc.) which *are independent of an actual individual*, and (ii) among these formal roles, *any kind* of network of relations – that is, e.g., hierarchy – is conceivable, starting with complete egalitarianism up to the strictest dictatorship. Importantly, only the *ability* enabling the creation of such roles and their relations is coded genetically, meanwhile the specific nature of the hierarchy is not. This latter one is coded culturally, and thus, in order to distinguish it from the dominance hierarchy, we shall call it *cultural hierarchy*. Table 3.1 shows a comparison regarding the most important features of the two types of hierarchies. Furthermore, cultural hierarchy is in close relation to our affinity towards rules and towards *following* these rules and punishing those who do not follow them appropriately.

In human groups both the dominance and the cultural hierarchies are present, and, although their origins and functioning are rather independent, they interact in a unique way with each other and this interaction, from time to time, can create conflict within (or among) the group(s).

**Table 3.1** Comparison of the two types of hierarchies present in human groups: Dominance hierarchy and cultural hierarchy.

|  <u>**Dominance hierarchy**</u> | <u>**Cultural hierarchy**</u> |
|---|---|
| • Genetically coded <br> → restricted variability: the basic features are the same within one species. | • Culturally coded <br> →can take *any* form from strict dictatorship to complete egalitarianism |
| • Controlled mainly by hormones (testosterone, stress hormones, etc.) →mostly instinctive | • Controlled mainly by the Neocortex →mostly conscious |
| • Its main purpose is to minimise the inner-group aggression by determining the access to the common resources | • Its main purpose is to harmonize the behaviour of the group members via political power. |

Distinguishing these two hierarchy types also resolves the "mystery" surrounding the "lost hierarchy" of hunter-gatherer groups. According to this view, the (mostly) egalitarian nature of hunter-gatherer groups is very difficult to explain, since from the animal world we have inherited a strong tendency towards hierarchical group organization, and after the settlement (around 10-15,000 years ago) hierarchies appeared again. So where did it disappear in between (Dubreuil 2010)? A likable answer is that it did not disappear anywhere; dominance hierarchy has been continually present throughout human history in all human groups, although often oppressed by culturally coded norms manifesting themselves – among others – in cultural hierarchy. In hunter-gatherer societies the cultural hierarchy is most often week (but not always), and is close to an egalitarian organization.

In the following, when we refer to human hierarchies, we mean *cultural hierarchy*, if not stated otherwise.

### 3.2.2 Large-scale hierarchies in societies

During the vast majority of the circa 200 thousand years of human history people lived in small-scale, mostly egalitarian hunter-gatherer communities comprising around 30-50, or at most a few hundred individuals. The beginning of the transition to larger scale and more complex societies happened only 10 to 12 thousand years ago, when humans firstly settled and engaged in agriculture and animal domestication. The first large, state level societies appeared around 5 thousand years ago (in Egypt and Mesopotamia) and the scale of the societies has been increasing ever since.

The deep reasons for these transitions – primarily the first one, marking the transition from hunter-gatherer life style to settled communities – are quite blurry. Although many theories have been proposed, none of them seems to be completely satisfactory or clinching. Furthermore, in accordance with the general nature of historical and anthropological interpretations, they are mostly descriptive, making it difficult to make predictions allowing testing against measured data. However, in recent years, there has been a gratifying increase in formal (mostly, but not exclusively agent based) models regarding historical and anthropological issues (Barceló and Castillo (eds) 2016, Grinin and Korotayev (eds) 2014, Pumain and Reuillon 2017). Agent based models combined with game theoretical approach have also been proposed (Boix 2015, Greif 2006). The book of Turchin (2003) has a relevant analytical treatment of the problem.

In the present subsection we shall overview two formal approaches describing large scale hierarchies in human societies. One approach, the work of Turchin et al., will be reviewed in more details while the work of Boix (2015) will be described only briefly. Although many important related works exist, here we selected those two of them which offer a quantitative theory with predictions which make testing against real-life data possible.

In the search for the main driving forces behind historical patterns, Turchin and Gavrilets identifies *warfare* and *multilevel selection* as the two main causes leading to complex, hierarchical societies (Turchin and Gavrilets 2009, Gavrilets et al. 2010). Their (here somewhat simplified) train of thought is the following:

- Throughout most of human history people lived in small-scale, mostly egalitarian societies.
- These tribes often engaged in warfare with each other, over various resources.
- Although selfish behaviour can be beneficial for the individuals within a group, when groups intensively compete with each other (for example during warfare) those groups have the advantage which have more cooperative and less selfish members. Thus, human societies are subject to multilevel selection.
- On the one hand, warfare has the following effects on social evolution:
  - Groups become internally more cohesive
  - It drives technological progress, including military and organizational applications
  - It triggers the enlargement of group sizes, since "God always favours the big battalions" – as formulated by either Napoleon or Turenne.
- On the other hand, the capacity of the human brain has its limits as it cannot handle the social relations in detail among more than around 150 people (known as the "Dunbar number" after Barton and Dunbar (1997)). In other words, there is a limit on the size of egalitarian, face-to-face human groups.

Given the above, there exists a pressure on the group size to grow while this enlargement has biologically rooted barriers. According to Turchin and Gavrilets, the evolutionary response to this dilemma was the appearance of

1. the ability to demarcate group membership based on cultural traits (language, dialect, clothing, etc.)
2. hierarchical organization, allowing groups sizes to grow basically *ad infinitum*, since each element within a given level of a strictly hierarchical system needs to have at most $n+1$ connections: $n$ is the "span of control", and the +1 accounts for its superior.

If this hypothesis holds, more intense warfare results in bigger polities (political entities) with more hierarchical levels. In order to test this hypothesis they conducted numerical experiments by constructing the following agent-based model: The modelled area is divided into hexagonal cells (each representing a village) as shown in Fig. 3.12. Each of these autonomous local communities are characterised by

- a base-line resource level, accounting for the heterogeneous environment, defining the productive/demographic potential of the region (a "tunable" parameter.)
- actual resource level: the base-line resource level minus the costs of the various actions in which the given community participates

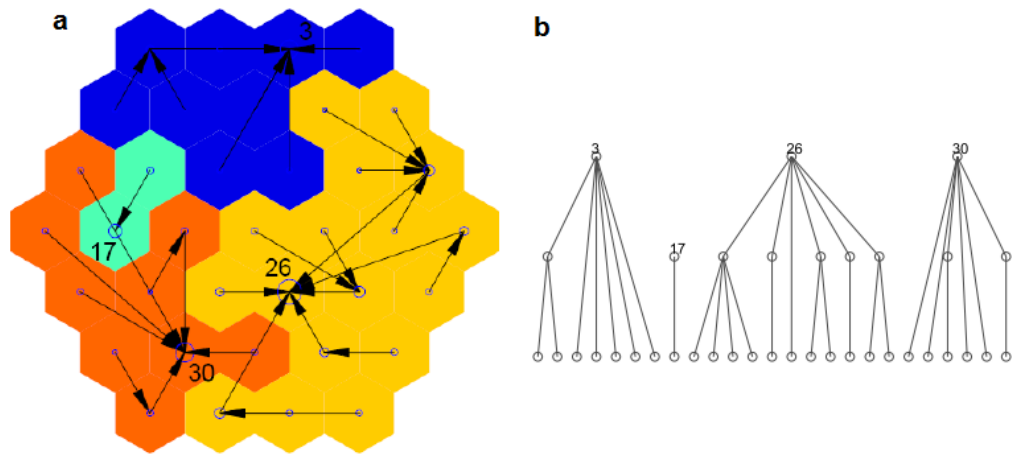

**Fig. 3.12** An example of the model's realizations. This system consist of 37 communities organising themselves into four polities The numbers in the hexagons mark the chief communities. **a** Spatial view. **b** The hierarchical structure being tree-like as in most of the agent based models of community formations. Reproduced from Gavrilets et al. (2010).

Each polity is organised in a hierarchical way (consisting of one or more villages/communities). Each subordinate community pays a fixed portion of its total resource to its chief village as tribute. Accordingly, the total resource level of a community is its base resource level minus the tribute it pays for its superior community plus the tribute it receives from its subordinates.

According to their estimated chances, polities may rebel or engage in warfare (conquest or being attacked) due to which their size permanently grows or decreases. A polity will attack its weakest neighbour if (i) it estimates that the attack will be successful, (ii) it is ready to pay the corresponding costs and (iii) it is not too devastated from previous wars. Quantitatively, the probability of an attack is:

$$A_{ij} = P_{ij} \cdot e^{-\beta c_{ij}} \cdot \frac{F_i}{F_{i,0}} \qquad (3.3)$$

where the terms are the following:

The first term, $P_{ii}$, which is the probability of the attack by community $i$ on community $j$ *to be successful*, increases the probability of the attack. The probability of success is

$$P_{ij} = \frac{F_i^a}{F_i^a + F_j^a} \qquad (3.4)$$

where $F_i$ is the power of polity $i$ and *'a'* is the 'success probability exponent'.

The second term, $e^{-\beta c_{ij}}$, accounting for the cost of warfare $c_{ij}$, decreases the probability of the attack. $\beta$ is a parameter and $F_{i,0}$ is the maximum possible power of polity $i$.

Each time step is considered to be a year. Each year the chief community of a polity decides whether to launch an attack on its weakest neighbour or not. If it decides to go on war then first it attempts to conquer the bordering communities, followed by a series of "battles" until it either suffers a defeat or the chief community of the victim polity falls. Annexing the conquered communities may require restructuring the hierarchical organization of the winner polity since the number of subordinates of any community has constraints (a parameter varying between 4 and 10).

When launching an attack, the direct subordinates of the aggressor chief community might decide to secede if they estimate that the attack will be unsuccessful. In this case the chief polity attempts to supress the rebellion, which, if successful, leads to spatial separation from the master state. All the subordinate communities of the rebelling village will secede too.

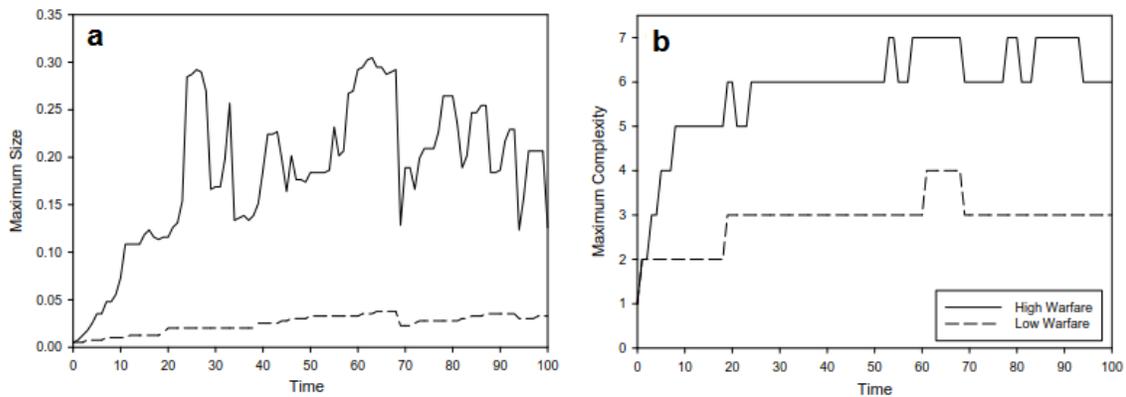

**Fig. 3.13** The size (**a**) and the hierarchical complexity (**b**) of the polities under low and high pressure of war. According to the predictions of the model, intense warfare results in larger and more complex polities. Reproduced from Turchin and Gavrilets (2009).

This model provides a fission-fusion cycle reminiscent of the dynamics characterising early states of humans. More importantly, the effect of the intensity of warfare on the complexity and sizes of polities can be analysed. As it can be seen in Fig. 3.13, low level of warfare results in considerably smaller polities regarding the extent of their territories (Fig. 3.13 **a**) with less hierarchical levels (Fig. 3.13 **b**). In contrast, high level of warfare results in larger and more complex polities, in accord with the case of historical societies (Turchin and Gavrilets 2009).

The final purpose of such models is to gain a deeper understanding of the dynamics of historical events by reproducing (at least some) properties of historical polities. Using an improved version of the above model, Turchin et al. (2013) tested its predictions for a realistic landscape of the Afro-Eurasian landscape against real historical data. The main proposition to be tested was that costly institutions enabling large human groups to stay together within one political unit and function as one society evolved as a result of warfare. More specifically, they compared the model's predictions with a large dataset documenting the spatiotemporal distribution of historical (large-scale) societies appearing in the Afro-Eurasian lands between 1,500 BCE and 1,500 CE (See Fig. 3.14).

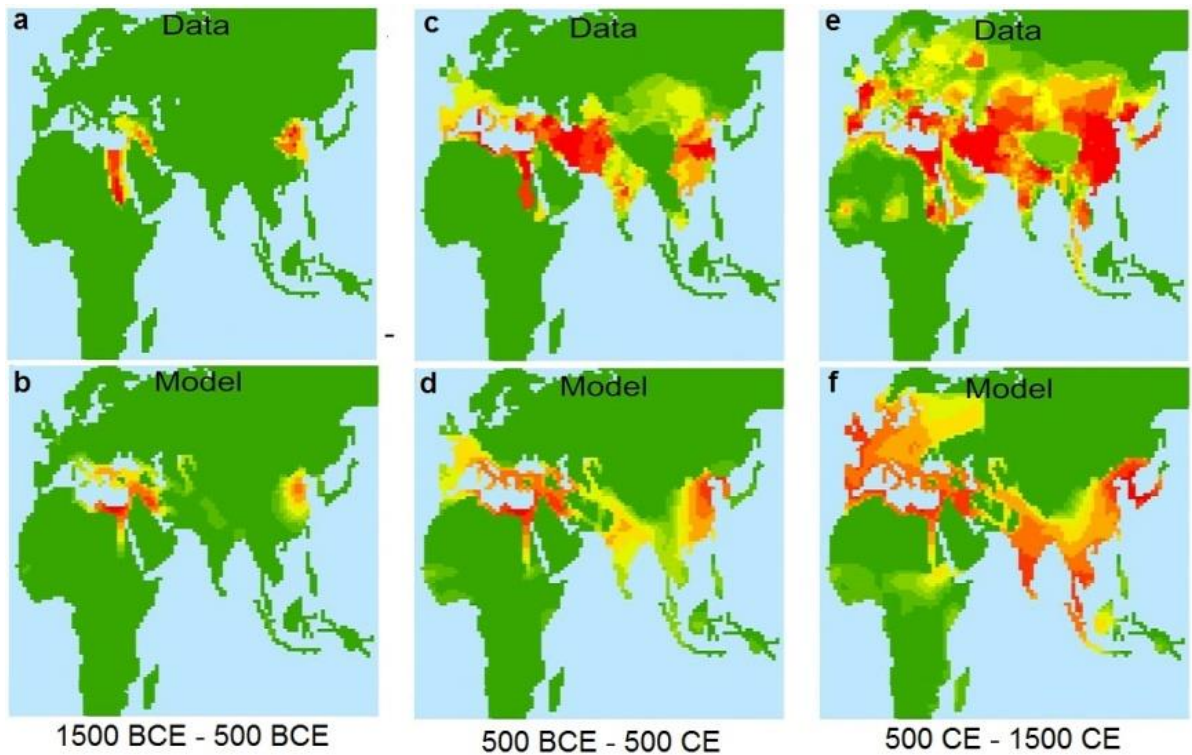

**Fig. 3.14** Comparison of the predicted (**b**, **d**, **f**) and "measured"(**a**, **c**, **e**) data. The model predictions ("Data") are averages of 20 runs. The colour codes are the following: green: the absence of large-scale polities within the given time period, yellow: moderate density of large polities, and red: regions where large polities arose frequently. Reproduced from Turchin et al. (2013)

Until now in this subsection we have seen "political hierarchy" formed by superior and subordinate communities. But this is not the only type of hierarchy that is manifested in large-scale societies. *Inequality*, the wealth distribution among members is another type, the one that is triggering much more attention and, accordingly, has much more extensive literature. However, quantitative models are scarce in this field as well. Here we shall mention only one quantitative study briefly.

By using a game theoretical approach, Carles Boix (2015) argues that voluntary cooperation exists only in small scale egalitarian communities, while economic growth inevitably brings about inequality (hierarchy), breaking down spontaneous cooperation. Economic growth usually results from technological development, such as plant domestication or the appearance of a new agricultural tool. According to his model, these changes *precede* the formation of state – which statement is in contrast with the main-stream view holding that the formation of state was first, and inequality appeared as a result of it. (In his work institutions play a much more limited role than it is usually assumed.) According to his framework, group members benefiting from the technological change can be considered as "producers", while those group members who are not, will be the "looters". The state emerges as a result of this situation. The type of a given state depends on the primary military technology: if the technology favours looters (e.g., those who use horses), monarchies tend to rise, while military machineries favouring producers (e.g. navy) call forth republican polities. The early states were mostly monarchies, but some of them were republican or mixed. Inequality tends to be higher in monarchical polities, but it depends on political institutions and other endowments as well.

### 3.2.3 Nested hierarchy structure of human societies

Up to this point, when discussing hierarchy, we have been considering *flow* hierarchy. However, the structure of human societies exhibits a *nested* nature as well with a preferred scaling ratio between 3 and 4. (For the definition of nested hierarchy see Sect. 2.) In other words, instead of a continuous spectrum of group sizes, it was argued that a geometrical series of 3-5, 12-20, 30-45, etc. individuals (Zhou et al. 2005) per group size can be observed.

Various anthropological studies report that both human (Kottak 1991) and non-human primate (Dunbar 1988) societies consist of a series of nested groups classified as (Dunbar and Spoor 1995, Hill and Dunbar 2003, Zhou et al. 2005):

- *Support clique* is the smallest one with the strongest emotional ties, "defined as the set of individuals from whom the respondent would seek personal advice or help in times of severe emotional and financial distress; its mean size is typically 3–5 individuals." (Zhou et al. 2005)
- *Sympathy group*, characteristically containing 12-20 individuals, those with special ("friendship") ties, contacting each other at least once every month.
- *Bands* (or *overnight camps*), reported in the ethnographic literature on hunter-gatherer societies (Dunbar 1993), refer to those more or less unstable groups of 30-50 individuals whose members belong to the same clan.
- *Clan* (or *regional group*) contains ca. 150 individuals. This formation is very typical in small-scale traditional societies, and this number, 150, has become known as the "*Dunbar number*" referring to the biological limits of human cognition. According to the *Social brain hypothesis* (Byrne and Whiten 1988; Barton and Dunbar 1997) the main evolutionary force acting on the formation of primate brain has been the need to remember, coordinate and manage the complex social relations within a group. Since the stability of a group depends on the intimate knowledge of each other (meaning the ability to keep count of the social ties based on which the behaviour of the group mates' can be predicted), the size of the brain (assumed to be proportional to its computational ability) imposes a limit on the group size. Above this limit, the group inevitably becomes unstable and splits up.

- **Mega-band** is the next level identified in the literature, comprising about 500 individuals, and finally the
- **Tribe** unites ca. 1000-2000 individuals, those belonging to the same linguistic unit.

Zhou et al. (2005) searched the sociological and other related literatures for quantitative data on such human groupings, based on which they constructed a dataset consisting of 61 groups. The $3^{rd}$ column of Table 3.2 indicates the mean values of the sizes for all the six group types.

Next they calculated the ratio of the sizes of the successive groups $S_{i-1}$ and $S_i$, and found that

$$\frac{|S_i|}{|S_{i-1}|} = 4.58, 3.12, 2.98, 3.11, 4.28, 3.05 \tag{3.5}$$

for $i$=1,…,6. |….| denotes the average group size ($3^{rd}$ column in Table 3.2), and the mean value of these ratios is 3.52. Based on these values, they concluded that "humans form groups according to a discrete hierarchy with a preferred scaling ratio between 3 and 4." (Zhou et al. 2005)

**Table 3.2** Human (and other primate) societies tend to have a well-defined inner structure consisting of nested communities with a scaling ratio between 3 and 4. The table shows the names of these communities ($2^{nd}$ column) and their mean sizes ($3^{rd}$ column). Reproduced from Zhou et al. (2005).

|       | Group type         | Mean group size |
|-------|--------------------|-----------------|
| $S_0$ | Ego (individual)   | 1               |
| $S_1$ | Support clique     | 4.6             |
| $S_2$ | Sympathy group     | 14.3            |
| $S_3$ | Band               | 42.6            |
| $S_4$ | Clan               | 132.5           |
| $S_5$ | Mega-band          | 566.6           |
| $S_6$ | Tribe              | 1728            |

Being interested in similar questions, Hamilton et al. (2007) reviewed the ethnographic literature on hunter-gatherer societies and analysed the data of 1189 social groups belonging to 339 hunter-gatherer societies. They also found a self-similar structure with a scaling ratio close to 4. Importantly, their database contained groups from highly distinct cultures from five continents.

These findings all indicate that there exists a kind of biologically rooted attribute defining fundamental features of human (and primate) social self-organization. A further support for this view came from the statistical analysis of a large-scale, high-precision, internet-based social network formed by the players of a massive multiplayer online game (MMOG) called *Pardus* (Fuchs et al. 2014). In this game (http://www.pardus.at), more than 400,000 players control avatars living in a virtual, futuristic world. Their interaction is based on an internal, private (one-to-one) messaging system through which they can communicate without restrictions. Among others, they can make friendships, express their sympathy or revulsion, they can trade, cooperate or defect. As a result, a superposition of dense social networks of various types (trade, friendship, communication) is spontaneously created.

The most important result is that even though the game is purely virtual – and as such it does not allow face-to-face human interactions – a highly structured social system emerges, one that strongly resembles those found among hunter-gatherers and other "real-life" human societies.

On the low levels of the friendship and communications networks, the support cliques appear, containing 5.1 individuals on average. The next level contains the "clubs" (corresponding to the sympathy groups) with 11.5 members in average. On the middle scale "alliances" appear: this type of groups is a formal establishment in the game. The average size of this formation is 24.7, and the biggest alliance contains 136 members – a value pretty close to the Dunbar number. In fact, by analysing the same game, Pardus, Szell and Thurner (2010) found that the upper limit for friendship and other communication communities is exactly the Dunbar number, a value that does not seem to change by the usage of digital media (Dunbar 2012).

The next organizational level, the "political factions" are pre-defined by the game designers, so at this level we finish the discussion of spontaneously emerging communities. However, it is important to note, that even under such conditions the group sizes on these higher levels remain of the same order of magnitude than the ones observed in real-life human societies.

## 3.2.4 Phenomenological theory of collective decision-making

This subsection addresses a common situation in which hierarchy manifests itself in a specific way. The problem to be solved is complex (or "multidimensional") and, as such, needs several "specialists" to be solved efficiently. Thus, there is a simple two-level hierarchy involved: within a given field involving distinct/particular knowledge, specialists are significantly better at solving problems than non-specialists. Furthermore, as it will be shown, in an optimally working group specialists will have at least some level of knowledge/insight related to other sub-problems. (Zafeiris et al. 2017) Thus, the situation has some analogies with that of Sect. 3.1.4 where - in order to find the best solution - the abilities of the group members were distributed according to an order hierarchy. But, here, a similar condition should hold simultaneously for a number of sub-problems.

The process of collective decision-making has generated great scientific interest for a long time (Clearwater et al. 1991, Forsyth 2006, Surowiecki 2005, Planas et al. 2015), since it is a highly relevant aspect of social group behaviour. In particular, it has been measured, argued and shown analytically that the "wisdom of crowds" can go qualitatively beyond that of the individuals (Surowiecki 2005). This statement is true for both animal and human communities (Conradt and List 2009, Nagy et al. 2010, Couzin et al. 2011). An essential, but rarely considered case is when the problem has many "dimensions", i.e., it has many facets and aspects. Under such conditions the performance of the group (the quality of the collective solution) is highly influenced by the composition of the group. Apparently, if the group members are identical, the performance of the group cannot go beyond the performance of its members. However, if the problem to be solved is complex, i.e., has a number of different aspects or dimensions (Vicsek 2002) then a group with members specialized in their own respective fields of expertise is expected to be much more efficient in providing a high quality answer than a uniform community. The stress is on the independent nature of the sub-problems, making the problem multidimensional. In a way, solving multidimensional problems can be considered as a quantitative approach to the problem of labour division (Smith 1970, Durkheim 1997) in the context of collective decision-making: the task is to bring about a decision, and the division is made among the specialists who work out the solutions for the various sub-problems.

In spite of the above almost trivial observation regarding the necessity of specialists in heterogeneous/diverse groups, a quantitative demonstration of its validity needs a carefully constructed framework. Prior works involving quantitative analysis primarily focused on problems that could be regarded as "one-dimensional" (Surowiecki 2005, Page 2010, Hong and Page 2001, Zafeiris and Vicsek 2013) in the sense that the problem to be solved needed a single kind of ability (for example, navigational skill). In the case of one-dimensional problems it has been demonstrated – by using approaches from theory (Page 2010, Hong and Page 2001), agent based simulations (Guttal and Couzin 2010), genetic optimization (Zafeiris and Vicsek 2013) and observations (Hamilton et al. 2003, Ruderman et al. 1996) - that diverse groups can outperform homogeneous ones.

Some scenarios covered by the model presented below include (i) a board of directors for a large company, (ii) groups of animals searching for resources, (iii) a government, (iv) a scientific team for interdisciplinary research, etc. In the example of the board of directors a potential candidate for a problem can be the question of where to build the next factory. The various aspects of this problem are quite diverse, each of them requiring specific knowledge, like the history of the given country, characteristics of the labour force, education, local taxation laws, geographical and logistic conditions, potential markets in the region, and so on. Importantly, the members of the group cannot get any information about the quality of their proposals from an "outsider" knowing the optimal solution *ab ovo*.

The formal description of the collective decision-making process is the following: We consider groups of $N$ individuals solving a problem $P$ having $M$ sub-problems $P_j$ ($j$=1,2,…$M$) so that each sub-problem needs a unique (specific) skill/ability to be addressed (it should be pointed out that $P$ is not specified further). Thus, a set of $N \times M$ abilities, $A(i,j)$ ($i$=1,2,…$N$) or levels/degree of skill is to be considered. $A(i,j)$ corresponds to the ability of an individual $i$ to give the best answer for the $j$th sub-problem ( - in other words, ability corresponds to competence).

Next it is assumed that $A(i,j)$ takes values from the unit interval [0, 1]. It is important to note that the cost of obtaining an ability $A$ is typically not a linear function of $A$, since achieving the capacity of perfect knowledge ($A$=1) is much more costly than achieving a partial knowledge (e.g., $A$=0.5). For the sake of simplicity we assume that the cost $C$ for obtaining ability $A$, is $C = f(A)= Const\ A^x$ (where 1< $x$, and *Const* is a constant corresponding to the relative weight of the costs).

In this framework, the optimization is done with a genetic algorithm, where the evolving parameters are the $A(i,j)$ values. The initial distribution is random, and the fitness function (to be maximized) is

$$F = Q - C \tag{3.6}$$

where $Q$ is the quality of the final collective decision.

In other words, high fitness values correspond to distributions providing the best possible solution for the smallest possible (or for a given prefixed) cost. Then, the process of collective decision-making is divided into four basic stages.

1. Each member $i$ suggests a solution for each sub-problem $P_j$ in a way that the quality of the given proposition $Q_{ij}$ depends only on $i$'s competence, $A_{ij}$. This assumption, in the simplest case means that $Q_{ij} = A_{ij}$, although, importantly, the addition of noise did not change our results.

2. Perhaps this is the most essential step of the algorithm: group members, one after another, provide an evaluation of the proposals of the other members. If a member has zero ability to evaluate the proposal for a given sub-problem, then the contribution of this member to choose an otherwise excellent proposition becomes totally erratic. Conversely, even a relatively small ability to estimate the right quality of a proposal results in a decreased level of randomness in the evaluation and, in this way, provides a more accurate estimate of the quality of the proposition. Formally this step is described as next: The evaluation of member $i'$ regarding the quality $Q_{ij}$ is denoted by $E_{ij}^{i'}$ and it is proportional to $Q_{ij}$ (the quality of that given proposal). The *accuracy* of the evaluation $E_{ij}^{i'}$ is distorted by a stochastic factor representing that those members who have small abilities to evaluate a proposal belonging to a given field of expertise $j$ (that is, $A_{i'j}$ is small), tend to make mistakes during their evaluation with an amplitude involving randomness.

$$E_{ij}^{i'} = Q_{ij} \times A_{i'j} + \left(1 - A_{i'j}\right) \times Rand \qquad (3.7)$$

where *Rand* is a random number from the (0, 1) interval.

3. "Round table discussion". This step refers to the stage when somebody (most often, but not always an expert of the given field) tries to convince other members of the group about her/his opinion by sharing her/his ideas. Formally, $X\%$ ($X$=10, 20 or 30) of the $N$ members is chosen in proportion to their ability in the given field to present their evaluations, to which the evaluations of others converge.

4. Finally, the quality of the solution for a given $P_j$ is obtained by accepting the proposal of member $i^*$ receiving the highest average evaluation. The quality $Q$ of the solution for the original problem $P$ is then obtained by aggregating these proposals (having the highest evaluations) for all the $j=1...M$ fields, after the last round.

The most important result is summarized in Fig.3.15 **c**, where each column represents a sub-problem (specialty), and each row refers to an individual (group member). The colour of the square in the $i^{th}$ row and $j^{th}$ column corresponds to $A_{ij}$, according to the colour bar (Fig. 3.15 **d**). As it can be seen, there is exactly one red square in each column, meaning that exactly one expert is needed for each field. By and large, these results overlap with the general intuition (holding that a well performing group needs a specialist for all fields, but no more "extra knowledge" is required from other members). What is less intuitive is that the rest of the squares are not homogeneously dark blue (corresponding to close-to-zero knowledge), but they are all shades of blue, meaning that in a group, optimal decision can be made if the members have at least some idea of other members' field of expertise. This result is likely to be due to better flow of relevant information among the group members.

Figure 3.15 **a** and **b** depicts the course of optimization: the fitness function $F$, the quality of the solution $Q$, the cost $C$ and the diversity $D$, as a function of the generation number $G$. In all cases we find that the optimal distribution of the abilities is highly diverse. Diversity is calculated according to (3.8)

$$D = \frac{\sum_{i,j} \left[ \left( \max_i A_{ij} \right) - A_{ij} \right]}{M \cdot (N-1)} \tag{3.8}$$

because this definition – motivated by Freeman (1978) – differentiates appropriately among the diversity of distributions in a way being both in agreement with the intuition and sensitive enough in the range determined by the actual distributions of $A_{ij}$-s.

The plots in Fig. 3.15 display the behaviour of the model for $N$=10 and $M$=14, but these features of the optimal ability distribution do not differ qualitatively for other sets of $N$ and $M$. Random initial conditions correspond to relatively low fitness values and high costs. The efficiency/fitness of a group quickly increases at the first stage of the optimization. It is to be noted that higher fitness values ($F$) correspond to higher diversity values ($D$).

On the left (sub picture **a**) the ability cost is fixed, referring to a more general situation where a fixed amount of resource can be distributed among the members. Sub picture **b** belongs to the case when there is no pre-defined limit regarding the growth of the ability values, except for the fitness function in which it appears as a cost. (Note that the *average ability* is not the same as the *cost of ability* because $C$ is not a linear function of $A$.)

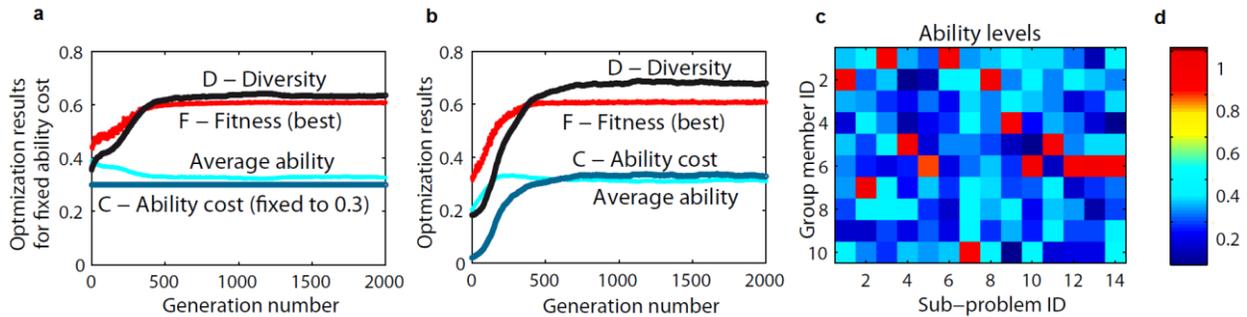

**Fig. 3.15** The process (**a**, **b**) and the result (**c**) of calculating the optimal distribution of abilities, $A^{OPT}(i,j)$, using genetic algorithm as optimization method. The course of the optimization: the fitness ($F$), the diversity ($D$) and the average ability level as a function if the generation number ($G$). In sub figure **a** the ability *cost* ($C$) is fixed to 0.3 (hence the fitness function $F$ depends only on the performance of the group $Q$). On sub figure **b** the fitness $F$ is calculated as $F=Q-C$, that is, $C$ is not fixed. The averaging is made over a population size of 2000 groups. The corresponding diversity, $D$, is indicated by black line. The groups consisted of $N$=10 members and the problem $P$ had $M$=14 sub-problems. Sub figure **c** displays the optimal ability matrix visualized by colours - the scale being indicated in **d**. These results describe a generic case, into which a few plausible assumptions are incorporated: the sub-problems have equal importance (weight) and $X$=30% of the members take role in the round-table discussion. The most relevant message of **c** is that there is one specialist for each sub-problem and, perhaps rather intriguingly, the specialists are found to have a clearly non-negligible competence concerning several of the other sub-problems. If we add some cost for the case when a single person is a specialist of more than one sub-problem, the solution ceases to have multiple specialities per person. Reproduced from Zafeiris et al. (2017)

Up to this point a certain ratio of evaluators (commenters) were assumed during the discussion phase. By fixing various values $X_e$ ($e=1,2,\ldots$) for this ratio and assuming a related cost function $C_t = \mathrm{Const}_t f(X_e)$ the full cost function becomes the sum of the formerly introduced $C$ and $C_t$.

Figure 3.16 shows how the efficiency (fitness) of a group depends on the number of contributors for various time-dependent cost coefficients. It is interesting to note that for a range of parameters the optimal number of contributors is scattered in a range around 30% of the size of the group. In this way we can support a widely observed phenomenon as well regarding the relation of the efficiency of a meeting and its duration.

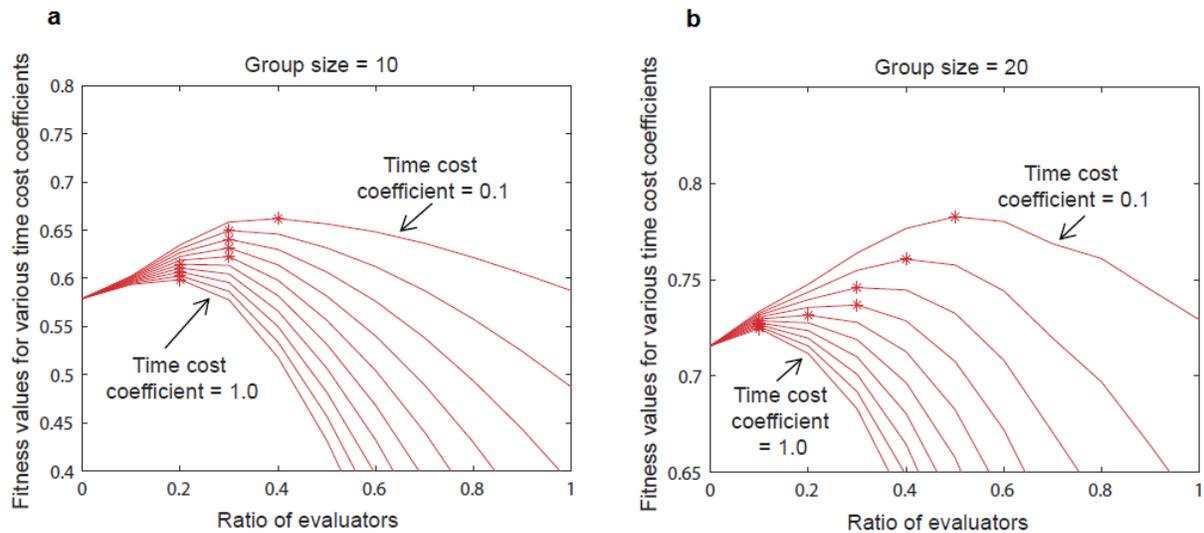

**Fig. 3.16** Efficiency of a group as a function of the number of commenters during the negotiation phase, for various time cost values, in a group consisting of **a**: 10 and **b**: 20 members. For a range of the parameters the maximum is scattered around 0.3% of $N$ where $N$ is the size of the group.

The above results are also exemplified by a number of studies on collaboration, especially on the creative groups formed by scientists, working on solving increasingly complex problems. At a recent meeting (Ball 2015) on interdisciplinary science it was concluded that productive interdisciplinary researchers have a deep knowledge of at least one field but also a working awareness of others. In other words, during broad collaborations individuals' breadth is as important as their depth of knowledge within their own field of expertise. In fact, Uzzi and collaborators have shown using huge bibliographic data sets (see, e.g., Wuchty et al. 2007, Uzzi et al. 2013) that papers of high impact tend to be produced by larger collaborations involving a broader scope of knowledge. One interpretation of these observations is likely to be related to the growing relevance of interdisciplinary research, requiring various kinds of specific scientific inputs.

The formalism discussed in the present subsection covers applications to more specific cases corresponding to various real-life situations. It can be considered as the decision-making equivalent of the "division of labour" concept. It can also be easily generalised to cases with various relative weights/influences of the members of a group (depending, e.g., on their social status in an organization). Additional future research could address further interesting questions such as, e.g., the effect of "overlapping" problems, the optimal size of a group for a given number of sub-problems, the most reasonable time interval spent on discussions, etc. Furthermore, the bilateral relations among the members (which may be interpreted as an underlying network) can also play an important role in finding the best solution.

## *Reference list*

Adler NE, Boyce T, Chesney MA et al (1994) Socioeconomic status and health: The challenge of the gradient. Am Psychol, 49:15-24

Ákos Z, Beck R, Nagy M et al (2014) Leadership and Path Characteristics during Walks Are Linked to Dominance Order and Individual Traits in Dogs. PloS Comput Biol 10(1): e1003446

Ball P (2015) Private communication

Ballerini M, Cabibbo N, Candelier R. et al (2008) Interaction ruling animal collective behavior depends on topological rather than metric distance: Evidence from a field study. PNAS 105: 1232

Barceló JA, Castillo FD (eds) (2016) Simulating Prehistoric and Ancient Worlds (Computational Social Sciences). Springer, Cham, Switzerland

Barton RA, Dunbar RIM (1997) Evolution of the social brain. In: Whiten A, Byrne RW (eds) Machiavellian intelligence II. Cambridge Univ Press, Cambridge, p 240–263

Bíró D, Sumpter DJT, Meade J et al (2006) From compromise to leadership in pigeon homing. Curr Biol 16:2123-2128

Boix C (2015) Political Order and Inequality. Cambridge Univ. Press, New Jersey

Boos M, Pritz J, Lange S et al (2014) Leadership in Moving Human Groups. PLoS Comput Biol 10(4):e1003541. doi:10.1371/journal.pcbi.1003541

Byrne RW, Whiten A (eds) (1988) Machiavellian intelligence. Oxford Univ Press, Oxford

Clearwater S, Huberman B, Hogg T (1991) Cooperative solution of constraint satisfaction problems. Science 254:1181-1183

Conradt L, List C (2009) Group decisions in humans and animals: a survey. Phil. Trans. R. Soc. B. 364:719-742

Conradt L, Roper TJ, (2003) Group decision-making in animals. Nature 421:155-158

Couzin ID, Ioannou CC, Demirel G et al (2011) Uninformed Individuals Promote Democratic Consensus in Animal Groups. Science 334:1578-1580

Couzin ID, Krause J, Franks NR et al (2005) Effective leadership and decision-making in animal groups on the move. Nature 433:513-516

Creel S (2001) Social dominance and stress hormones. Trends Ecol Evolut 16(9):491-497

Dubreuil B (2010) Human Evolution and the Origins of Hierarchies: The State of Nature. Cambridge University Press, Cambridge

Dunbar RIM (1988) Primate social systems. Chapman & Hall, London

Dunbar RIM (1993) Coevolution of neocortex size, group size and language in humans. Behav Brain Sci 16:681–694

Dunbar RIM (2012) Social cognition on the internet: testing constraints on social network size. Philos Trans R Soc Lond B Biol Sci 367:2192-2201

Dunbar RIM, Spoor M (1995) Social networks, support cliques and kinship. Hum. Nature 6:273–290

Durkheim E (1997) The Division of Labor in Society. The Free Press, New York

Eibl-Eibesfeldt I (1990) Dominance, Submission, and Love: Sexual Pathologies from the perspective of Ethology. In: Feierman J R (ed) Pedophilia - Biosocial Dimensions. Springer New York, pp 150-175

Fischhoff IR, Sundaresan SR, Cordingley J et al (2007) Social relationships and reproductive state influence leadership roles in movements of plains zebra, Equus burchellii. Anim Behav 73:825-831

Forsyth DR (2006) Decision making. In Forsyth DR, Group Dynamics, 5th edn. Thomson Wadsworth, Belmont CA, p 317-349

Freeman LC (1978) Centrality in social networks: Conceptual clarification. Soc. Networks 1(3):215-239

Fuchs B, Sornette D, Thurner S (2014) Fractal multi-level organisation of human groups in a virtual world. Sci Rep 4:6526. doi:10.1038/srep06526

Gavrilets S, Anderson DG, Turchin P (2010) Cycling in the Complexity of Early Societies. Cliodynamics 1:58-80

Gerencsér L, Vásárhelyi G, Nagy M et al (2013) Identification of Behaviour in Freely Moving Dogs (Canis familiaris) Using Inertial Sensors. PloS one 8(10): e77814

Gesquiere LR, Learn NH, Simao MCM et al (2011) Life at the Top: Rank and Stress in Wild Male Baboons. Science 333:357-360

Greif A (2006) Institutions and the Path to the Modern Economy: Lessons from Medieval Trade. Cambridge Univ. Press, New York

Grinin L, Korotayev A (eds) (2014) History & Mathematics: Trends and Cycles. Uchitel, Volgograd

Guttal V, Couzin ID (2010) Social interactions, information use, and the evolution of collective migration. Proc. Natl. Acad. Sci. U.S.A. 107:16172-16177

Hamilton MJ, Milne BT, Walker RS et al (2007) The complex structure of hunter-gatherer social networks. Proc R Soc B 274:2195-2202

Hamilton BH, Nickerson JA, Owan H (2003) Team incentives and worker heterogeneity: an empirical analysis of the impact of teams on productivity and participation. J Pol Econ 111:465-497

Harari YN (2013) Sapiens: A Brief History of Humankind. Random House, London

Hill RA, Dunbar RIM (2003) Social network size in humans. Hum Nature 14:53–72

Hong L, Page SE (2001) Problem solving by heterogeneous agents. J. Econ. Theory 97:123-163

Hölldobler B, Wilson EO (2008) The superorganism: The Beauty, Elegance and strangeness of Insect Societies. W.W. Norton & Co., New York

Huber P (1802) Observations on several species of the genus apis, known by the name of humble bees, and called bombinatrices by linneaus. Trans Linn Soc Lond 6:214-98

King AJ, Cowlishaw G (2009) Leaders, followers and group decision-making. Commun Integr Biol 2:1-4

King AJ, Douglas CMS, Huchard E et al (2008) Dominance and affiliation mediate despotism in a social primate. Curr Biol 18: 1833–1838

Kottak CP (1991) Cultural anthropology, 5th edn. McGraw-Hill, New York

Mazur A, Lamb ThA (1980) Testosterone, status and mood in human males. Horm. Behav.14:236-246

Miller N, Garnier S, Hartnett AT et al (2013) Both information and social cohesion determine collective decisions in animal groups. PNAS 110(13): 5263–5268

Muller MN, Wrangham RW (2004) Dominance, cortisol and stress in wild chimpanzees (*Pan troglodytes schweinfurthii*). Behav. Ecol. Sociobiol. 55(4):332-340

Nagy M, Akos Z, Biro D et al (2010) Hierarchical group dynamics in pigeon flocks. Nature 464(7290): 890-893

Nagy M, Vásárhelyi G, Pettit B et al (2013) Context-dependent hierarchies in pigeons. *PNAS* 110(32):13049-13054. doi:10.1073/pnas.1305552110

Ozogany K, Vicsek T (2015) Modeling the emergence of modular leadership hierarchy during the collective motion of herds made of harems. J Stat Phys 158: 628-646

Page SE (2010) Diversity and Complexity Primers in Complex Systems. Princeton Univ. Press, New Jersey

Pasquaretta C, Levé M, Claidière N et al (2014) Social networks in primates: smart and tolerant species have more efficient networks. Sci Rep 4: 7600, doi:10.1038/srep07600

Pérez-Escudero A, Vicente-Page J, Hinz RC et al (2014) Nat. Methods 11:743-748. doi:10.1038/nmeth.2994

Petit O, Bon R (2010) Decision-making processes: The case of collective movements. Behav Processes 84: 635-647

Pettit B, Ákos Zs, Vicsek T et al (2015) Speed Determines Leadership and Leadership Determines Learning during Pigeon Flocking. Curr Biol 25: 3132-3137

Planas I, Deneubourg J-L, Gibon C et al (2015) Group personality during collective decision-making: a multi-level approach. Proc. R. Soc. B. 282:20142515. Doi:10.1098/rspb.2014.2515

Pumain D, Reuillon R (2017) Urban Dynamics and Simulation Models (Lecture Notes in Morphogenesis). Springer, Cham, Switzerland

Reebs SG (2000) Can a minority of informed leaders determine the foraging movements of a fish shoal? Anim Behav 59:403-409

Ruderman M, Hughes-James M, Jackson S (eds) (1996) Selected Research on Work Team Diversity. Am. Psychol. Assoc., Washington DC

Sapolsky RM (1983) Endocrine aspects of social instability in the olive baboon. Am. J. Primatol. 5:365-372

Sapolsly RM (2004) Social status and health in humans and other animals. Annu. Rev. Anthropol 33:393-418

Sapolsky RM (2005) The Influence of Social Hierarchy on Primate Health. Science 308(5722):648-652. Doi: 10.1126/science.1106477Sárová R, Spinka M, Arias

Sarova R, Spinka M, Panama JLA et al (2010) Graded leadership by dominant animals in a herd of female beef cattle on pasture. Anim Behav 79:1037–1045

Schaerf TM, Herbert-Read JE, Myerscough MR et al (2016) Identifying differences in the rules of interaction between individuals in moving animal groups. arXiv:1601.08202

Smith A (1970) The Wealth of Nations. Penguin Books, Baltimore

Strandburg-Peshkin A, Farine DR, Couzin ID et al (2015) Shared decision-making drives collective movement in wild baboons. Science 348(6241) 1358-1361

Sueur C, Petit O (2008) Organization of group members at departure is driven by social structure in macaca. Int J Primatol 29:1085-1098

Surowiecki J (2005) The wisdom of crowds. Anchor, New York

Szell M, Thurner S (2010) Measuring social dynamics in a massive multiplayer online game. Soc. Networks 32:313-329

Turchin P (2003) Historical Dynamics: Why States Rise and Fall. Princeton Univ. Press, New Jersey

Turchin P, Currie TE, Turner EAL et al (2013) War, space, and the evolution of Old World complex societies. PNAS 110(41): 16384-16389

Turchin P, Gavrilets S (2009) Evolution of Complex Hierarchical Societies. Soc. History Evol. 8(2): 167-198

Uzzi B, Mukerjee S, Stringer M et al (2013) Atypical Combinations and Scientific Impact. Science 342:468-472

Vicsek T (2002) Complexity: The bigger picture. Nature 418:131-131

Vicsek T, Czirok A, Ben-Jacob E et al (1995) Novel type of phase transition in a system of self-driven particles. Phys Rev Lett 75(6): 1226-1229

de Waal F (2007) Chimpanzee politics: power and sex among apes. 25[th] ed. John Hopkins Univ. Press, Baltimore

Weisfeld GE, Beresford JM (1982) Erectness of posture as an indicator of dominance or success in humans. Motiv Emotion 6:113-131

Whallon R, Lovis WA, Hitchcock R (eds) (2011) Information and its Role in Hunter-Gatherer Bands. CIoA Press, Los Angeles

Wuchty S, Jones B, Uzzi B (2007) The Increasing Dominance of Teams in the Production of Knowledge. Science 316:1036-1039

Zafeiris A, Kován Zs, Mones E et al (2017) Phenomenological theory of collective decision-making. Phys A 479:287-298

Zafeiris A, Vicsek T (2013) Group performance is maximized by hierarchical competence distribution. Nat. Commun. 4:2484

Zhou WX, Sornette D, Hill RA et al (2005) Discrete hierarchical organization of social group sizes. Proc. R. Soc. B 272:439-444. Doi:10.1098/rspb.2004.2970

# 4 Experiments on the emergence and function

By the nature of the subject, it is very difficult to obtain data about the emergence of hierarchy in actual living systems. In most of the cases the process is too slow, and the documentation of the relations among the organism is too difficult for being available in the required details. For example, it is clear that evolution resulted in hierarchies both concerning a single organism as well as a whole social community of them. The available data is more like a timeline than a set allowing a deeper insight into the process itself. In this Chapter we discuss two experiments which were used/designed to track down how a hierarchy of leader-follower relationships can emerge in a group of humans over a week, or even about an hour.

## *4.1. The Liskaland camp experiment*

### 4.1.1. The Liska model of economy

In his theory, Hungarian economist Tibor Liska introduced a model of a "trans-capitalistic" socio-economic system (Liska 2006, Liska 2008). This system would be trans-capitalistic as it would operate in a way that is self-regulating through a "pure" market and unlimited competition to a higher degree than present day capitalism. In this model, property itself is fully open to competition as gaining control over property in open competition is regarded as a fundamental human right. The model allows the state to have only the role of a „referee". Accordingly, this system needs a drastic reduction of the role of the public sector and it must be totally self-controlled. The self-controlled economy would also manage redistribution, education, environmental problems and all other socio-economic subsystems much more efficiently than present-day economies. The theory envisions a society without taxation, where all income is fully personal and all property (that is, means of production) is social but is in personal stewardship. The research produced substantial results.

In the model the means of production should be owned by those who can generate the maximum profit from them, so properties would be openly and freely competed for. Everybody must have the right and the option to become a runner in this competition starting from a base level and, depending on his/her performance, can move forward. To achieve this goal, a new form of ownership was proposed: it is termed *personal social ownership*. It stands between private ownership and tenancy-type holding of property, while its status is significantly closer to private ownership. On the basis of personal social ownership, the bidder who guarantees the maximum of long-term profits will be selected for the position as owner, under the condition that he/she can keep this position only as long as he produces the maximum of long-term profits.

The basic principle is that anyone has the right to make decisions and his further possibilities should depend on the result. According to Liska, "…systems in which one decides and others bear the cost of the decision are not desirable". The model assumes that people want to arrange their own things and do not like if others tell them what to do. While the other expectations of the model are justifiable relatively easily, this one is an exception. There are numerous people who do not like to make decisions even concerning their own interests. They prefer to follow a leader who tells them what is right and what is wrong.

From among the elements of the model, the two most essential have been elaborated in precise detail.

- *Social inheritance* means that the income produced by one generation will be redistributed among the members of the next generation, which would also guarantee the most efficient way of spending money.
- *Personal social ownership* would meet the selection criteria of putting the right man in the right place, the right man being the person who is able to manage the property in the most profitable way.

## 4.1.2 The experiment

Due to the above mentioned tendency of people trying to follow those who they think are better at making good decisions, we expect that in a simple realization of the "Liska economy" a system of leader-follower relationships is gradually built up. Such a highly simplified version took place in a 2012 summer camp called "Liskaland" for participants coming with various backgrounds (the related results – see below – have not been published). The camp is a mini-society following the rules of the Liska model. Participants had to make financial decisions and competed to gain more "öki" – the currency of the camp, with those turning out after a week to be the most successful ones winning prizes. In the Liska model, certain rules of the game are set, but otherwise there is an extremely minimal role of the state and the public sector, and an emphasis on unlimited market competition (Cropanzano and Mitchell 2005). The participants of the camp could bid for ownership of certain enterprises on the first day of the camp. During the camp these enterprises provided basically all of the services, such as accommodation, food, etc. Nothing was for free – participants had to buy these services. They also had to organize the functioning of these services (as owners of the services or as paid helpers of the owners). The camp lasted for 8 days. Each day of the camp simulated 1 year, so events happened fast, with enterprises switching often between owners. Financial decisions had to be made in situations where there were several uncertain factors (what will be the income of the enterprise, etc.). Further details can be found in Liska (2011).

*Dataset* — 96 people (mainly university students) participated in the camp, 82 of them were competitors, the rest took part as organizers and as workers in the "state" companies, such as the bank. On the first day the competitors were asked to fill out a questionnaire, and 73 of them filled it out. Average age of the respondents was 22 (Mones et al. 2013).

In case of a range of *monetary transactions* competitors had to check into a *computer system* with username and password to manage their transactions. At this point they were presented with questions on the computer monitor regarding their decision making. They were offered pen-drives as incentives to answer the questions and the camp leaders also repeatedly asked participants to help the research. As additional incentive, those who cooperated were offered information on how many other people marked them in their own replies to these questions. In this paper, we concentrate on answers to the following query: "Whose decisions did you follow (by making similar decisions) – when making economic and other decisions?" Participants were given this information linked to the question: "Following someone's decision means that you have reached a decision, because you found out about that persons decision and you chose to do in similar fashion." As answer they could choose one or more people from the participants of the camp. 72% of the competitors answered the questions at least once during the camp.

*Local reaching centrality* — A directed network was then built based on the answers. In the network, each node represents a participant and there is an edge pointing from *A* to *B* if participant *B* followed the decision of participant *A* (i.e., *B* marked *A* in the questionnaire), see Fig. 4.1. As a simple filtering procedure, the participants who were chosen by a single participant but did not choose anyone were removed, thus reducing the effect of incidental answers. In order to describe the influence of each participant, the local reaching centralities, $C_R(i)$ were calculated (see Sect. 2.1.2).

### 4.1.3 Results

*Hierarchical layout* — Visualization of the hierarchical relations inside the network of decisions based on following the others is based on the distribution of $C_R(i)$ (see Sect. 2.1.2) .

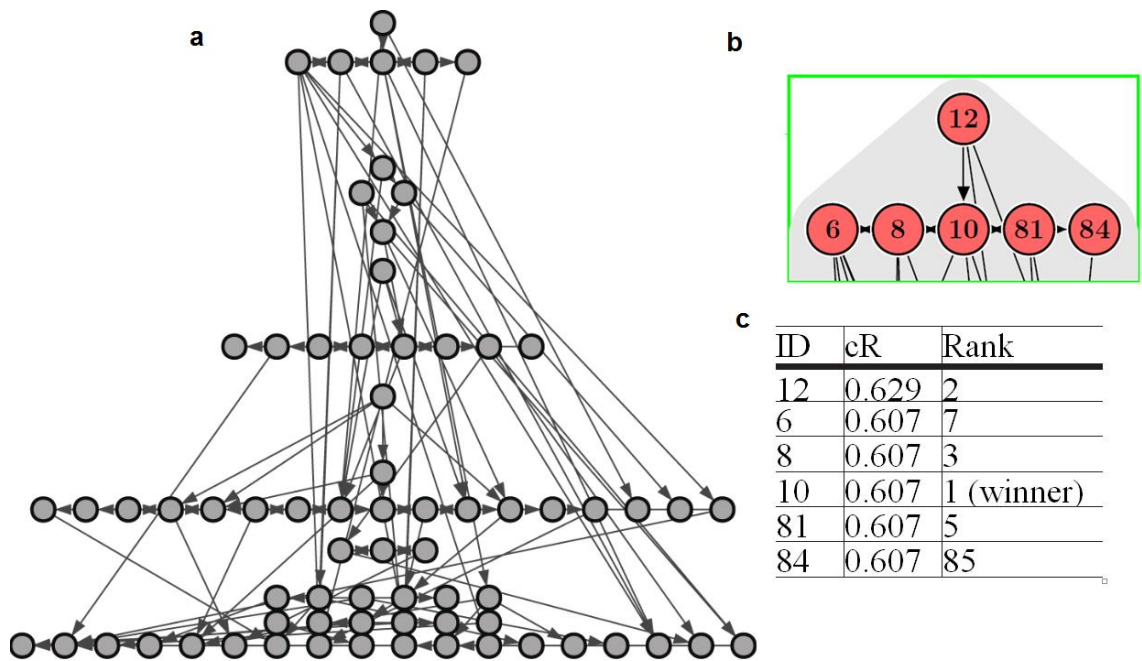

| ID | cR | Rank |
|----|-------|------------|
| 12 | 0.629 | 2 |
| 6 | 0.607 | 7 |
| 8 | 0.607 | 3 |
| 10 | 0.607 | 1 (winner) |
| 81 | 0.607 | 5 |
| 84 | 0.607 | 85 |

**Fig. 4.1** Network of leader-follower relationships corresponding to the directed interactions between the participants of the "Liskaland" experiment. The graph in **a** was generated by the approach reviewed in the visualization part. Reproduced from Nepusz and Vicsek (2013). In **b** the uppermost part of the whole network is shown, while **c** shows the corresponding ID-s, reaching centralities and final ranks of the participants. Visualization was made according to Sect. 2.2.2.

In Fig. 4.2, we plot the histogram of the local reaching centrality of the Liskaland network in comparison with those of the random network. The random network (Erdős–Rényi graph) is generated by taking *N* number of nodes and adding *M* number of directed edges between randomly chosen nodes (with uniform distribution), where *N* and *M* are the number of nodes and edges in the Liskaland network. The corresponding distribution of the random network is determined from the average of $10^6$ realizations. It is obvious that the distribution of $C_R(i)$ in the Liskaland network is more heterogeneous, having a high peak at very small values and decreasing slowly. This means that most of the participants do not have influence on the others. However, there is a small fraction of the participants that have large local reaching centrality. On the contrary, the distribution of $C_R(i)$s in the random graph is dominated by the peak at large values, meaning that most of the nodes can reach a very large portion of the graph. This suggests that the role of participants in the Liskaland network is more heterogeneous compared to the random graph, which is a sign of the hierarchical organization. The above observation is also confirmed by the global reaching centrality values: for the Liskaland network, $G_R^{(Liskaland)} = 0.628$ while for the random network, $G_R^{(Random)} = 0.194$ with a standard deviation of $0.057$. The statistics on the random graph is obtained by a sample of the size $10^6$. This means that the experimentally observed network is more hierarchical than a corresponding random graph with a significance level of more than 99.9%.

**Fig. 4.2** Distribution of the reaching centrality values for the network of interacting participants and for a random network of the same size

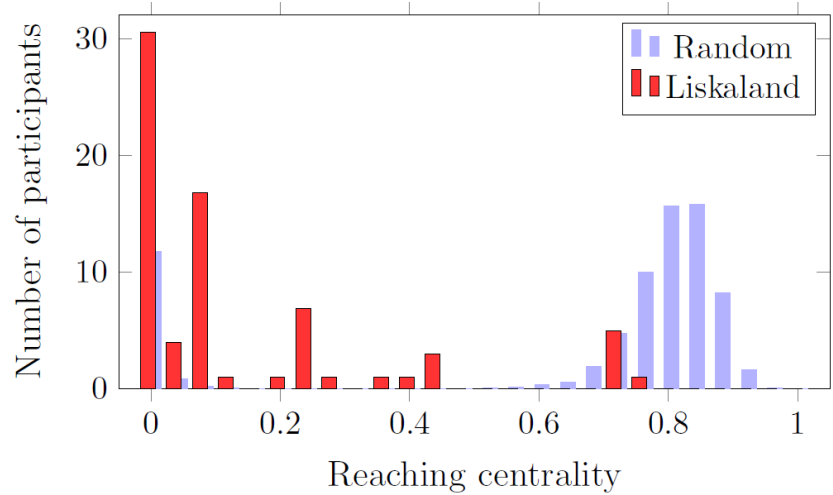

## 4.2 Picturask

### 4.2.1 The "game"

Next we discuss an experiment named as Picturask (Mones et al. 2015) which was conducted over the Internet. The subjects participated in a game, where everyone had to take part in a common task (estimating the number of circles in an image), and the players were able to see the guesses of the others by clicking on the other players' tiles (displayed on the screen in order to inform the players about their performance). The player's actions through the game had been recorded, and these records were later used to inference the behaviour of the players. After the game, deep interviews were conducted with some of the players to see if the quantitative evaluations were in agreement with the players' intentions.

From social psychological aspects the players in the game form a small group structure which can be characterized by interdependence where individuals act in a common interpersonal space, while they influence each other's actions in a special way (Johnson and Johnson 2005). In this Picturask, this space is somewhat artificial because it is determined by the rules of the game; for example direct communication was not allowed: players could only see the others' anonymized guesses. It depended on the player whether he/she considered and/or accepted the tip of a number of selected other players as an advice.

This setting is similar to the one considered in social exchange theory (Cropanzano and Mitchell 2005) where the subject of interaction is the information about each other's tips. Players gain this information from the colour of the tiles therefore tiles can be identified as heuristics. The theory of heuristics describes the stereotyping pattern of human being where a complex question may be often answered with a simplifying method – like players tend to judge a tip based on the colour of tiles (darker is better).

Altogether 170 users signed up for the game of which 96 participated in more than 90% of the turns. The majority (89%) of the players were of age 18 to28, thus 65% of them received at least bachelor's degree or equivalent, and33% accomplished high school. 56% of the players were female. Subjects were divided into three groups to play the game, each group playing on different weeks, for three consecutive days. Participation was encouraged by a 2000HUF reward. (Equivalent to 6.7 EUR, approximately 1% of average salary in Hungary). Four players with the best results were assured to get special awards as motivation.

## 4.2.2 Methods

A player participated in the experiment on-line, from home, or from any other place, they could have access to the World Wide Web. The game was based on standard LAMP architecture, and players were able to participate through a standard Internet browser. The competition involved making 45 decisions/guesses and took less than an hour per session. Sessions were arranged 3 times per week and each starting at 7p.m. over a time interval of 3 weeks.

During the game the participants were asked to make a simple estimation: they had to guess the number of bubbles on randomly generated images (Fig. 4.3). As the game advanced, the correct answer changed with a relatively slow frequency in order to mimic the collective decision making process modelled by Nepusz and Vicsek (2013). The actual number of circles changed once in five turns, even though this was unknown for the players. Whilst answering, the player could reveal the previous answer of a maximum of 10 other players in each turn by clicking on its tile (Fig. 4.4). This act was considered as "asking for advice", and the player being asked was called advisor. One of the motivations for this process was to see if the results of the related agent-based simulations (Nepusz and Vicsek (2013), Sect. 5.1) could be reproduced in real life.

**Fig. 4.3** This is a sample picture illustrating the one shown to the players who have to guess the interval between two numbers (see Fig.4.4) into which the number of coloured disks falls.

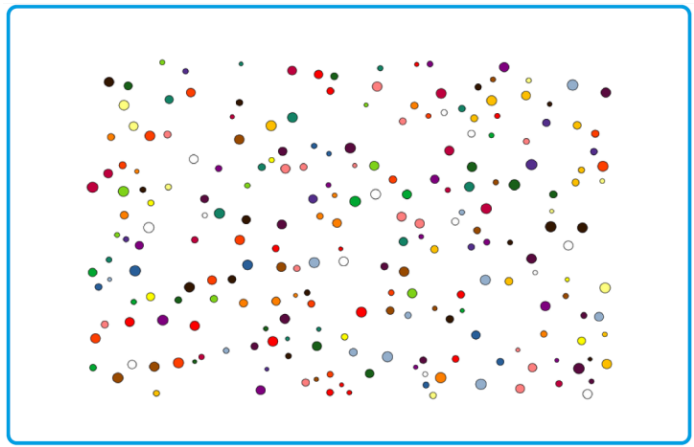

Depending on the ratio of advices that turned out to be correct, the tiles of the rivals became darker, indicating those advisors who gave the player the best advices. The estimated knowledge (the ratio of correct advices) was also displayed on the rival tile, but no other information was available. In fact, the rivals were displayed in a random permutation for each player in order to avoid potential biases introduced by the order of the tiles (Fig. 4.4).

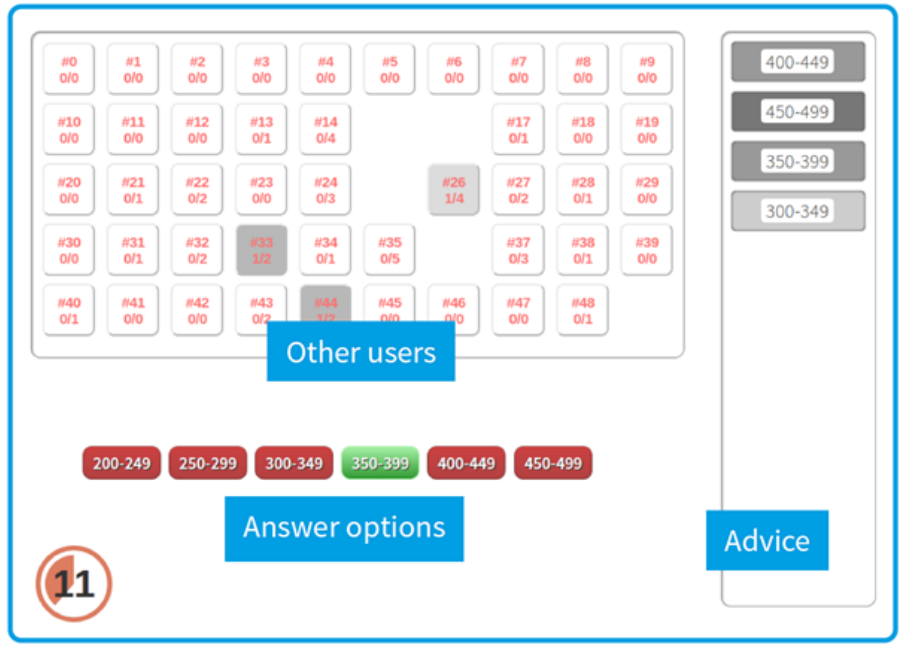

**Fig. 4.4** Snapshots of the Picturask interface. The task was to estimate the number of circles shown for a limited time (5 seconds). On the answering interface (above), users were allowed to find out the previous answer of selected (by them, and at most 10) other users. They were provided by a panel of the others (Other users) which also displayed the success ratio of the previously inquired users encoded in the colour of each tile. Inquired advices were displayed in the Advice panel.

For the analysis of the game logs various methods were used in order to represent the social structure at the end of the game by a graph. In this graph nodes represented the players, and a directed edge indicated a leader-follower type relation between the players: in the case the source node often asked advice from the target node the direction of the edge pointed from the target to the source node (Fig. 4.5). Global reaching centrality (*GRC*) was used to quantitatively describe the level of hierarchy in the resulting networks. Importantly, *GRC* was significantly higher in real setups than in those with a randomized version of the corresponding networks or the randomized control experiments.

## 4.2.3 Results

The primary conjecture about the experiments was that since the less well performing players would prefer to ask the advices of better performing individuals, they would voluntarily arrange themselves into a hierarchical network. Therefore, out of the nine games, two were so-called control experiments, where the advices provided were randomized thus eliminating apparent differences in performance between players.

It is worth mentioning that although the resulting structures in the real and control experiments were rather different, the majority of the players did not notice any difference between the true and the scrambled games/guesses. This is likely to indicate the existence of a characteristic time that is necessary to discover differences in performance; otherwise the social structure remains random.

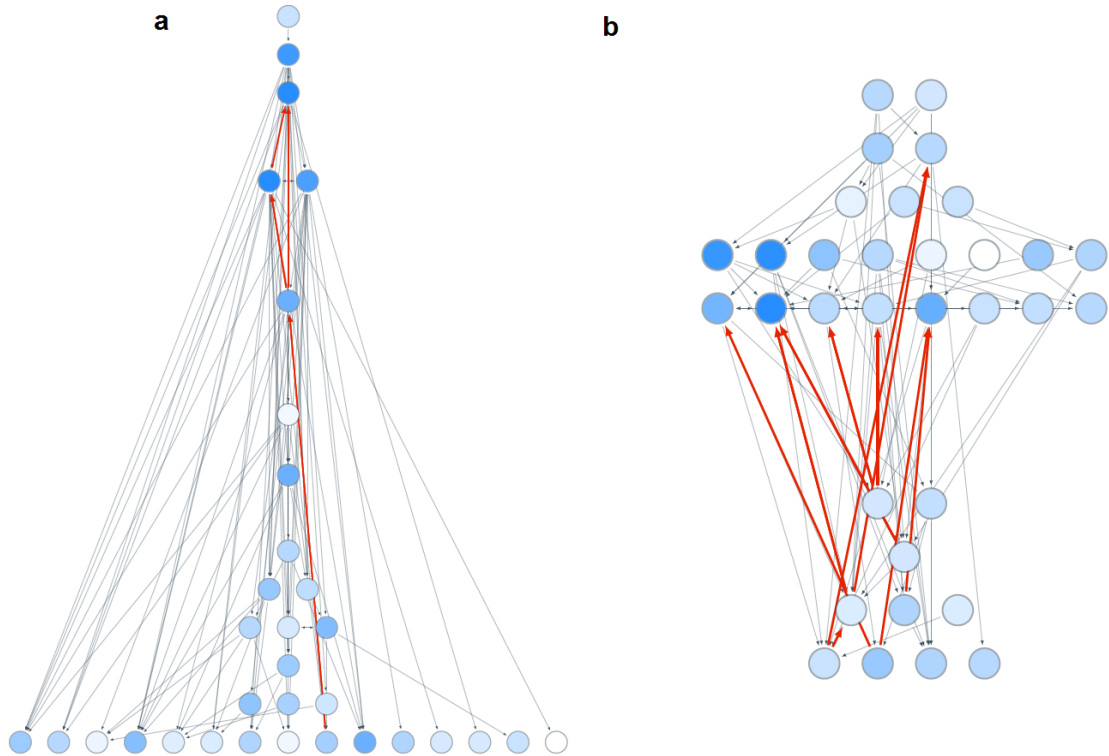

**Fig. 4.5** Hierarchical structure of the leader-follower relationships. Two illustrative examples of the structures that emerged during the game as players queried advice from others: **a** Game 6 (regular experiment) and **b** Game 9 (control experiment). Colour intensity indicates the performance of the users, and edges pointing upward in the network are drawn in red colour. As the structure of the consecutive levels indicates, the network of influences in the regular experiments features hierarchical characteristics in contrast to the one found in the control experiment. Visualization was made according to Sect. 2.2.2.

Based on the interviews made after the 9 sessions had been over, it became clear that there had been three main stages of decision making in Picturask. First, the player makes an initial guess, then he/she collects information from others, at the end of the 1 minute given for an answer the player makes a final decision using the information collected from others. It is important to separate information gathering from the decision making process.

The only information a player new about the others was the colour of the tile corresponding to the ratio of correct advices by the person represented by the tile. Even if the players did not understand what the exact meaning of the tile was, they used it as a heuristic: linking good performance with the colour of the tile. Thus, the colour had two functions: in most cases, it indicated the pool of possible advice givers and the order of asking advice (the darker is better), but it also could give more weight to the advice of some players.

## *Reference List*

Cropanzano R, Mitchell MS (2005) Social exchange theory: An interdisciplinary review. J. Manag 31(6): 874-900

Johnson DW, Johnson RT (2005) New developments in social interdependence theory. Genet Soc Gen Psychol Monogr 131(4):285-358.

Liska T (2006) The Liska Model. http://www.liska.hu/fliska/seconomy.htm. Accessed 12 Jan 2016

Liska T (2008) The Liska model. Societly and Economy 29(3):363-381. Doi: 10.1556/SocEc.29.2007.3.5

Liska T (2011) Experimental Economy. http://www.liska.hu/fliska/experiment.htm . Accessed 19 May 2017

Mones E, Liska T, Vicsek L and Vicsek T (2013) Leader-follower relationships in a Liskaland camp in 2012. Unpublished

Mones E, Tóth B, Havadi G, Pollner P, Vicsek T (2015) Picturask: an on-line game for studying the spontaneous emergence of leader-follower relationship network in a group of humans. Working paper.

Nepusz T, Vicsek T (2013) Hierarchical Self-Organization of Non-Cooperating Individuals. PLoS ONE 8(12):e81449

# 5. Modelling emergence and control

In this Chapter we shall consider some of the dynamical aspects of hierarchical systems as obtained from simulations of the related models. Although the emergence of hierarchy and the optimal ways of controlling the processes in a hierarchical structure represent two of the most relevant aspects of the subject of hierarchy, the related results are far from being complete. Further research in these directions is of essential importance.

*Relation to game theory:* In the related works emergence is typically considered as a result of optimizing a quantity which is called by various names (e.g., performance, effectiveness) but is analogous to the notion of fitness. And, much like in game theory, fitness has a positive ingredient (benefits) and a negative part associated with disadvantages (costs). Although the models we discuss can be mapped onto games, our preference will be using the language and the techniques of networks, agent based modelling and statistical mechanics.

## 5.1 Emergence of hierarchy in model systems

This section is about approaches involving simple models that are capable of reproducing the emergence of multi-level network structures based on the degree to which the units (individuals) are able to contribute to the efficiency (capacity to operate on a high level) of the system. We shall adopt terminologies that are, on one hand, used in statistical mechanics and network science, while, on the on the other hand, being typically used in the context of organizations and the underlying networks of collaborations. However, we expect this framework to be applicable to a significantly larger class of systems. Thus we consider the groups of humans as a paradigm, but our approach is so general that it is expected to be applicable to simpler systems such as groups of collaborating animals (apes, wolfs, etc.) as well as complex machines constructed by people.

There are only relatively few works on how a hierarchical network structure emerges. We first briefly discuss two works that describe the emergence of networks which have undirected edges only, but can be considered to be hierarchical from the point self-similarity. Mengistu et al. (2016) investigate the changing structure of networks using evolutionary arguments. They evolved graphs which can be regarded as computational abstractions of animal brains. Such structures are commonly called artificial neural networks (ANNs) and can solve hierarchical Boolean logic problems (Fig. 5.1). Evolving the ANNs with or without a cost for network connections leads to qualitatively different results. Specifically, the experimental treatment selects for maximizing performance and minimizing connection costs (*performance and connection cost*, P&CC), whereas the control treatment selects for performance only (*performance alone, PA*)

A comparison of the evolved networks (under varying conditions) resulted in the conclusion that the P&CC networks are significantly more hierarchical, modular, than those of the P&A ones and solve significantly more sub-problems.

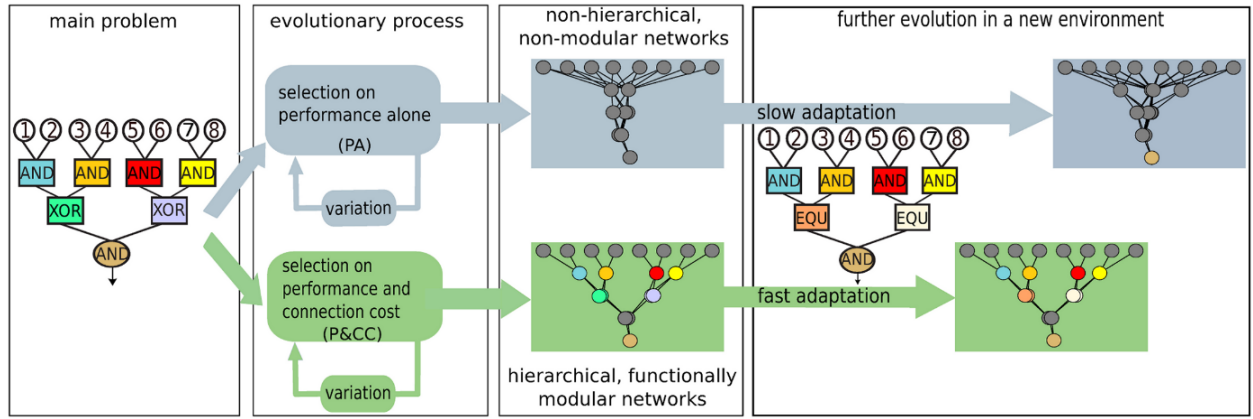

**Fig 5.1** The main hypothesis. Evolution with selection for performance only results in non-hierarchical and non-modular networks, which take longer to adapt to new environments. Evolving networks with a connection cost, however, create hierarchical and functionally modular networks that can solve the overall problem by recursively solving its sub-problems. These networks also adapt to new environments faster. Reproduced from Mengistu et al. (2016)

Lee et al. (2011) used a game theoretical type model to investigate how social structures emerge. In their approach, a number of feedback couplings from the behaviour of the agents to their environment was assumed, and this is why it could be considered as a multiadaptive game. The expressions they used were relatively simple – and somewhat arbitrary – still the resulting behaviour of the network of agents was very rich since even the strategies of the agents were evolving as a function of their interaction network configurations. In one of the phases the simulated networks had the scaling of the distribution $P(k)$ of the degrees $k$ of the nodes as well as the clustering coefficient $C(k)$ as a function of the degree of the nodes. Here $C(k)$ is the average of the ratio of the triangles around a given node to the total number of potential triangles). Such a simultaneous scaling was showed by Ravasz et al. (2002) to correspond to a self-similar structure. Since the edges were undirected we consider this type of self-similarity as a less pronounced manifestation of hierarchy then the one based on directed flows.

Now we turn to discussing directed hierarchical networks. Corominas-Muntra et al. (2013) do not discuss realistic criteria leading to emergence, but their work is still interesting from the point of our subject. They consider a set of possible mechanisms leading to a very wide set of potentially realizable hierarchical structures with directed edges. Four major kinds of hierarchies were identified by analysing the large voids in the morphospace defined by the authors. Two of them matched those structures what were expected from random networks with similar connectivity, thus suggesting that nonadaptive factors were at work. Ecological and gene networks define the other two domains, indicating that their topological order is the result of functional constraints. We presented more features of these results in a different context in Sect. 2.1.3.

Next we describe in more detail a model which was designed to investigate how advantageous leader-follower relations result in the emergence of hierarchy in the presence of a changing environment (Nepusz and Vicsek 2013)

The assumptions of the model were the following:

i) A group of people is embedded in a changing environment. Better adaptation (by which we mean the ability of finding out about the new conditions as quickly as possible) represents one of the core advantages an individual can have.

ii) The individuals possess different abilities to hold on in such a constantly varying environment on their own,

iii) The actors constantly monitor the decisions of their group mates which observations alter their own decisions. The effect on their own decisions is proportional to the degree to which they trust the judgment of that certain other group member, as compared to their own level of competence. The degree of trust is dependent on the prior success of the observed group mates.

iv) Maintaining a connection with another group member has a cost (effort).

Once the above assumptions are integrated into a set of rules that are corresponding to a game theoretic-like, stochastic model, a collaboration structure emerges in which the leader-follower relationships manifest themselves in the form of a multi-level, hierarchical network. This network is at the same time both stable and sensitive to the changes in the environment, according to which it is capable to re-wire itself in a dynamic fashion. Omitting any of the above four assumptions leads to the loss of the emergence of the multi-level, hierarchical structure.

The main steps of the decision making process are:

1. The state of the environment is simply given by a value of either 1, 2 …or *l*. After the state assumes one of these values it sticks to it and changes to a randomly selected other state only with a probability *p*. where *p* is in the range of 0.05-0.2.

2. The actors have different abilities to find out the state of the environment (that is, to "adopt" to the environment), described by a pre-defined parameter taking values from the [0, 1] interval according to some distribution. In each turn, the guess of each individual is based on the weighted average of (i) its own estimation, and (ii) its interactions with the other *k=1, 2,…,m* most trusted other actors. (*k* typically ranges between 2 and 7).

3. In each round, after each agent has finished with the decision making, the actual state of the environment is revealed, letting the actors finding out which one of them has made the correct decision and which one of them made the incorrect estimate.

4. Based on the above information the so called 'trust matrix' (*T(i,j)*)is updated, in which the values correspond to the degree to which actor *i* trusts actor *j*. This trust is proportional to the number of rounds agent *i* made use of the estimate of agent *j* in a way that the guess of agent *j* contributed positively to the guess of agent *i*. Accordingly, the trust-level of an individual is based on his/her prior performance. Agents that are more trusted are "listened to" more frequently.

By iterating the above steps, the dynamics of the system typically converges to a trust matrix in which the values depend on the original abilities of the actors in a non-trivial way. A typical run starts with a uniform trust matrix (except for the values in the diagonal positions) converging to a state that corresponds to a much better performing set of interactions and a hierarchical structure (See Fig. 5.2).

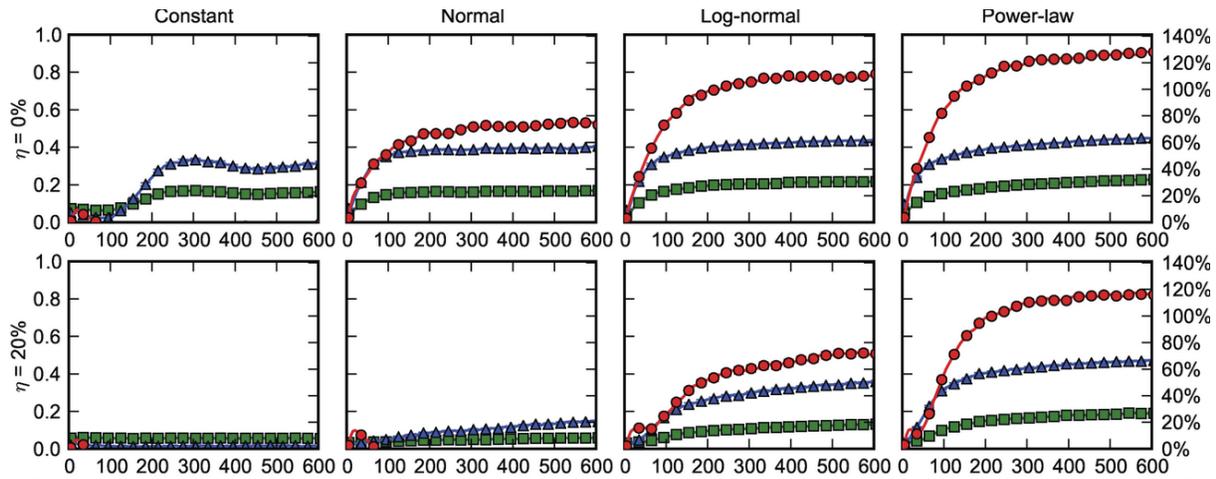

**Fig. 5.2** The dynamics of the model as a function of time and noise for various ability distributions. Each column represents a different ability distribution with the same mean and a variance. The distributions are: (i) constant,(ii) normal, (iii) log-normal and (iv) power-law respectively. The upper row corresponds to the noiseless case; whereas the middle one corresponds to 20% relative noise, where noise stands for randomly perturbing the decisions directly following from the rules of the model. The middle and the bottommost curves correspond to two different hierarchy measures: (1) fraction of forward arcs and (2) global reaching centrality (GRC). In both cases the hierarchy level is expressed as numbers between 0 and 1, where 0 corresponds to "no hierarchy", and 1 corresponds to "maximal hierarchy". The topmost lines (small red circles) show the improvement of the overall performance. Reproduced from Nepusz and Vicsek (2013).

To create a graph from the trust matrix the following procedure is made: each agent is a node in the graph and the weight of the edge (reflecting how much actor *i* trusts actor *j*) is the element (*i,j*) in the matrix. Only the strongest ties are taken into account in the network. The resulting graphs have a hierarchical structure with multiple levels in them (Fig. 5.3). These structures emerge in time, as depicted in Fig. 5.2 and, in which two complementary hierarchy measures are shown: the 'global reaching centrality', GRC, proposed by (Mones et al. 2012), and the normalized fraction of forward arcs, defined by (Eades et al. 1993). Both measures are discussed in detail in Sect. 2.1.2.

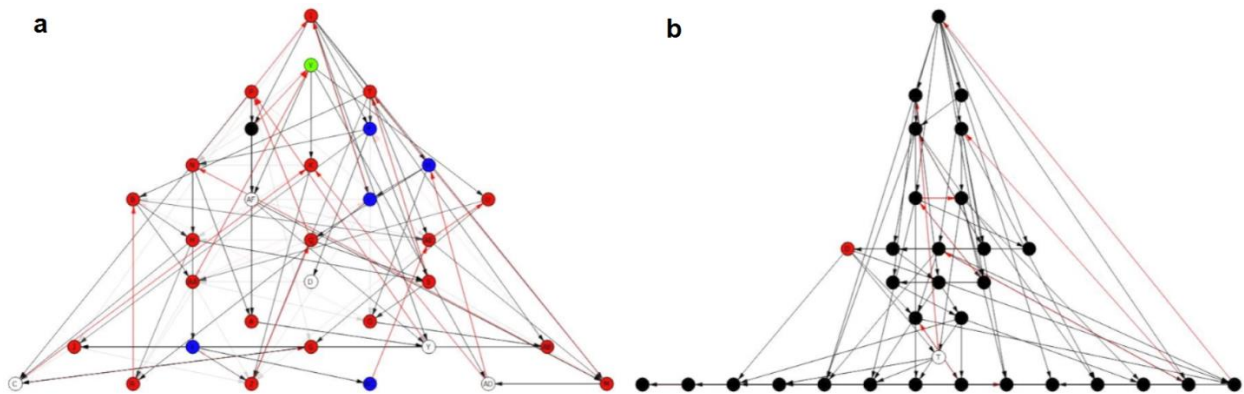

**Fig. 5.3** An early, less stable state of the emerging hierarchical network (**a**) and a more stable (**b**), persistent solution corresponding to the local maximum of performance of the network showing the nodes copying decisions (indicated by arrows) of the other members of the group. The numbers (environment) to be guessed are denoted by colours. Network (**b**) is reminiscent of the experimental result displayed in Fig 4.1. **a**.

Thus, the actors in the model show a strong tendency to structure themselves into a multilevel hierarchical organization that – apart from being a commonly seen, is an "intuitively natural" form of self-organization – which has recently gained support from a human experiment as well, called "Liskaland" (see Sect. 4.1)

## 5.2 The complex efficiency landscape of hierarchical organizations

In this section the emergence of optimal network structures is discussed using an approach which is reminiscent of the one introduced by statistical physicists in order to interpret complex systems using relatively simple rules of units and interactions. The original – so called spin-glass – approach assumes spin states (up or down) in the nodes of a network. The interaction along the edges is randomly ferromagnetic or antiferromagnetic and the configuration of the network is fixed and ranges from lattices (Edwards and Anderson 1975), through scale-free (Kim et al. 2005) to full graph (Sherrington and Kirckpatrick 1975).

The essential new feature of the treatment we describe below is that instead of optimizing by looking for a locally optimal state of the spins in the nodes of a pre-defined network, optimal networks are searched with the states of the nodes being fixed. Thus, the approach represents searching for extrema – as a function of the underlying network topology – in the complex efficiency landscape. In addition to the above (ferromagnetic or antiferromagnetic links), in contrast to spin-glass models, the edges in the underlying network of interactions have a direction.

The sign associated with an edge corresponds to collaboration (positive) or antagonistic (negative) relations. Searching for optimal states then is carried out by modifying the network topology so that both the collaborating partners (within an organization) and the flow of influences result in a maximal efficiency.

Within such an approach it is possible to address the question of the spontaneous emergence of hierarchical networks displaying behaviours some of which are analogous to those of glasses. By glassy behaviour we mean that while we are searching for a stable state, our structures do not converge to a unique network with a well-defined extremal value of their efficiency (an analogue of the energy in the physics literature). Instead, glassy systems "freeze" into various disordered structures representing local extrema full of strains or frustrations.

### 5.2.1 Modelling organizations

The relations in an organization can be represented by a network made of *directed edges* corresponding to the leader-follower relations in the system. In this approach the ability of member $i$ to contribute to the effectiveness of the organization is denoted by $a_i$. In an ideal case the direction of an edge between members $i$ and $j$ would point from $i$ to $j$ if $a_i > a_j$ (it is advantageous and is typically indeed the case that agents with higher abilities can enforce their decisions on agents with smaller abilities, i.e., occupy a higher position within the organizational hierarchy). However, with some finite probability, in a realistic case a proportion $p$ of all of the links between two members points from the less knowledgeable to the more knowledgeable person (Zamani and Vicsek 2017).

Next, it is assumed that on an absolute value scale the contribution of the members is between 0 and 1. In addition, the joint contribution of two members $E_{eff,ij}$ is linearly proportional to their abilities, i.e., $E_{eff,}(i,j)= a_i\, a_j$. However – and this is an essential point, when one considers the relations of sophisticated creatures – the interaction between two individuals can be both harmonic and antagonistic with probabilities (1-$q$) and $q$, respectively. In the "harmonic" case, the contribution of the two members is positive, on the other hand, if they are in an antagonistic relation their interaction will result in a decrease of the total efficiency, thus their interaction enters the expression for the efficiency as negative contribution.

Assuming that the total performance of the organization can be represented as the contribution of the pairwise interactions it follows that

$$E_{eff}(p,q) = 1/N \sum_{ij}^{M} J_{ij}(p,q)a_i a_j \qquad (5.1)$$

with the summation running over nodes that have at least one incoming or outgoing edge. (Remark: in the original publication $1/N$ was used for normalizing the efficiency, however, $1/M$ is a more appropriate quantity for this purpose.) According to the above arguments about the possible relations between two interacting members, we assume that $J_{ij}$ can be equal to 1 or -1. For the $a_i$ values it is quite natural to use randomly generated numbers on the unit interval following a bounded *log-normal distribution* (which can be argued to be characteristic for the outputs of complex entities).

The sign of $J_{ij}$ and the direction of the edge $ij$ are decided by two factors: 1) whether the $ij$ edge points from the larger to the lower ability of the participants $i$ and $j$ and 2) whether these participants are compatible or antagonistic. Thus, (with the corresponding probabilities)

i) $J_{ij}$ =1 if the $ij$ edge points from a node with larger ability to a smaller ($a_i > a_j$ and the two individuals cooperate (and $J_{ij}$ = - 1 otherwise)

ii) $J_{ij}$ = - 1 if the $ij$ edge points from a node with smaller ability towards a larger one and the two individuals are antagonistic (and $J_{ij}$ = 1 otherwise)

iii) If there is no edge between $i$ and $j$ then $J_{ij}$ = 0.

iv) A further essential restriction has to be taken into account to make the system more realistic (much like as it was described in Sect. 5.1). In addition to the above, it should also be required that the total number of edges of a node cannot exceed a pre-defined value.

## 5.2.2 Simulations and results

The numerical experiments start out with a full graph with $N$ nodes each associated with a constant ability $a_i$ and with edges pointing towards lower ability sites from larger ability ones with a probability 1-$p$. In addition, 1 or - 1 is associated with each edge, independent of their direction (however, because of the term 1-*p*, the number of negative edges for small *p* will occur in larger overall number for the $a_i > a_j$ cases (than for $a_i < a_j$) so that the efficiency values and the structure of the graph become coupled.

The next step is searching for locally optimal networks. This will be a particular subgraph containing *M* edges within the full graph of *N* nodes, where *N* is the total number of possible members and *M* is the actual one. The initial configuration is a random connected subgraph of 3*N* edges. Throughout the calculations, the number of edges within the subgraphs satisfy the criteria that *in average* the ratio of edges in them pointing from a site with larger to a smaller ability will be equal to 1-*p* and the number of antagonistic interactions $J_{ij} = -1$ will be, again, in average, *qM* ( for $a_i < a_j$). The searching is much like a Monte Carlo simulation, where efficiency plays the role of energy times -1. In each step a randomly selected edge is eliminated and next two random nodes are chosen which are not yet connected by one of the *M*-1 edges. The sign and the direction of the new edge, $J_{ij}$ is chosen according to conditions i)-iv) outlined in the Sect. 5.2.1. A randomly generated new edge is accepted if it results in an increasing efficiency and is also accepted with a small probability of it decreases the efficiency

Repeating the procedure results in a distribution of the individually obtained locally optimal efficiencies. The corresponding histogram (probability density function – PDF) can be constructed for various *N* values to see the size effect. The properties of the networks corresponding to the locally optimal states were also investigated from the point of their hierarchical nature using the GRC measure (see Fig. 5.4).

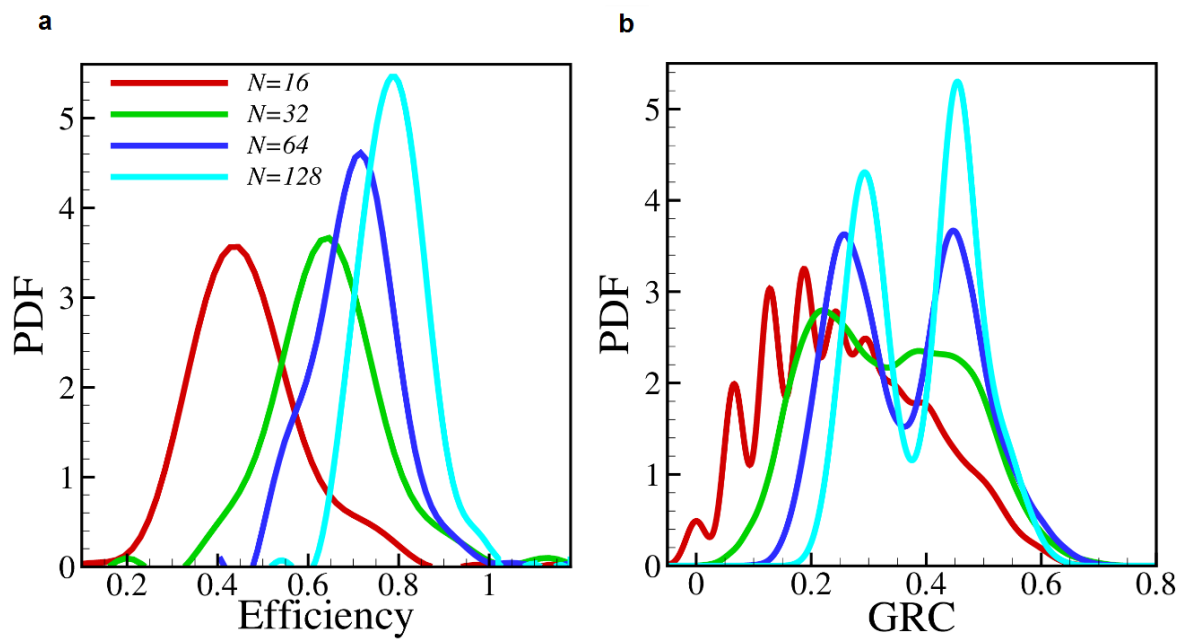

**Fig. 5.4** Distribution of the local maxima of the efficiency values (**a**) and Global Reaching Centrality (GRC) values (**b**) for the locally optimal states. Averaging over the initial full graph (250 initial full graphs and for each initial full graph we have 250 local optimal states) of *N* nodes and the initial subgraphs of 3*N* edges and for *p*=*q*=0.2 was carried out. There is an overall tendency of the PDF-s as a function of the system size. GRC and the average efficiency grows with increasing *N*. Reproduced from Zamani and Vicsek (2017).

In Fig. 5.4 **a** few characteristic dependences of the related networks are depicted. Fig. 5.4 **a** shows that larger systems are likely to be more efficient, while Fig. 5.4 **b** shows that for larger networks the optimal configurations seem to fall into two classes with one having a smaller and another one a distinctly more hierarchical structure. In order to illustrate the variety of the optimal structures, in Fig 5.4 a number of typical examples are shown. These include smaller and larger networks ($N$=16, $N$=128), networks for smaller or larger GRC for $p$=$q$=0.2. For visualization we use the method described in detail in Sect. 2.2.2.

According to the above results, the structures of the obtained networks (Fig 5.5) are such that they possess the two, perhaps most important features of complex systems: a simultaneous presence of adaptability and stability. Stability is associated with the presence of a local optimum. Only significant perturbations can "kick out" a given arrangement of the participants from this favourable state. However, if the perturbation is large enough (the external conditions change significantly) the network can adapt itself and settle into an alternative, more optimal configuration that suits the new conditions better. The efficiency of the hierarchical structure is higher than a randomly chosen sum of the contributions of the pairwise interactions. These features are in an analogy with those of the glasses including spin glasses.

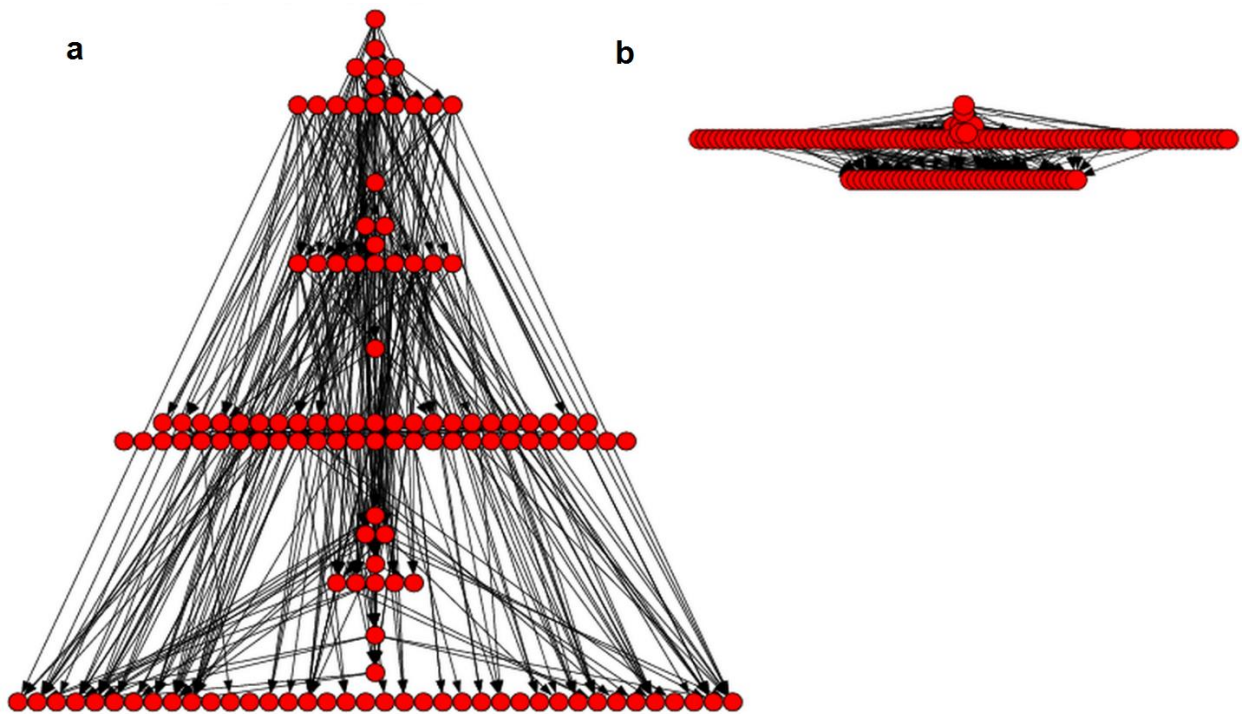

**Fig. 5.5** Hierarchical graphs in selected local optimal states of networks. Number of nodes is $N$=128, the other two parameters are $p$=$q$=0.2, **a**: $GRC$ = 0.62, **b**: $GRC$=0.25. Reproduced from Zamani and Vicsek (2017).

### *5.3 Controlling hierarchical networks*

In this section we shall overview two approaches from a field which – although very important – for some reason have been overlooked by researchers in the last decades: the *controllability properties* of complex networks. This field is about the study of the conditions under which a network can be driven from any initial state into any final state within finite number of steps.

This definition implies that the nodes have a state, for example the amount of traffic flowing through a node in, say, a traffic or communication network, or the transcription factor concentration in a gene regulatory graph. The question is that how and where one has to intervene in order to drive the system into a desired (pre-defined) condition.

### 5.3.1 Structural controllability – controlling nodes

The first approach, proposed by Liu et al. (2011), investigates the controllability properties of weighted directed networks. The main idea is to identify a set of the so called *driver nodes* (a set of vertices through which the dynamics of the entire system can be controlled). The nodes of the network are assumed to behave according to non-linear processes, but their behaviour is approximated by the following linear dynamics:

$$\frac{d\vec{x}(t)}{dt} = A\vec{x}(t) + B\vec{u}(t) \tag{5.2}$$

where $\vec{x}(t) = (x_1(t), \dots, x_N(t))^T$ is the state vector of the $N$ nodes (in which $x_i(t)$ describes the state of node $i$ at time $t$), $A$ is the $N{\times}N$ adjacency matrix, capturing the interaction strength among the elements of the systems (which are the nodes of the graph), and $B$ is an $N{\times}M$ matrix defining the *driver nodes*: these are the vertices which are to be controlled from the outside in order to drive the system to the desired state. Finally, $\vec{u}(t) = (u_1(t), \dots, u_M(t))^T$ is the time-dependent input signal, the vector controlling the system. The justification of this approximation is that according to Slotine and Li (1991), the controllability of a nonlinear system is in many aspects *structurally similar* to that of a linear system.

A dynamics $(A, B)$ is said to be "structurally controllable" if it is possible to choose the non-zero elements in $A$ and $B$ in a way that the network can be driven from any state to any other final state by appropriately choosing the elements of $\vec{u}$. This property, structural controllability, is important, because in real-life complex systems the weights of $A$ (the link weights of the network) are usually unknown, or just partly known. A structurally controllable system can be shown to be controllable for almost all weight combinations, thus this property helps to overcome the incomplete knowledge of the link weights in $A$. Then, the minimum number of driver nodes is determined by the *maximum matching* in the graph, which is a maximal set of links that do not share start or end vertices. A node is *matched*, if an edge in the maximum matching points at it, otherwise it is *unmatched*. Then, from here, the task is basically solved, because according to Yu el al (2010), full control can be gained over a directed network if and only if each unmatched node is directly controlled and there are directed paths from the input signals to all matched vertices.

The key results of this study are the following: (See also Table 5.1)

- The number of driver nodes within a complex network is mainly determined by the network's degree distribution.
- Sparse inhomogeneous networks (a type of graph very often seen in relation to real-life complex systems, often the ones that have evolved to control another underlying process such as a transcriptional regulatory network) are the most difficult to control (they need many input signals.)
- In contrast, dense and homogeneous graphs need only a few driver nodes in order to be controlled
- And finally, the most counterintuitive result is that the driver nodes tend to avoid the high-degree nodes ("hubs"), both in model and real-life systems. In other words, control signals control the hubs only indirectly, which is, according to Nepusz and Vicsek (2012), due to the fact that in this approach the driver nodes are not able to control their subordinate vertices independently from each other.

Importantly, these results apply for linear nodal dynamics.

## 5.3.2. Switchboard dynamics – controlling edges

Nepusz and Vicsek (2012) proposed a dynamics that takes place on the edges, instead of the nodes, and leads to significantly different controllability properties for the same real-life networks. The motivation is visualized in Fig. 5.6 demonstrating that hierarchical networks – in case the nodes are directly controlled by one of their neighbours or by driver nodes – need a disproportionally large number of driver nodes. In this model the state variables correspond to the edges of a directed complex network, and the vertices of the network act as linear operators that map state variables of inbound edges to outbound edges. It is called *switchboard dynamics*, exactly because of this property: each node acts as a "switchboard-like device" mapping the signals coming from the inbound edges to the outbound edges, by applying a linear operator called mixing, or switching matrix. That is, each node has a separate switching matrix, $\boldsymbol{M}_i$, enabling the nodes to control their subordinates separately.

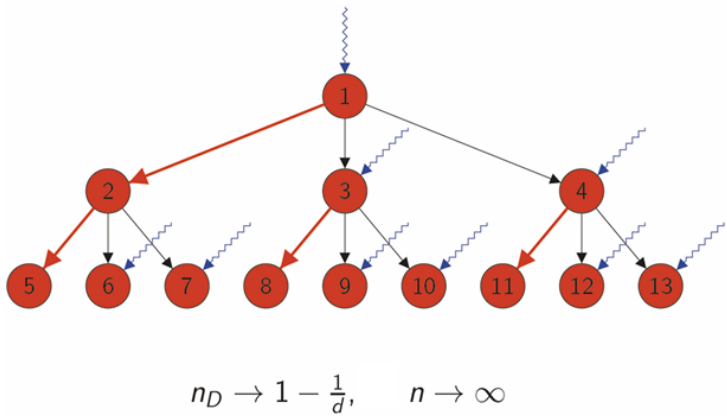

$$n_D \to 1 - \frac{1}{d}, \qquad n \to \infty$$

**Fig. 5.6** When node controllability is considered, a given node can influence the state of only one of its neighbours. This needs a relatively large number of nodes to be controlled from "outside" (denoted by wavy lines). Thus, an alternative approach may be more efficient. Reproduced from Nepusz and Vicsek (2012).

This arrangement is reminiscent of the characteristics of many real-life networks in which each node constantly processes information coming from its inbound edges and forwards them in a differentiated way via its outbound edges. A plausible example can be an arbitrary social communication network in which the nodes are the persons who constantly receive and forward messages, but in ways depending on the recipient.

In this switchboard dynamics framework (SBD), each *edge* has a *state*, which is denoted by the $\vec{x} = [x_j]$ state vector. For each $i$ node belongs a $\vec{y}_i^+$ and $\vec{y}_i^-$ vector pair, consisting of those $x_j$ edge-state values that correspond to the incoming and outgoing edges, respectively, of node $i$. For example in Fig. 5.7 **a**, the inbound edges of node $i$ are $c$ and $d$, whose states are defined by the values of $x_c$ and $x_d$, whereas its outbound edges, $e$, $f$, and $g$, are in the state described by the values $x_e$, $x_f$ and $x_g$, respectively. The switching matrix of this node, $M_i$, has three rows (out-degree, number of out-going edges) and two columns (in-degree, number of incoming edges). The dynamics is controlled from the outside by adding an *offset vector* (or *control signal*) $\vec{u}_i$ to the state vectors of the outgoing edges of node $i$, marked with red undulate arrows on Fig 5.7 **a**.

The dynamics of the network is described by the following equation:

$$\frac{d\vec{y}_i^+}{dt} = M_i \vec{y}_i^-(t) - \tau_i \otimes y_i^+(t) + \sigma_i \vec{u}_i(t) \qquad (5.3)$$

where $\sigma_i$ is 1 if node $i$ is a driver node (see Sect. 5.3.1), and 0 otherwise. The vector $\tau_i$ includes damping terms corresponding to the edges in $\vec{y}_i^+(t)$, and finally $\otimes$, denotes the entry-wise product of two vectors being of the same size.

**Fig. 5.7 a** The dynamics of the system in the switchboard dynamics framework is controlled by adding an *offset vector* $\vec{u}_i$ to the state vectors of the outgoing *edges* of node $i$ (marked with red undulate arrows). In case this vector is not a null-vector, node $i$ is a *driver node*. **b** The state of an arbitrary edge $j$ originating in node $r$ and terminating in vertex $s$ depends only on itself, $x_j$, and on the states belonging to the inbound edges of node $r$, that is, on the set $\vec{y}_r^-$.

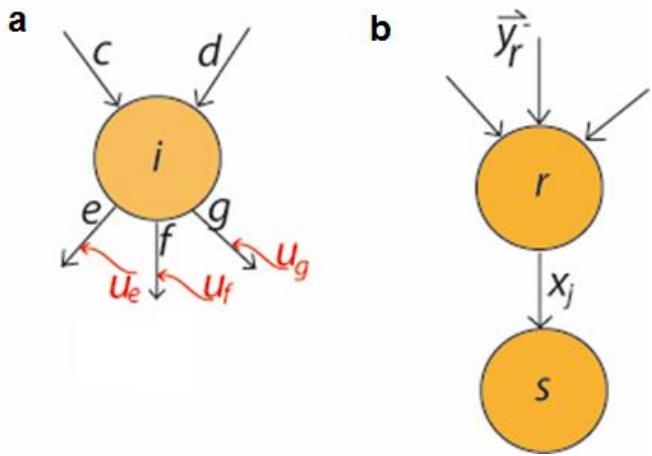

In order to simplify the equation, the state variables and control signals are implicitly considered as time-dependent, even if the time variable is omitted. By re-writing (5.3) in terms of $x_i$, the dynamics of the system yields a more simplified form (5.4):

$$\frac{dx_j}{dt} = \sum_{k \in \Gamma_j^-} w_{kj} x_k - \tau_j x_j + \sigma_s u_j \qquad (5.4)$$

where $w_{kj}$ is an element of the switching matrix $\boldsymbol{M}_r$ belonging to node $r$ defining the information process between the inbound edge $k$ and the outbound edge $j$. Note that the set $\Gamma_j^-$ includes all the state variables on which the derivative of the state variable $\dot{x}_j$ depends on, since it is effected only by itself, $x_j$, and on the states of the edges ending on node $r$, that is, on $\vec{y}_r^- \colon \Gamma_j^- = \{\vec{y}_r^-, x_j\}$.

By defining all the values in $\boldsymbol{M}_r$ which do not affect the state of $x_j$ as zero (that is, those $w_{kj}$ values which are not in the set $\Gamma_j^-$), we get (5.5)

$$\dot{x} = (W - T)\vec{x} + H\vec{u} \qquad (5.5)$$

where the $\boldsymbol{W}$, $\boldsymbol{T}$ and $\boldsymbol{H}$ matrices are:

- $\boldsymbol{W}=[w_{kj}]$, where $w_{kj}$ can be non-zero if and only if the end-point of edge $k$ is the staring node of edge $j$.
- The diagonal matrix $\boldsymbol{T}=[\tau_{jj}]$ contains the damping terms related to the edges, and
- $\boldsymbol{H}$ is also a diagonal matrix in which the $j$th diagonal element is $\sigma_{s,}$ if edge $j$ originates in node $s$.

(5.5) basically describes a simple linear time-invariant dynamical system in which $\boldsymbol{W}$ is the adjacency matrix of the line graph $L(G)$ of the original graph $G$. This means that each node in $L(G)$ corresponds to an edge in $G$, as it is demonstrated on an example graph on Fig. 5.8 **a** and **b**.

By applying the *maximum matching theorem*, in the spirit as it was done in the structural controllability framework by Liu et al. (2011), we get a set of *control paths and driven nodes* in the line graph $L(G)$. Note that these are at the same time *driven edges* in the original graph $G$, that is, a set of edges whose state should be modified in order to gain control over the network. Since edges can only be controlled from the nodes they are originated, this set of edges define the set of *driver nodes* as well: these are the vertices from which at least one driven edge originates from. As it is proven in Nepusz and Vicsek (2012), the *minimal* set of driver nodes in a graph $G$ can then be determined by selecting those vertices in $G$ for which $d_v^+ > d_v^-$, and one arbitrary vertex from each '*balanced component*'. (Balanced component is a connected component consisting only of nodes for which $d_v^+ = d_v^-$, and above this, contains at least one edge.)

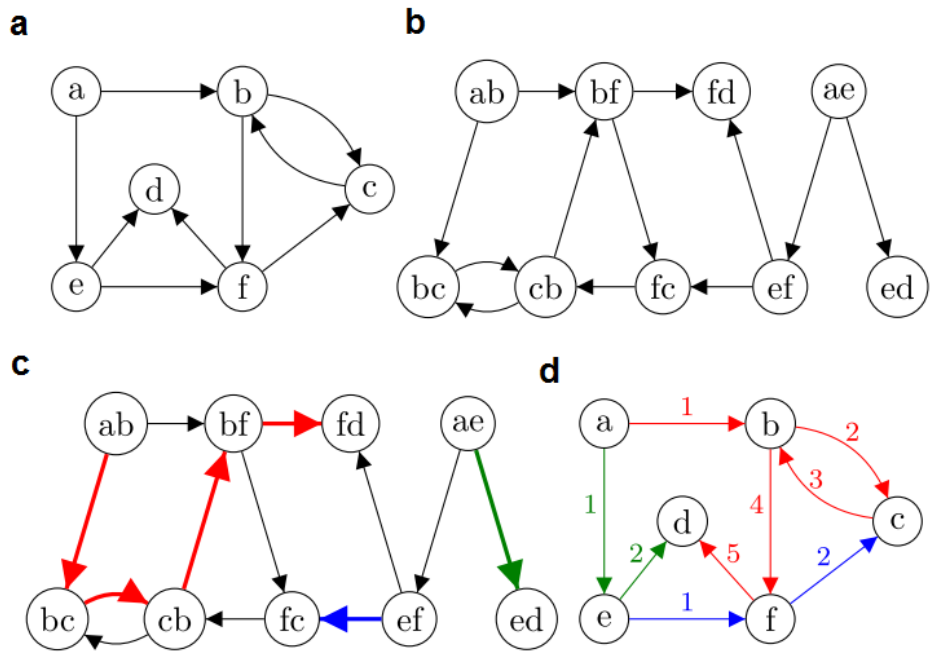

**Fig. 5.8** Demonstration of how the switchboard dynamics approach pinpoints the driver nodes. **a** a simple input network $G$ with six nodes and nine edges. The control applies to the edges of the network, instead of the nodes. **b** The line graph $L(G)$ corresponding to $G$. A linear time-invariant dynamics on the nodes of $L(G)$ is equivalent to the switchboard dynamics in $G$. Node labels refer to the endpoints of the edges in $G$. **c** The maximum matching theorem applied to $L(G)$ returning disjoint control paths. **d** The control paths in $G$, inferred from the results yielded on $L(G)$. Note how each path in the line graph $L(G)$ became an edge-disjoint walk in $G$. Numbers on the lines represent the order in which the edges have to be traversed in the walks. The two driver nodes are **a** and **e** since each walk starts from either one of them. Reproduced from Nepusz and Vicsek (2012)

The minimum number of driver nodes (nodes that are driven from the outside in order to gain control over the entire network) is found to be largely determined by the joint degree distribution of the network.

The following survey of 38 real-life networks, shown in Table 5.1, reveals that under this approach, transcriptional regulatory networks are well-controllable with a small number of driver nodes and also that most real-world networks are easier to control than random Erdős–Rényi networks with the same number of nodes and edges (last 3 columns).

Note that this is in deep contrast with the findings of Liu et al. (2011), who have found that regulatory networks need a high fraction of driver nodes and that randomized Erdős-Rényi networks are easier to control than the real-world ones.

The differences are very spectacular in highly hierarchical, tree-like networks as well, in which the presence of central out-hubs rapidly increase the required number of driver nodes within the framework of Liu et al., while the same out-hubs can efficiently control many subordinate nodes in the switchboard dynamics – and thus decrease the required number of driver nodes. This is a central result, since such hierarchies are ubiquitous in nature and society, from scales as small as gene regulatory networks through leader-follower relationships, up to large-scale organizations.

**Table 5.1** Controllability properties of real networks. First five columns: (1) type of the network, (2) its number, (3) name, (4) number of nodes and (5) number of edges, respectively. 6$^{th}$ column: $n_D^{SBD}$, fraction of driver nodes in the switchboard dynamics framework. 7$^{th}$ column: $n_D^{Liu}$, fraction of driver nodes in the structural controllability framework, (overviewed in Sect. 5.3.1), and 8$^{th}$ column: $n_D^{ER}$, fraction of driver nodes in the switchboard dynamics framework in randomized networks using the Erdős–Rényi model and $n_D^{Dgr}$ the degree-preserving configuration model. References to the real systems data can be found in Table 1. and the reference list of the Supplementary Material to Nepusz and Vicsek (2012).

*: Networks in which the edges have been reserved compared to the original publication.

†: Results calculated from the degree distribution. From Nepusz and Vicsek (2012).

| Type | # | Name | Nodes | Edges | $n_D^{SBD}$ | $n_D^{Liu}$ | $n_D^{ER}$ | $n_D^{Dgr}$ |
|---|---|---|---|---|---|---|---|---|
| Regulatory | 1 | Ownership-USCorp | 7,253 | 6,726 | 0.160 | 0.820 | 0.339 | 0.085 |
| | 2 | TRN-EC-2 | 418 | 519 | 0.222 | 0.751 | 0.366 | 0.148 |
| | 3 | TRN-Yeast-1 | 4,441 | 12,873 | 0.034 | 0.965 | 0.415 | 0.033 |
| | 4 | TRN-Yest-2 | 688 | 1,079 | 0.177 | 0.821 | 0.381 | 0.137 |
| Trust | 5 | Collage* | 32 | 96 | 0.344 | 0.188 | 0.418 | 0.315 |
| | 6 | Epinions* | 75,888 | 508,837 | 0.336 | 0.549 | 0.445 | 0.448 |
| | 7 | Prison* | 67 | 182 | 0.403 | 0.134 | 0.411 | 0.451 |
| | 8 | Slashdot* | 82,168 | 948,464 | 0.323 | 0.045 | 0.458 | 0.392 |
| | 9 | WikiVote* | 7,115 | 103,689 | 0.281 | 0.666 | 0.463 | 0.620 |
| Food web | 10 | Grassland | 88 | 137 | 0.318 | 0.523 | 0.381 | 0.297 |
| | 11 | Little Rock | 183 | 2,494 | 0.639 | 0.541 | 0.463 | 0.649 |
| | 12 | SeaGrass | 49 | 226 | 0.449 | 0.265 | 0.436 | 0.433 |
| | 13 | Ythan | 135 | 601 | 0.304 | 0.511 | 0.432 | 0.337 |
| Metabolic | 14 | C. Elegans | 1,173 | 2,864 | 0.182 | 0.302 | 0.409 | 0.309 |
| | 15 | E. coli | 2,275 | 5,763 | 0.182 | 0.382 | 0.409 | 0.309 |
| | 16 | S. cerevisiae | 1,511 | 3,833 | 0.185 | 0.329 | 0.409 | 0.313 |
| Electronic circuits | 17 | s208a | 122 | 189 | 0.451 | 0.238 | 0.381 | 0.431 |
| | 18 | S420a | 252 | 399 | 0.456 | 0.234 | 0.385 | 0.440 |
| | 19 | S838a | 512 | 819 | 0.459 | 0.232 | 0.381 | 0.442 |
| Neuronal and brain | 20 | C. elegans | 297 | 2,359 | 0.549 | 0.165 | 0.449 | 0.499 |
| | 21 | Macaque | 45 | 463 | 0.333 | 0.022 | 0.446 | 0.457 |
| Citation | 22 | arXiv-HepPh* | 34,546 | 421,578 | 0.356 | 0.232 | 0.459 | 0.577 |
| | 23 | arXiv-HepTh* | 27,770 | 352,807 | 0.359 | 0.216 | 0.460 | 0.569 |
| WWW | 24 | Google | 15,763 | 171,206 | 0.670 | 0.337 | 0.457 | 0.612 |
| | 25 | Polblogs | 1,490 | 19,090 | 0.509 | 0.471 | 0.460 | 0.501 |
| | 26 | nd.edu | 325,729 | 1,497,134 | 0.271 | 0.677 | 0.433 | 0.301 |
| | 27 | Standford.edu | 281,904 | 2,312,497 | 0.665 | 0.317 | 0.450 | 0.653 |
| Internet | 28 | P2p-1 | 10,876 | 39,994 | 0.334 | 0.552 | 0.425 | 0.344 |
| | 29 | P2p-2 | 8,846 | 31,839 | 0.344 | 0.578 | 0.423 | 0.344 |
| | 30 | P2p-3 | 8,717 | 31,525 | 0.343 | 0.577 | 0.424 | 0.344 |
| Social communication | 31 | Twitter*† | 41.7*$10^6$ | 1.47*$10^9$ | 0.402 | - | 0.476 | 0.434 |
| | 32 | UCIOnline | 1,899 | 20,296 | 0.216 | 0.323 | 0.456 | 0.375 |
| | 33 | WikiTalk | 2,394,385 | 5,021,410 | 0.022 | 0.968 | 0.399 | 0.026 |
| Organizational | 34 | Consulting* | 46 | 879 | 0.522 | 0.043 | 0.458 | 0.460 |
| | 35 | Freemans-1* | 34 | 645 | 0.412 | 0.088 | 0.441 | 0.476 |
| | 36 | Freemans-2* | 34 | 830 | 0.588 | 0.029 | 0.439 | 0.465 |
| | 37 | Manufacturing* | 77 | 2,228 | 0.597 | 0.013 | 0.468 | 0.424 |
| | 38 | University* | 81 | 817 | 0.519 | 0.012 | 0.451 | 0.532 |

Thus, the central corollary of the above research is that the presence (or absence) of hierarchical structure appears to be an important factor in the controllability properties of large dynamical systems.

## Reference list

Corominas-Murtraa B, Goñid J, Solé RV et al (2013) On the origins of hierarchy in complex networks. PNAS 110:13316-13321

Eades P, Lin X, Smyth WF (1993) A fast and effective heuristic for the feedback arc set problem. Inf.Process.Lett. 47(6):319-323

Edwards SF, Anderson PW (1975) Theory of spin glasses. J Phys F (5):965-974

Kim DH, Rodgers GJ, Kahng B et al (2005) Spin-glass phase transition on scale-free networks. Phys Rev E (71):056115. doi: 10.1103/PhysRevE.71.056115

Lee S, Holme P, Wu Z-X, Emergent hierarchical structures in multiadaptive games, *Phys. Rev. Lett.* **106**, 028702 (2011).

Liu Y, Slotine J, Barabási AL (2011) Controllability of complex networks. Nature 473(7346):167-73

Mengistu H, Huizinga J, Mouret J-B et al (2016) The Evolutionary Origins of Hierarchy. PLoS Comput Biol 12(6): e1004829. doi:10.1371/journal.pcbi.1004829

Mones E, Vicsek L, Vicsek T (2012) Hierarchy measure for complex networks. PLoS One 7(3):e33799

Nepusz T, Vicsek T (2012) Controlling edge dynamics in complex networks. Nat Phys 8(7):568-573

Nepusz T, Vicsek T (2013) Hierarchical self-organization of non-cooperating individuals. PLoS One 8(12):e81449

Ravasz E, Somera AL, Mongru DA et al (2002) Hierarchical organization of modularity in metabolic networks. Science 297:1551-1555

Sherrington D, Kirkpatrick S (1975) Solvable model of a spin-glass. Phys Rev Lett 35 (26):1792-1796

Slotine J-J, Li W (1991) Applied Nonlinear Control. Prentice-Hall, New Jersey

Yu W, Chen G, Cao M et al (2010) Second-order consensus for multi-agent systems with directed topologies and nonlinear dynamics. IEEE Trans. Syst. Man Cybern. B 40:881-891

Zamani M, Vicsek T (2017) Glassy nature of hierarchical organizations. Sci Rep 7:1382

# 6. Conclusions

In this Chapter we summarize the main lessons one can learn about hierarchy by considering the results presented in the previous Chapters. The first point we would like to make is that we concentrated on works involving quantitative results. There is a huge literature on the vast qualitative or "narrative" interpretation of hierarchies, but the number of studies based on calculus is rather limited. As it was already mentioned in the introduction, in this book we describe studies related to hierarchy in general and the particular cases we consider have been related to that domains of nature, which can be described as assemblies of organisms that can communicate by processing information as unique individuals.

Before going into some details, we would like to point out a very general aspect of hierarchy. According to the studies we presented, hierarchy and complexity (as it is understood when the expression of complex systems is used) are intimately related. Complex systems are usually associated with many units displaying a widely varying behaviour, but, more importantly for us, a property, which is summarized by the following statement: a complex system exhibits a qualitatively different behaviour (as a whole) from that of its units. Now, this is true for hierarchical systems as well with an addition that this emergence of new qualitative behaviour can be associated with the existence of "hierarchical levels" of an underlying network in the systems.

## 6.1 General features of hierarchical structures.

Hierarchy has several manifestations. We classified these as order, embedded and flow hierarchies. The most compelling and complex of these is flow hierarchy that assumes directed or undirected interactions among its units. In order to characterise quantitatively the structure of a system having an underlying flow hierarchy is far from being trivial. This is true for the visualization of the hierarchical nature of the flow of information in a complex system. Correspondingly, in Chap. 2 we give many related details. As it turns out even the level of hierarchy is a problem that cannot be defined in a unique way. This is also true for visualizing the hierarchical nature of a system. We overviewed a number of suggestions to quantify and make hierarchy visible even for the case of flow hierarchies.

In all cases a hierarchical system has "levels". Sometimes these levels are well defined, but not always. Each level has its own behavioural patterns and may contain groups/units made of closely related organisms and separated from the other groups.

We argue that the most complex representation of a hierarchical system can be achieved by considering an underlying network of directed and undirected edges. In fact, we also conclude that in a system of organisms interacting by exchanging and evaluating information, an approach based on flow hierarchy is the most suitable and this is what we mostly consider in the book.

## 6.2. Origins of flow hierarchy

Our main observation is that complex hierarchical systems are usually implying the relevance of the flow of information. This is so in part because the units are not fully informed about their environment. The above flow can have several manifestations: for example, the less informed units copy information (either by freely provided or simply obtained by "watching), or a person on a higher level giving orders to subordinates.

Based on the observations and models presented in this book our main conclusions are that the origins of hierarchies in systems of organisms are related to two main factors:

      i) optimizing the functioning and
      ii) limits concerning the resources or costs involved.

Before going into a bit more detail, we point out that in practice both i) and ii) become dominant factors due to the incomplete flow of incomplete information. If every participant was aware of the exact information all the time, hierarchy would not be needed in most of the real-life cases.

Next, we shortly discuss point i). First of all, optimization (searching for the best performing state/structure of the system) typically involves finding a "synchronized" regime of behaviour of the agents/actors. We assume that complex systems of organisms are optimal from the point of their structure: this is due to a natural competition/selection principle in the Darwinian sense of the process. There are two possible main variants of this aspect. In the first case the individual units are "selfish" they are trying to optimize only their own advantage from the interaction with the others. In this "soft" hierarchy there is no external, global condition that would force the hierarchy to emerge. We could also associate these systems with a bottom up structure. All this can be studied both experimentally and by modelling.

The other main version of hierarchy is due to the simultaneous action of both the above points i) and ii). It usually involves an external pressure (in the context of which the optimization, i.e., i) has to take place). In this "hard" hierarchy optimizing in the presence of this pressure is the interest of the whole group of organisms and it becomes the main determinant of the hierarchy. Examples include armies or even universities. There is a global goal which has to be achieved (win the battle, educate in many areas as efficiently as possible). Such hierarchies usually involve that the direction of the ties between the units determine the behaviour of a subordinate as a function of the decision of the "boss". They are clearly organized from top to bottom.

The above points can also be approached from a more practical approach. Points i) and ii) can be best demonstrated by recalling two specific examples which, on the other hand, bear the essential features of most of the other possible examples. (A) Let us consider a system, in which the participants are trying to optimize their own performance by copying the decisions of those group mates who are better at making good decisions. Without limitations this process would inevitably lead to a star-like network with everyone following the decisions of the agent with the highest ability to make the right choices. However, if the number of connections an agent can manage (as it is in real life) is limited, the above structure cannot be maintained and rather, a cascade of information flow appears along a hierarchically organized network (also leading to a better average performance than independent decisions). (For example, if 3 is the maximum number of edges an agent can maintain, a simple tree-like structure – with one incoming and two outgoing edges – can optimally result in the best information flowing from the top to the many agents at the bottom of the hierarchy in an optimal way). (B) The simplest everyday example one can mention is that of an army (already mentioned above). If the decisions concerning many people have to be made on a short time scale, a hierarchical organization has a great advantage. Here the limiting cost is time. Imagine an "egalitarian" army fighting against a "dictatorial" one. In short: until the egalitarian army – following many rounds of discussions – make the decision which tactics to choose, the dictatorial (hierarchical) army – using a simple strategy – can override them.

## 6.3 Emergence of hierarchy

Obviously, this is one of the most interesting questions one can raise in the context of hierarchies. Most of the related studies describe historical processes without calculations. However, there exist by now a few quantitatively treatable models giving insight into the "abstracts" process of emergence.

The emergence of hierarchies can be discussed in terms of evolutionary biology and/or game theory, with a number of relevant differences. If one aims at incorporating realistic assumptions and the corresponding real-life like behaviours, than exact treatment is not possible. However, computer simulations are feasible (see, e.g., Chap 5). In addition, numerical experiments aimed at shedding light on particular aspects of emergence can be carried out which is hardly realizable in the context of evolution.

Thus, we present in the book a few models that involve such notions (borrowed from game theory and evolutionary theory) as fitness, benefits and costs. These models include specific definitions/assumptions for the above notions and give new insight into the mechanisms or the reasons why hierarchy emerges from an originally random set of connections among the units of a system.

Finally, our book, by the nature of the subject, is far from being complete. There are many more, and we expect, there will be much more works on the topic since the pool of phenomena related to hierarchy is inexhaustible. We wish that our book would stimulate works concentrating on quantitative treatments of this exciting field.